\documentclass[11pt]{iopart}

\usepackage{hyperref}
\usepackage{graphicx}
\usepackage{amssymb}
\usepackage[dvipsnames]{xcolor}
\usepackage{mathrsfs}
\expandafter\let\csname equation*\endcsname\relax
\expandafter\let\csname endequation*\endcsname\relax
\usepackage{amsmath}% to avoid conflict with iopart.cls
\usepackage[normalem]{ulem}
\newcommand{\bee}{\begin{enumerate}}
\newcommand{\eee}{\end{enumerate}}
\newcommand{\bei}{\begin{itemize}}
\newcommand{\eei}{\end{itemize}}
\def\mn{{\mu\nu}}
\def\cR{{\mathcal{R}}}
\def\tcR{{\tilde{\mathcal{R}}}}
\def\tf{{\tilde{f}}}
\def\tF{{\tilde{F}}}
\def\tFp{{\tilde{F}^{\prime}}}

\def\fcr{{f_\mathcal{R}}}

\def\boxx{\Box}
\def\mn{{\mu\nu}}

\def\p#1{{\partial}_{#1}}

\def\Om0L{\Omega_{\rm m0}^{\Lambda}}
\def\Or0L{\Omega_{\rm r,0}^{\Lambda}}
\def\Ol0L{\Omega_{\Lambda,0}^{\Lambda}}
\def\H0L{H_{0}^{\Lambda}}
\def\Omf0{\Omega_{\rm m0}^{f(R)}}
\def\Orf0{\Omega_{\rm r,0}^{f(R)}}
\def\Olf0{\Omega_{\Lambda,0}^{f(R)}}
\def\Hf0{H_{0}^{f(R)}}
\def\OmL0i{\Omega_{i,0}^{\Lambda}}
\def\Omf0i{\Omega_{i,0}^{f(R)}}
\usepackage{enumerate}
\usepackage{comment}
\usepackage{graphicx}
\usepackage{cleveref}% for referring multiple figures, tables etc

\newcommand{\multiref}[2]{\ref{#1}-\ref{#2}}
\usepackage[toc,page]{appendix}
\usepackage{import}
\usepackage{longtable}
\usepackage{pdfpages}
\let\svthefootnote\thefootnote
\newcommand\freefootnote[1]{%
  \let\thefootnote\relax%
  \footnotetext{#1}%
  \let\thefootnote\svthefootnote%
}
\def\imgtwo#1{
\begin{minipage}[b][.48\linewidth]{.48\linewidth}\centering
\includegraphics[width=\linewidth,height=1.0\linewidth,keepaspectratio]{#1}\end{minipage}}
\usepackage{afterpage}
\usepackage{longtable}
\usepackage{lscape}
\usepackage{caption}
\begin{document}

\title[Constraints on $f(R)$ models in Palatini formalism]{Investigating the accelerated expansion of the
Universe through updated constraints on viable
$f(R)$ models within the Palatini formalism}

\author{Kumar Ravi}
Ramakrishna Mission Vivekananda Educational and Research Institute, Belur Math 711202, Howrah, West Bengal, India
\ead{cimplyravi@gmail.com}

\begin{abstract}
The observed accelerated expansion of the Universe at present epoch can
be explained by some of the $f(R)$ models without invoking the existence of dark
energy or any other such exotic component in cosmic fluid. The $f(R)$ models in
Palatini formalism is relatively less explored in recent times with respect to their
counterpart in metric formalism. We study seven $f(R)$ models in Palatini formalism:
Hu--Sawicki (two cases), Starobinsky, exponential, Tsujikawa, $f(R) = R -\beta /R^ n$, and
$f(R)= R + \alpha \ln(R) - \beta$. Following standard statistical procedure and utilizing data
sets: type Ia supernovae data, cosmic chronometer observations, baryonic acoustic
oscillations data, data from H \textsc{ii} starburst galaxies, local measurements of the \emph{Hubble}
parameter ($H_{0}$), and distance priors of cosmic microwave background radiation data,
we obtain constraints on the model parameters. When compared with the standard
`lambda-cold dark matter model', for many data set combinations, the support for
$f(R)$ models is significant. We obtain the relevant quantities for characterizing the
accelerated expansion of the Universe, and these quantities are consistent with those
obtained in a model-independent way by others. The curve of effective/total equation-
of-state parameter, obtained from parameter constraints, clearly shows correct phases
of the expansion history: the radiation-dominated epochs and the matter-dominated
epochs, of the past, and the current accelerated expansion epoch eventually evolving to
de-Sitter phase in the distant future. 
Overall, our results advocate in favour of pursuing $f(R)$ models in Palatini formalism as a potential alternative for explaining accelerated expansion of the Universe. % [words: 248]
\end{abstract}

\section{Introduction}
\label{intro_Pa}
By the end of last millennium, it was evident that the Universe is currently going through an accelerated phase of expansion. The discovery from the observations of type Ia supernovae (SNIa) \cite{SupernovaCosmologyProject:1997zqe, SupernovaCosmologyProject:1998vns, SupernovaSearchTeam:1998bnz, SupernovaSearchTeam:1998fmf}, has been further validated by subsequent cosmological observations, for instance, observation of baryonic acoustic oscillations (BAOs) \cite{SDSS:2005xqv, Percival:2006gs}, analysis of cosmic microwave background (CMB) radiation \cite{WMAP:2006bqn,WMAP:2008ydk,Planck:2013pxb,Planck:2015fie}, and examination of the power spectrum of matter distributions in the Universe \cite{2dFGRS:2005yhx,Percival:2006gt}. Any attempt to explain the accelerated expansion within the framework of Einstein's gravity, requires inclusion of a hypothetical fluid (generally called `dark energy') with negative pressure, which can counteract the gravitational pull to provide accelerated expansion.
\par
%By including $\Lambda$ term in Einstein's equation we get the `lambda-cold dark matter ($\Lambda$CDM) model'. 
With re-inclusion of, formerly abandoned $\Lambda$ term by Einstein, in his field equation of gravity, we get the `lambda-cold dark matter ($\Lambda$CDM) model' of Cosmology.
Although this simplest model provides explanation of above
phenomena and also fits to all major observed cosmological data very well (i.e. with respect to goodness of fit analysis), nevertheless, this model faces at least two serious issues, namely,
the \emph{cosmological constant problem}, and the \emph{cosmic coincidence problem}. With regards
to particle physics, it is expected that the cosmological constant should be identified
with the vacuum-energy. Then the quantum electrodynamical calculations estimate the
density of this vacuum-energy is to be $\sim 10^{72}\,\rm{GeV}^{4}$, whereas the measurements from
cosmological data yield a value $\sim 10^{-48}\,\rm{GeV}^{4}$, that is, a discrepancy of 120 orders of
magnitude! This embarrassing discrepancy between measured value and theoretical
estimate, is called as the cosmological constant problem \cite{Zlatev:1998tr}. In the $\Lambda$CDM model, the
density of matter contents of the Universe evolves as $\propto (1+z)^{3}$, where $z$ denotes redshift,
while the density corresponding to the $\Lambda$ term remains constant throughout. And yet the cosmological measurements yield their current fractional densities to be of same order, $\sim 0.3$ for the former and $\sim 0.7$ for the latter. To many researchers this coincidence
appear puzzling for various reasons, and is termed as the cosmic coincidence problem \cite{Martin:2012bt}.
\par
This inadequacy of the $\Lambda$CDM model has ended-up motivating researchers to
explore many alternative models or theories. We list some broad categories of
theories/models which have been in practice in this regards. (i) A first simpler
class of alternative models proposed, consists of replacing $w(z) = -1$ of the $\Lambda$CDM
model with variable, parameterized, $w(z)$. For instance the popular Chevalier--
Polarski--Linder model \cite{Chevallier:2000qy,Linder:2002et} has $w(z) = w_{0} + w_{a}z/(1 + z)$ with model parameters
$(w_{0},\, w_{a})$. There are many such proposals with its pros and cons (see \cite{Yang:2021flj}, and references
therein). These class of models are dubbed as `parametric dark energy models' in the
literature. (ii) Derived from incorporating a scalar field to the energy-momentum tensor
of the Einstein equation, the so-called scalar field models, are also being researched for
explaining the accelerated expansion. The two subclasses of these models, quintessence
models \cite{Tsujikawa:2013fta} and k-essence models \cite{Armendariz-Picon:2000ulo}, differ in a sense that unlike the former, the
latter are derived from Lagrangians with noncanonical kinematic terms. (iii) Shortly after
the discovery of accelerated expansion of the Universe, in 1999, it was also proposed
to relax the standard assumption of the so-called Copernican Principle (CP, i.e. the
assumption that at large enough scales, the Universe is homogeneous and isotropic)
and consequently, in this approach, inclusion of large-scale nonlinear inhomogeneities,
resulted in explaining away the effects of dark energy \cite{Pascual-Sanchez:1999xpt,Celerier:1999hp}. These shorts of theories
involve working with the spherically symmetric Lema\^itre--Tolman--Bondi (LTB) metric  rather than the usual Friedmann--Lema\^itre--Robertson--Walker (FLRW) metric. A
limitation or drawback of this approach, apart from violating of the CP, is often quoted
as the requirement of placing the observer at a centre of 1-3 Gpc wide void. For
a review or more on this field of research, known as `inhomogeneous cosmology', one
can see \cite{Krasinski:1997yxj,Bolejko:2011jc,Marra:2011ct,Enqvist:2006cg,Enqvist:2007vb}. (iv) There are also modified gravity theories,  derived from 
including higher order corrections to the Einstein--Hilbert (EH) action. The modified gravity
theories provide acceleration through geometry itself. One of the simplest of these
theories involve replacing the Ricci scalar, $R$, in the EH action, by an arbitrary function $f(R)$, known as `$f(R)$ gravity'. Originally proposed in context of cosmic inflation, with
the discovery of accelerated expansion of the Universe, $f(R)$ gravity models have evolved into an active area of research for the latter context too.
\par
Within the framework of this so-called $f(R)$  gravity, one can assume that the
Christoffel connections are derived from the metric itself and subsequently vary the
Einstein--Hilbert action with respect to metric, only, to obtain the field equations. This
approach is called `metric formalism' of the $f(R)$ gravity. Alternatively, one can also
treat the Christoffel connections and the metric as independent, and vary the Einstein--
Hilbert action with respect to each to get the governing field equations. This latter
approach is termed as `Palatini formalism' of the $f(R)$ gravity. For the $\Lambda$CDM model
where $f(R) \equiv R - 2\Lambda$, both of these approaches provide same sort of field equations,
whereas for any other $f(R)$, they provide quite different sets of governing equations. To
be specific, while in the metric formalism the resultant modified Friedmann equations include fourth order differential equation, in the Palatini formalism it is only of second order.
\par
In this work, we provide updated constraints on viable $f(R)$ gravity models using
the latest cosmological data sets, including SNIa from Pantheonplus compilation, cosmic
chronometer (CC) observations, BAO data, H \textsc{ii} starburst galaxies (H\textsc{ii}G) data, local
measurements of the \emph{Hubble} parameter ($H_{0}$), and the distance priors of CMB radiation
(CMBR). We study seven viable $f(R)$ models in the Palatini formalism: the Hu--
Sawicki model (two cases), the Starobinsky model, the exponential model, the Tsujikawa
model, plus two more models which are only viable in the Palatini formalism, namely,
$f(R) = R -\beta/R^{n}$ and $f(R) = R + \alpha \ln(R) - \beta$. These models in their original forms
give a false impression that they are non-reducible to the $\Lambda$CDM model, so we chose
to re-parameterize these models in terms of `deviation/distortion parameter (b)'. In
latter forms, with $b \rightarrow 0$, we can clearly see, a $f(R)$ model tends to the $\Lambda$CDM model. 
%For our investigation, we adopt the Palatini formalism techniques in the context of a homogeneous and isotropic Universe.
\par 
There are fewer past works which have constrained $f(R)$ models in the Palatini
formalism using observed data. The Hu--Sawicki model in its original form has been
constrained in \cite{Santos:2012vs}, using data of the gas mass fraction in X-ray luminous galaxy clusters,
CMB/BAO peaks ratio data, CC data and SNIa data from Union 2.1 catalogue. In \cite{Campista:2010jb},
the exponential model has been constrained using SNIa data from Union 2 catalogue
and CC data. Constraints on the model $f(R) = R - \beta/R^{n}$ have been obtained in
\cite{Amarzguioui:2005zq,Fay:2007gg,Carvalho:2008am}, and \cite{Santos:2008qp} using various data sets. The $f(R) = R + \alpha\ln(R) - \beta$ model has
been constrained in \cite{Fay:2007gg} using a total of 115 SNIa data points, BAO peak data and CMB
shift parameter data. We could not find any reference for other three models which we
have also included in this work within context of Palatini formalism.
\begin{comment}
\label{intro_Pa}
\label{fRcosmo_Pal}
\label{fRmodels_Pal}
\label{ocd}
\label{results_Pal}
\label{comparison}
\label{accel}
\label{concl}
\end{comment}
\par
This work is organized as follows. In Section \ref{fRcosmo_Pal} we derive the modified Friedmann 
equations and other required equations from the Einstein--Hilbert action for $f(R)$ gravity. A brief introduction to $f(R)$ models investigated in this work is provided
in Section \ref{fRcosmo_Pal}. We introduce the cosmological data sets and the corresponding equations
that establish connection between theory and data, in Section \ref{ocd}, along with discussing the
statistical procedures employed to obtain constraints on the model parameters. Resultant constraints on the model parameters, for all the $f(R)$ models, are presented in Section \ref{results_Pal}.
Assessment of performance of different models using statistical tools, is provided in Section \ref{comparison}. Section \ref{accel} is dedicated to deriving the relevant quantities that characterize expansion
history of the Universe based on the model constraints. Finally, we  provide the concluding remarks for this work in Section \ref{concl}.
\par
Unless otherwise specified, we have set $c=1$ (where, $c$ denotes speed of light in vacuum), and value of the \emph{Hubble} parameter is expressed in the unit km\,s$^{-1}$Mpc$^{-1}$.
%%%%%%%%%%%%%%%%%%%%%%%%%%%%%%%%%%%%%%%%%%%%%%%%%%%%%%%%%%%%%%%%%%%%%%%%%%%%
%%%%%%%%%%%%%%%%%%%%%%%%%%%%%%%%%%%%%%%%%%%%%%%%%%%%%%%%%%%%%%%%%%%%%%%%%%%%
%%%%%%%%%%%%%%%%%%%%%%%%%%%%%%%%%%%%%%%%%%%%%%%%%%%%%%%%%%%%%%%%%%%%%%%%%%%%
%%%%%%%%%%%%%%%%%%%%%%%%%%%%%%%%%%%%%%%%%%%%%%%%%%%%%%%%%%%%%%%%%%%%%%%%%%%%
\section{$\lowercase{f}(R)$ cosmology in the Palatini formalism}
\label{fRcosmo_Pal}
The modified version of Einstein--Hilbert action for $f(R)$ gravity can be expressed as %is given by
\begin{equation}
S = \frac{1}{2\kappa}\int\,d^{4}x\,\sqrt{-g}\,f(\cR) + S_{\rm m} + S_{\rm r},
\label{action_Pal}
\end{equation}
where $\kappa=8\pi{G}$, $g$ is the determinant of the metric tensor ($g_{\mu\nu}$), 
$S_{\rm m}$ is the action for matter fields and $S_{\rm r}$ is the action 
for radiation fields. The symbol $\cR$ stands for 
generalized Ricci scalar which is derived from the generalized Ricci 
tensor $\cR_{\mn}$. We will also be using symbols $R$ and $R_{\mn}$ for 
Ricci scalar and Ricci tensor, respectively, derived solely from metric. Unlike metric formalism
the connection and the metric are treated as independent fields in Palatini formalism, and
so correspondingly, the action has to be varied with respect to (w.r.t.) the metric, and again separately, w.r.t. the connection, to get the governing field equations. Varying the action Eq. \ref{action_Pal} w.r.t. metric gives
\begin{equation}
F\cR_{(\mn)} - \frac{1}{2}fg_{\mn} = {\kappa}T_{\mn}\,,
\label{FE1_Pal}
\end{equation}
where $F\equiv\fcr \equiv df/d\fcr$ and $T_{\mn}$ is the energy-momentum 
tensor of the matter and the radiation from actions $S_{\rm m}$ and $S_{\rm r}$. Variation of action Eq. \ref{action_Pal} w.r.t. 
connection gives
\begin{equation}
\nabla_{\alpha}\left(\sqrt{-g}\,\fcr\,g^{\mn}\right) = 0\,,
\end{equation}
where $\nabla_{\alpha}$ represents covariant derivatives. The independent connection is given by
\begin{equation}
\Gamma^{\lambda}_{\mu\nu} = 
\begin{Bmatrix}\lambda{\,\,}\\\mu\nu\\\end{Bmatrix}
+\frac{1}{2\fcr}\left(\delta^{\lambda}_{\mu}\p{\nu}\delta^{\lambda}_{\nu}\p{\mu} - g_{\mn}g^{\lambda\rho}\p{\rho}\right)\fcr,
\label{GammaI_Pal}
\end{equation}
where $\begin{Bmatrix}\lambda{\,\,}\\\mu\nu\\\end{Bmatrix}$ 
denotes the Christoffel symbols of the metric $g_{\mn}$. The generalized Ricci tensor constructed from the independent connection Eq. \ref{GammaI_Pal} is given by 
\begin{equation}
\cR_{\mn} = R_{\mn} + \frac{3}{2}\frac{\left(\nabla_{\mu}F\right)\left(\nabla_{\nu}F\right)}{F^{2}} -\frac{1}{F}\left(\nabla_{\mu}\nabla_{\nu}+\frac{1}{2}g_{\mn}\boxx\right)F,
\label{RicciTI_Pal}
\end{equation}
where $\boxx\equiv{g}^{\mn}\nabla_{\mu}\nabla_{\nu}$ denotes the covariant D'Alembertian. After contraction of the generalized Ricci tensor Eq. \ref{RicciTI_Pal} w.r.t. the metric $g_{\mn}$ we get the expression for generalized Ricci scalar as
\begin{equation}
\cR = R + \frac{3}{2}\frac{\left(\nabla_{\mu}F\right)\left(\nabla^{\mu}F\right)}{F^{2}} -\frac{3}{F}\,\boxx{F}\,.
\label{RicciSI_Pal}
\end{equation}
After substitution of Eqs. \ref{RicciTI_Pal} and \ref{RicciSI_Pal} into Eq. \ref{FE1_Pal} we get,
\begin{eqnarray}
G_{\mn} &=& \frac{\kappa}{F}T_{\mn} - \frac{1}{2}g_{\mn}\left(\cR - \frac{f}{F} \right)\nonumber\\
&&-\frac{3}{2F^{2}}\left[\left(\nabla_{\mu}F\right)\left(\nabla_{\nu}F\right) - \frac{1}{2}g_{\mn}\left(\nabla_{\alpha}F\right)\left(\nabla^{\alpha}F\right) \right]\nonumber\\
&&+\frac{1}{F}\left(\nabla_{\mu}\nabla_{\nu} - g_{\mn}\boxx\right)F,
\label{FE2_Pal}
\end{eqnarray}
where $G_{\mn}\equiv \cR_{\mn} - \frac{1}{2}g_{\mn}\cR$ is the Einstein tensor. We 
assume that the Universe is described by the spatially flat Friedmann--Lema\^itre--Robertson--Walker (FLRW) metric, given by
\begin{equation}
ds^{2} = -dt^{2} + a(t)^{2}\left[dr^{2} + r^{2}(d\theta^{2}+\sin^{2}\theta\,d\phi^{2}) \right]\,.
\end{equation}
We assume the general consideration that the Universe is made up of perfect-fluid with energy density $\rho$ and pressure $p$, and so correspondingly, the 
enery-momentum tensor simplifies to $T^{\mu}{}_{\nu}=\text{diag}\left(-\rho,\,p,\,p,\,p\right)$.
For FLRW metric, the $`00'$\,-\,component of tensorial equation (Eq. \ref{FE2_Pal}) gives
\begin{equation}
3H^{2} = \frac{\kappa \rho}{F}+\frac{1}{2}\left(\cR-\frac{f}{F}\right)-\frac{3\dot{F}^{2}}{4F^{2}}  - \frac{3H\dot{F}}{F},
\label{G00_Pal}
\end{equation}
and `$ij$'\,-\,component (where $i=j=1,2,3$) gives
\begin{equation}
-2\dot{H} -3H^{2} = \frac{\kappa p}{F}-\frac{1}{2}\left(\cR -\frac{f}{F}\right) -\frac{3\dot{F}^{2}}{4F^{2}} + \frac{2H\dot{F}}{F} + \frac{\ddot{F}}{F}\,.
\label{Gij_Pal}
\end{equation}
The overhead dots in above two equations represent derivative w.r.t. `$t$'. Combining Eqs. \ref{RicciSI_Pal}, \ref{G00_Pal} and \ref{Gij_Pal} we get the generalized as
Friedmann equation
\begin{equation}
\left(H + \frac{1}{2}\frac{\dot{F}}{F} \right)^{2} = \frac{1}{6}\left[\frac{\kappa(\rho + 3p)}{F} + \frac{f}{F} \right]\,.
\label{MFE_Pal}
\end{equation}
Assuming the matter content to be pressureless ($p_{\rm m}=0$) dust with density $\rho_{\rm m}$ and radiation with density $\rho_{\rm r}$ and pressure 
$p_{\rm r}=\rho_{\rm r}/3$, in the absence of any interaction between matter and radiation after the epoch of decoupling, the conservation equation $T^{\mn}{}_{;\nu}=0$ holds and can be written  as
\begin{equation}
\dot{\rho}_{m}+3H{\rho}_{m}=0,\quad \dot{\rho}_{r}+4H{\rho}_{r}=0\,,
\label{ConsEq_Pal}
\end{equation}
and their solutions can be obtained as $\rho_{\rm m}=\rho_{\rm m0}/a^{3}$ and $\rho_{\rm r}=\rho_{\rm r0}/a^{4}$. Defining $\Omega_{\rm m0}=\kappa\rho_{\rm m0}/(3H_{0}^{2})$ and $\Omega_{\rm r0}=\kappa\rho_{\rm r0}/(3H_{0}^{2})$, we get 
$\kappa\rho_{\rm m}= 3H_{0}^{2}\Omega_{\rm r0}(1+z)^{3}$ and 
$\kappa(\rho_{\rm r} + 3p_{\rm r})= 6H_{0}^{2}\Omega_{\rm r0}(1+z)^{4}$, where $z$ the redshift is related to the scale factor$(a)$ via $a=1/(1+z)$.
\par
Trace of Eq. \ref{FE1_Pal} provides us
\begin{equation}
F\cR - 2f = -{\kappa}{\rho}_{m} = -3H_{0}^{2}\Omega_{\rm m0}(1+z)^{3}\,,
\label{TE1_Pal}
\end{equation}
since the trace of radiative fluid vanishes. 
%We should also note that $\Omega_{m0}$ is completely determined from model parameters of $f$ and $\tcR(z=0)$. 
Combining the conservation Eq. \ref{ConsEq_Pal} with time derivative of Eq. \ref{TE1_Pal}, we get 
\begin{equation}
\dot{\cR} = -3H\left(\frac{2f-\cR{F}}{F-\cR{F}_{\cR}}\right)\,.
\label{Rdot_Pal}
\end{equation}
We define dimensionless quantities $\tilde{\cR}=\cR/H_{0}^{2},\,\tilde{f}\equiv{f}/H_{0}^{2}$, and so correspondingly, we get $\tilde{F}\equiv{F},\,\tilde{F}_{\tilde{\cR}} = H_{0}^{2}\,F_{R}\, (\equiv\tF^{\prime})$.  Combining Eqs. \ref{MFE_Pal} and \ref{Rdot_Pal}, we get
\begin{equation}
E^{2}(z)\equiv \frac{H^{2}}{H_{0}^{2}} = \frac{\kappa(\rho+3p)+\tf}{6\tF\xi^{2}},
\label{Ez_Pal}
\end{equation}
where,
\begin{equation}
\xi = \left[1 - \frac{3\tFp(2\tf - \tcR\tF)}{2\tF(\tF- \tcR\tFp)} \right].
\label{xi_Pal}
\end{equation}
The derivative of Eq. \ref{TE1_Pal} w.r.t. $z$ is
\begin{equation}
\frac{{\rm d}\tcR}{{\rm d}z} = \frac{9\Omega_{m0}(1+z)^{2}}{\tF - \tcR\tFp}\,.
\label{Rode_Pal}
\end{equation}
To get $H(\tcR)$ or $E(\tcR)$, we need $z(\tcR)$, which can be obtained by solving the ordinary differential equation (ODE) \ref{Rode_Pal}. The initial condition for ODE Eq. \ref{Rode_Pal} (i.e., $\tcR(z=0)= \tcR_{0}$(say)) can be obtained from Eq. \ref{Ez_Pal} 
with the realization that 
\begin{equation}
E^{2}(z=0) = 1 = \frac{(3\tf - \tcR\tF) + 6\Omega_{\rm r0}}{6\tF\xi^{2}}\Big|_{\tcR=\tcR_{0}}\,,
\label{NLER0_Pal}
\end{equation}
where we have used $\Omega_{\rm r0}=4.5\times10^{-5}$. After solving the nonlinear 
equation \ref{NLER0_Pal} for $\tcR_{0}$ we can use that $\tcR_{0}$ in the nonlinear Eq. \ref{TE1_Pal} to obtain $\Omega_{\rm m0}$. Having obtained $\tcR_{0}$ and $\Omega_{\rm m0}$, the ODE Eq. \ref{Rode_Pal} can be solved from $z=0$ to desired $z$ to get $\tcR(z)$, and numerically inverting the latter solution provides us $z(\tcR)$.
\par 
When using observed data for model constraining we need many cosmological distances, all of which could be defined from the `comoving distance', defined as
\begin{eqnarray}
d_{\rm C}(z) &=& \frac{c}{H_{0}}\int_{0}^{z}\frac{dz'}{E(z')}\nonumber\\
&=& \frac{c\left(3\Omega_{m0}\right)^{-1/3}}{3H_{0}}\int_{\tcR_{0}}^{\tcR_{1}}\frac{\left(\tF - \tFp\tcR\right)d\tcR}{\left(2\tf - \tF\tcR\right)^{2/3}E(\tcR)},
\label{dC_Pal}
\end{eqnarray}
where $\tcR_{0}=\tcR(z=0)$ and $\tcR_{1}=\tcR({z=z})$. Thus, we can get 
$d_{_{\rm C}}(z)$ with the help of the solution of ODE Eq. \ref{Rode_Pal}.
\par
If we compare the modified Friedmann Eqs. 
\ref{G00_Pal} and \ref{Gij_Pal} to the usual Friedmann equations with a 
dark energy component characterized by energy density $\rho_{\rm DE}$  and pressure $p_{\rm DE}$, that is, with the equations 
$3H^{2} = \kappa\left(\rho_{\rm m}+\rho_{\rm r}+\rho_{\rm DE}\right)$ 
and 
$-2\dot{H} - 3H^{2} = \kappa\left(p_{\rm m}+p_{\rm r}+p_{\rm DE}\right)$, 
we can deduce the `effective (geometric) dark energy' with following density  and pressure for any $f(R)$ model in Palatini formalism:
\begin{equation}
\rho_{\rm{DE}} = \frac{1}{\kappa}\left[\frac{1}{2}\left(F\cR-f\right)-\frac{3\dot{F}^{2}}{4F}  - {3H\dot{F}} + 3H^{2}(1-F)\right],
\end{equation}
and,
\begin{equation}
p_{\rm{DE}} = \frac{1}{\kappa}\left[-\frac{1}{2}\left(F\cR -{f}\right) -\frac{3\dot{F}^{2}}{4F} + {2H\dot{F}} +{\ddot{F}}-(2\dot{H}+3H^{2})(1-F)\right].
\end{equation}
Correspondingly, we can define equation-of-state parameter for effective dark energy as
\begin{equation}
w_{\rm geo}(z) = \frac{w_{\rm eff}-\Omega_{r}/3}{1-\Omega_{m}-\Omega_{r}},
\label{Eq:wDE_Pal}
\end{equation}
where,
\begin{equation}
w_{\rm eff}(z) = -1+ \frac{2(1+z)}{3H}\frac{{\rm d}H}{{\rm d}z}\,,%-1 - \frac{2}{3}\frac{\dot{H}}{H}
\label{Eq:weff_Pal}
\end{equation}
is the total equation-of-state parameter, that is, for all the components of cosmic fluid. In the above relation
the fractional matter and radiation density functions are defined as: $\Omega_{m}(z)=\kappa\rho_{m}/(3H^{2})$ and $\Omega_{r}(z)=\kappa\rho_{r}/(3H^{2})$, which utilizes the fact that $p_{\rm{m}} = 0$ and $p_{\rm{r}} = \rho_{\rm{r}}/3$.

%\clearpage
%%%%%%%%%%%%%%%%%%%%%%%%%%%%%%%%%%%%%%%%%%%%%%%%%%%%%%%%%%%%%%%%%%%%%%%%%%%%
%%%%%%%%%%%%%%%%%%%%%%%%%%%%%%%%%%%%%%%%%%%%%%%%%%%%%%%%%%%%%%%%%%%%%%%%%%%%
\section{$\lowercase{f}(R)$ models}
\label{fRmodels_Pal}
A brief introduction to the seven $f(R)$ models investigated in this work is presented
here. Apart from presenting the $f(R)$ models in their original proposed forms, we also
transformed them to forms which make apparent their reducibility to the $\Lambda$CDM model.
For this we re-parameterized the models in terms of `deviation/distortion parameter ($b$)',
where with $b \rightarrow 0$, all $f(R)$ model tends to the $\Lambda$CDM model. Eventually, we obtain all the $f(R)$ models expressed in terms of 
$\Lambda$, $b$ and other parameters. However, from Eq. \ref{TE1_Pal} we can see that $\Lambda$, $b$, and $\Omega_{m0}$ are not all independent. Therefore, for reporting the results from data fitting, we choose a seemingly more meaningful parameter combination, that is, $\Omega_{m0}$, $b$, and other parameters.
%%%%%%%%%%%%%%%%%%%%%%%%%%%%%%%%%%%%%%%%%%%%%%%%%%%%%%%%%%%%%%%%%%%%%%%%%%%%
\subsection{The Hu--Sawicki model}
In its original form the Hu–Sawicki model \cite{Hu:2007nk}, is expressed as
\begin{equation}
f(R)_{\rm HS} = R - \mu^{2}\frac{c{1}(R/\mu^{2})^{n_{{\rm HS}}}}{1+ c{2}(R/\mu^{2})^{n_{{\rm HS}}}}\,,
\label{HSmodel0}
\end{equation}
where $c_{1}$, $c_{2}$, ${n_{{\rm HS}}}$ (assumed to be a positive integer) are dimensionless parameters with $\mu^{2}\approx\Omega_{\rm m0}H_{0}^{2}$. 
In this form it is hard to see how this model could be relatable to
the $\Lambda$CDM model, and with this need, later a re-parameterization was introduced by \cite{Basilakos:2013nfa} as
\begin{equation}
f(R)_{\rm HS} = R - 2\Lambda
\left[1 - \left\{1+\left(\frac{R}{b\Lambda}\right)^{{n_{\rm HS}}}\right\}^{-1}\right],
\label{HSmodel1}
\end{equation}
where $\Lambda =\mu^{2}c_{1}/2c_{2}$ and $b = 2\left(c_{2}^{1-1/{n_{{\rm HS}}}}\right)/c_{1}$. Although this model have been studied
with various cosmological data, very actively in the metric formalism (see \cite{Ravi:2023nsn}, and references therein), there is lack of studies of this model in the Palatini formalism \cite{Santos:2012vs}. 
In this work, apart from common practice of taking ${n_{{\rm HS}}}=1$, we will also constrain the case of ${n_{{\rm HS}}}=3$. Later we will see that in the re-parameterized versions, the case of ${n_{{\rm HS}}}=2$ here is equivalent to the Starobinsky model.
%%%%%%%%%%%%%%%%%%%%%%%%%%%%%%%%%%%%%%%%%%%%%%%%%%%%%%%%%%%%%%%%%%%%%%%%%%%%
\subsection{The Starobinsky model}

In Starobinsky model \cite{Starobinsky:2007hu} 
\begin{equation}
f(R)_{\rm ST} = R - {\lambda}R_{\rm S}\left[1 - \left(1 + \frac{R^{2}}{R_{\rm S}^{2}}\right)^{-n_{_{\rm S}}}\right]\,,
\label{STmodel0}
\end{equation}
where ${n_{_{\rm S}}}$ is a positive constant, $\lambda(>0)$ and $R_{\rm S}\approx R_{0}$ ($R_{0}$ denotes the Ricci scalar at present epoch) are free parameters. This model can also be represented in a more general form as \cite{Basilakos:2013nfa}
\begin{equation}
f(R)_{\rm ST} = R - 2\Lambda\left[1 - \left\{1+\left(\frac{R}{b\Lambda}\right)^{2}\right\}^{-{{n_{_{\rm S}}}}}\right],
\label{STmodel1}
\end{equation}
where $\Lambda = {\lambda}R_{\rm S}/2$ and $b=2/\lambda$. In this work, we specifically consider the case where
${{n_{_{\rm S}}}}=1$, with the justification for not exploring values of ${{n_{_{\rm S}}}}$ higher than 1 can be 
found in \cite{Ravi:2023nsn}. From Eqs. \ref{HSmodel1} and \ref{STmodel1}, it is clear that the 
Hu--Sawicki model with
${{n_{_{\rm HS}}}}=2$ and the Starobinsky model with ${{n_{_{\rm S}}}}=1$ are equivalent. The algebraic form of
$f(R)_{\rm ST}$ (Eq. \ref{STmodel1}) suggests that, irrespective of the data set, from MCMC fitting procedures,
one will certainly obtain the parameter $b$ distributed symmetrically around $b = 0$.
Without loss of generality, we adopt $b > 0$, as we are mainly interested in studying
deviation/distinguishabililty from the $\Lambda$CDM model. We could not find any earlier work
where this model has been constrained in context of the Palatini formalism, whereas
there are many such works in the metric formalism (see \cite{Ravi:2023nsn}, and references therein).

%%%%%%%%%%%%%%%%%%%%%%%%%%%%%%%%%%%%%%%%%%%%%%%%%%%%%%%%%%%%%%%%%%%%%%%%%%%%
%%%%%%%%%%%%%%%%%%%%%%%%%%%%%%%%%%%%%%%%%%%%%%%%%%%%%%%%%%%%%%%%%%%%%%%%%%%%
\subsection{The exponential model}

The exponential model from \cite{Cognola:2007zu}, widely constrained in context of metric formalism \cite{Linder:2009jz,Chen:2014tdy,Odintsov:2017qif,Leizerovich:2021ksf} is given by
\begin{equation}
f(R)_{\rm E} = R + \alpha\left[\exp(-\beta{R}) - 1\right]\,,
\label{expmodel0}
\end{equation}
where $\alpha>0$ and $\beta>0$ are the parameters of this model. Though with $\beta \rightarrow \infty$ and
the identification $\alpha = 2\Lambda$, it's reduction to the $\Lambda$CDM model is apparent, but still we re-parameterize it with the deviation parameter, $b$, as
\begin{equation}
f(R)_{\rm E} = R - 2\Lambda\left[1 - \exp\left(-\frac{R}{b\Lambda}\right)\right]\,,
\label{expmodel1}
\end{equation}
with $\Lambda = \alpha/2$ and $b=2/(\alpha\beta)$. In latter form, for finite values of the parameters, the exponential model tends to the 
$\Lambda$CDM model. From this, it is evident that when $b \rightarrow 0$,
$R \gg b\Lambda$, $f(R)_{\rm E} \rightarrow R - 2\Lambda$. In the past,  exponential model in Palatini formalism has been constrained by \cite{Campista:2010jb} for the purpose of dark energy studies.
%%%%%%%%%%%%%%%%%%%%%%%%%%%%%%%%%%%%%%%%%%%%%%%%%%%%%%%%%%%%%%%%%%%%%%%%%%%%
\subsection{The Tsujikawa model}
The Tsujikawa model \cite{Tsujikawa:2007xu} is given by
\begin{equation}
f(R)_{\rm T} = R - {\xi}R_{\rm T}\tanh\left(\frac{R}{R_{\rm T}}\right)\,.
\end{equation}
Here $\xi(>0)$ and $R_{\rm T}(>0)$ are the model parameters. By defining $\Lambda = {\xi}R_{\rm T}/2$ and $b=2/\xi$, this model too can be expressed in more general form as
\begin{equation}
f(R)_{\rm T} = R - 2\Lambda\tanh\left(\frac{R}{b\Lambda}\right)\,.
\label{tsujimodel1}
\end{equation}
Evidently, as $b \to 0$ (meaning $\xi \to \infty$, $R_{\rm T} \to 0$ but ${\xi}R_{\rm T}$ remains finite), the Tsujikawa model simplifies to $f(R)_{\rm T} = R - 2\Lambda$. We should also note that while doing numerical
computations for this model, it is better to express $\tanh(x) = (1 - e^{-2x})/(1 + e^{-2x})$
rather than $\tanh(x) = (e^{x} - e^{-x})/(e^{x} + e^{-x})$ and so on. The former avoids numerical
instability while the latter do not, especially for very large and very small values of $x$
(for $x > 0$). Although this model have been constrained enthusiastically in the metric
formalism (see \cite{Ravi:2023nsn}, and references therein), we could not find any literature which has constrained it in the Palatini formalism.

%%%%%%%%%%%%%%%%%%%%%%%%%%%%%%%%%%%%%%%%%%%%%%%%%%%%%%%%%%%%%%%%%%%%%%%%%%%%
%%%%%%%%%%%%%%%%%%%%%%%%%%%%%%%%%%%%%%%%%%%%%%%%%%%%%%%%%%%%%%%%%%%%%%%%%%%%
\subsection{Two more $\lowercase{f}(R)$ models}
We also investigate constraints on two alternative models described by
(i) $f(R) = R - \beta/R^{n}$ and  (ii) $f(R) = R + \alpha\ln(R) - \beta$.
Here $n$, $\alpha$ and $\beta$ represent the parameters of these models. 
Although these two models are found to be non-viable in the metric formalism, however, these models are of particular relevance in the Palatini formalism, as they yield a correct sequence of radiation-dominated, matter-dominated, and de-Sitter era in expansion history of the Universe \cite{Amarzguioui:2005zq,Fay:2007gg,Carvalho:2008am,Santos:2008qp}. These two models 
also can be re-parametrized in terms of $\Lambda$ and $b$, trivially, as follows:
\begin{equation}
f(R) = R - \beta/R^{n} = R - \frac{2\Lambda}{R^{b}}\,,
\label{Eq:RbetaRn}
\end{equation}
with identifications $\beta=2\Lambda$ and $n=b$; and 
\begin{equation}
f(R) = R + \alpha\ln(R) - \beta = R - 2\Lambda\left(1 - \frac{b}{2\Lambda}\ln(R)\right),
\label{Eq:RlnR}
\end{equation}
by recognizing $\alpha$ as $b$ and $\beta$ as $2\Lambda$.
%%%%%%%%%%%%%%%%%%%%%%%%%%%%%%%%%%%%%%%%%%%%%%%%%%%%%%%%%%%%%%%%%%%%%%%%%%%%
%%%%%%%%%%%%%%%%%%%%%%%%%%%%%%%%%%%%%%%%%%%%%%%%%%%%%%%%%%%%%%%%%%%%%%%%%%%%
%%%%%%%%%%%%%%%%%%%%%%%%%%%%%%%%%%%%%%%%%%%%%%%%%%%%%%%%%%%%%%%%%%%%%%%%%%%%
%%%%%%%%%%%%%%%%%%%%%%%%%%%%%%%%%%%%%%%%%%%%%%%%%%%%%%%%%%%%%%%%%%%%%%%%%%%%
\section{Observed cosmological data and Statistical analysis}
\label{ocd}
Here we briefly introduce the cosmological data sets utilized in this work to constraint $f(R)$ models. For any $f(R)$ model, we mainly obtain $H(z)$ numerically, using equations \ref{Ez_Pal}, \ref{xi_Pal}, \ref{Rode_Pal}, and \ref{NLER0_Pal}. So providing theoretical relations between $H(z)$ and various observed quantities becomes imperative, which we also do in this section. Towards the end of this section, we also discuss the statistical methods employed to extract model parameters from data sets.

%%%%%%%%%%%%%%%%%%%%%%%%%%%%%%%%%%%%%%%%%%% subsection %%%%%%%%%%%%%%%%%%%%%%%%%%%%%%%%%%%%%%%%%%%%%%%%%%%%
\subsection{Type Ia supernova data}%Pantheonplus SNe Ia
The type Ia Supernovae (SNIa) have been established as a standard candle class of astrophysical objects \cite{Phillips:1993ng}. In fact, it was the observations of SNIa that led to the discovery of accelerated expansion of the Universe in late 1990s \cite{SupernovaCosmologyProject:1997zqe, SupernovaCosmologyProject:1998vns, SupernovaSearchTeam:1998bnz, SupernovaSearchTeam:1998fmf}. In this work, we employ apparent
magnitude data for SNIa from the recently released PantheonPlus compilation \cite{Scolnic:2021amr}.
This latest compilation comprise of 1701 distinct light curves of 1550 uniquely
spectroscopically confirmed SNIa from 18 surveys. It offers a broader range of low-
redshift data compared to the earlier Pantheon compilation \cite{Pan-STARRS1:2017jku}, with redshift spanning
$0.00122 < z_{\rm HD} <2.26137$ (where $z_{\rm HD}$ denotes \emph{Hubble} diagram redshift). We utilize
the apparent magnitude at maximum brightness ($m_{\rm b}$), along with heliocentric redshift
($z_{\rm hel}$), cosmic microwave background (CMB) corrected redshift ($z_{\rm cmb}$), and the combined
covariance matrix (statistical + systematic) from this compilation \cite{Scolnic:2021amr}.
\par
The expression for luminosity distance
\begin{eqnarray}
d_{\rm L}(z) &=& (1 + z_{\rm hel})\,d_{\rm C}(z_{\rm cmb}),
\label{d_L_Pal}
\end{eqnarray}
%which can be transformed as a function of $\tcR$ with the help of 
which can be obtained with the help of comoving distance ($d_{\rm C}$) defined in Eq. \ref{dC_Pal}.
Theoretically, apparent magnitude ($m_{\rm th}$) is defined in terms of luminosity distance and expressed as
\begin{equation}
m_{\rm th} = M +5\log_{10}\left(\frac{c/H_{0}}{\text{Mpc}}\right) + 5\log_{10}\left(D_{\rm L}(z)\right) + 25\,,
\label{mth}
\end{equation} 
where $M$ is referred to as the absolute magnitude, and $D_{\rm L}(z) \equiv H_{0}d_{\rm L}(z)/c$ is termed
as the `dimensionless \emph{Hubble}-free luminosity distance'. In light of inherent degeneracy
between the absolute magnitude ($M$) and the current epoch \emph{Hubble} parameter ($H_{0}$) in the calculation of apparent magnitude (as revealed in Eq. \ref{mth}), we employ marginalization over parameters $M$ and $H_{0}$. The subsequent marginalized $\chi^{2}$ which requires to be minimized for model fitting, is defined as \cite{SNLS:2011lii}
\begin{equation}
\tilde{\chi}^{2}_{\rm sn} = A - \frac{B^{2}}{D} + \log\frac{D}{2\pi}\,,
\label{chiSN}
\end{equation} 
where $A = (m_{\rm b}-m_{\rm th})^{T}C^{-1}(m_{\rm b}-m_{\rm th})$, $B = (m_{\rm b}-m_{\rm th})^{T}C^{-1}\boldsymbol{1}$, and  $D = \boldsymbol{1}^{T}C^{-1}\boldsymbol{1}$.
Here $C$ represents the combined covariance matrix, while $\boldsymbol{1}$ constitutes an array of ones
matching the length of data set. Of course, here, the theoretical estimations of apparent magnitude ($m_{\rm th}$) are subject to the model under consideration.

%%%%%%%%%%%%%%%%%%%%%%%%%%%%%%%%%%%%%%%%%%% subsection %%%%%%%%%%%%%%%%%%%%%%%%%%%%%%%%%%%%%%%%%%%%%%%%%%%%

\subsection{Cosmic chronometers data and local measurement of $H_{0}$}

By observing the age difference ($\Delta{t}$) of passively evolving galaxies (where star formation
and interactions with other galaxies have ceased) at redshift difference ($\Delta{z}$), the cosmic
chronometers measurements employ the following so-called `differential age method' \cite{Jimenez:2001gg, Wei:2018cov, Simon:2004tf} to obtain \emph{Hubble} parameter at different redshifts:
\begin{equation}
H(z)\equiv\frac{\dot{a}}{a} = -\frac{1}{1+z}\frac{{\rm d}z}{{\rm d}t} \simeq -\frac{1}{1+z}\frac{\Delta{z}}{\Delta{t}}\,.
\end{equation}
So far there have been made a total of 32 such measurements by various researchers
\cite{Simon:2004tf,Moresco:2016mzx, Moresco:2015cya, Zhang:2012mp, Stern:2009ep, Moresco:2012jh, Ratsimbazafy:2017vga,Borghi:2021rft} in the redshift range of 0.07-1.965. In this study we use
these 32 data points comprising of $(z,\,H,\,\sigma_{H})$ from Table B2 of \cite{Ravi:2023nsn}.
\par
For parameter estimations of any model, one can minimize the following residual
\begin{equation}
\chi^{2}_{_{\rm CC}} = \sum_{i=1}^{32}\frac{\left(H_{\rm obs}(z_{i}) - H_{\rm th}(z_{i})\right)^{2}}{\sigma_{H,i}^{2}}\,,
\label{chiCC}
\end{equation}
where $H_{\rm obs}(z_{i})$ denotes the observed values of the 
\emph{Hubble} parameter function at redshift
$z_{i}$, and $\sigma_{H,i}$ denotes the associated uncertainties in these measurements. The theoretical
\emph{Hubble} function at redshift $z_{i}$, which is model-dependent and obtained from Eq. \ref{Rode_Pal}, is
denoted by $H_{\rm th}(z_{i})$ in the above Eq. \ref{chiCC}.
\par
There remains a debatable issue regarding current value of the \emph{Hubble} parameter, $H_{0}$. The \emph{Planck} constraints suggest $H_{0}=67.4\pm 0.5$ \cite{Planck:2018vyg} (which is inferred
from the CMB data assuming standard cosmological model) while the locally
measured value from the SH0ES collaboration is $H_{0}=73.04\pm 1.04$ \cite{Riess:2021jrx} (which is obtained
from SNIa and Cepheid data in a cosmological model independent way). The two
measurements show a tension of $4\sigma-5\sigma$ indicating a notable discrepancy between them.
Considering this apparent \emph{Hubble} tension, we have incorporated both scenarios, with
and without the SH0ES prior for $H_{0}$ in our investigation. We use following Gaussian
prior on $H_{0}$ for $\chi^{2}$ minimization corresponding to the SH0ES data
\begin{equation}
\chi^{2}_{\rm SH0ES} = \frac{(H_{0}-73.04)^{2}}{1.04^{2}}\,.
\label{chiSH0ES}
\end{equation}

%%%%%%%%%%%%%%%%%%%%%%%%%%%%%%%%%%%%%%%%%%%%%%%%%%%
\subsection{${\rm H}$ $\textsc{\textmd{ii}}$ starburst galaxies data}
H \textsc{ii} starburst galaxies (H\textsc{ii}Gs) are compact and massive regions where stars form
actively and are also surrounded by ionized hydrogen gas. Their optical spectra reveal
narrow Balmer lines, such as H$\alpha$ and H$\beta$, amidst a faint continuum. The observed
empirical correlation between the luminosity ($L$) of the H$\beta$ line and the velocity
dispersion ($\sigma$)--a measure of spectral line width--underscores the significance of H\textsc{ii}G data
for cosmological investigations. This correlation is understandable, since an increase in
the mass of the starburst component leads to heightened emission of ionization-induced
photons (thereby enhancing luminosity $L$) and also augmented turbulence in velocity
(resulting in higher velocity dispersion $\sigma$, see \cite{Gonzalez-Moran:2021drc}, and references therein).
\par
This study utilizes a total of 181 data points taken from \cite{Gonzalez-Moran:2021drc,Erb:2006tg, Maseda:2014gea, Masters:2014gna}, and \cite{Gonzalez-Moran:2019uij}
of the Balmer H$\beta$ line emission. The redshift coverage of these data points is
$z \in [0.0088, 2.5449]$. One advantage of using H\textsc{ii}G data set could be that it explores
higher redshifts, along with relatively denser coverage at higher redshift, when compared
with those of SNIa data \cite{Ravi:2023nsn}.
\par
The aforementioned empirical correlation between luminosity $L$ and velocity
dispersion $\sigma$ can be expressed as follows:
\begin{equation}
\log\left[\frac{L}{\mathrm{erg\,s^{-1}}}\right] = \beta\log\left[\frac{\sigma}{\mathrm{km\,s^{-1}}}\right]+\alpha\,.
\end{equation}
To obtain luminosity from observed velocity dispersion $\sigma$, we take $\beta = 5.022\pm0.058$ and $\alpha=33.268\pm0.083$ from \cite{Gonzalez-Moran:2021drc,Gonzalez-Moran:2019uij}. Having obtained the luminosity, we can derive
the luminosity distance using observed value of flux ($F$) associated with H$\beta$ line, as
$d_{\rm L} = \left[L/(4{\pi}F)\right]^{1/2}$. Consequently, the distance modulus of H\textsc{ii}G data set can be obtained as
\begin{equation}
\mu_{o} = 2.5\left(\alpha + \beta\log\left[\frac{\sigma}{\mathrm{km\,s^{-1}}}\right] - \log\left[\frac{F}{\mathrm{erg\,s^{-1}\,cm^{-2}}}\right] - 40.08\right).
\label{mu_HIIG}
\end{equation}
\par
With theoretical definition of distance modulus  $\mu_{\theta} = m_{\rm th}-M$, given by Eq. \ref{mth},
and above derived $\mu_{o}$ from observations, we can constraint any cosmological model by
minimizing following $\chi^{2}$
\begin{equation}
\chi^{2}_{_{\mathrm{H}\textsc{ii}\mathrm{G}}} = \sum_{i=1}^{181}\frac{\left[\mu_{o} - \mu_{\theta}\right]^{2}}{\epsilon^{2}}\,.
\label{chiHIIG}
\end{equation}
The combined variance $\epsilon^{2}$ in the above equation, is comprised as
\begin{equation}
\epsilon^{2} = \epsilon_{\mu_{o,\mathrm{stat}}}^{2} + \epsilon_{\mu_{\theta,\mathrm{stat}}}^{2} + \epsilon_{\mathrm{sys}}^{2}\,,
\end{equation}
where,
\begin{equation}
\epsilon_{\mu_{o,\mathrm{stat}}}^{2} = 6.25\left(\epsilon_{\log{F}}^{2} +  \beta^{2}\epsilon_{\log{\sigma}}^{2} + \epsilon_{\beta}^{2}(\log{\sigma})^{2} + \epsilon_{\alpha}^{2}\right)\,,
\end{equation}
and,
\begin{equation}
\epsilon_{\mu_{\theta,\mathrm{stat}}}^{2} = \left[\frac{5}{\ln10}\left(\frac{c(1+z)}{d_{L}(z)H(z)} + \frac{1}{1+z} \right)\sigma_{z}\right]^{2},
\end{equation}
and $\epsilon_{\mathrm{sys}}=0.257$ \cite{Chavez:2016epc,Gonzalez-Moran:2019uij,Gonzalez-Moran:2021drc}. With simple error propagation theory we can derive
$\epsilon_{\mu_{o,\mathrm{stat}}}$ from Eq. \ref{mu_HIIG}, and $\epsilon_{\mu_{\theta,\mathrm{stat}}}$, stemming from redshift uncertainties ($\sigma_{z}\sim10^{-4}$), from
the standard relation of $\mu$ in terms of $d_{\rm L}$ and $d_{\rm L}$ in terms of $H(z)$.
%%%%%%%%%%%%%%%%%%%%%%%%%%%%%%%%%%%%%%%%%%%%%%%%%%%
\clearpage
%%%%%%%%%%%%%%%%%%%%%%%%%%%%%%%%%%%%%%%%%%% subsection %%%%%%%%%%%%%%%%%%%%%%%%%%%%%%%%%%%%%%%%%%%%%%%%%%%%
\subsection{CMB data and BAO data}
Shortly after Big Bang, the Universe was made-up of hot, dense plasma, consisting
mainly of protons, electrons, and photons, and was almost uniformly distributed but
with small fluctuations. With Thomson scattering, baryons and photons were tightly
coupled. The subsequent expansion caused decrease in temperature and density, and
also amplification of fluctuations due to gravity. With pull due to gravitation and push due to radiation pressure, tightly coupled baryon-photon mixture condensed in regions with
higher densities and thus creating compressions and rarefactions in the form of acoustic
waves referred to as BAOs. Approximately 380,000 years after the Big Bang, the Universe
had cooled sufficiently for protons and electrons to combine and form neutral hydrogen
atoms. Consequently photons decoupled from baryons. After this transition, known
as recombination, photons travelled freely through space without scattering off charged
particles. The CMB radiation we observe today are those early decoupled photons,
providing a snapshot of the Universe's state at the epoch of recombination/decoupling
($z_{\star}$). The CMB or CMBR observed today, is still a blackbody radiation, distributed
almost uniformly throughout the Universe at a temperature of 2.73 K, with anisotropies
of order of $\Delta{T}/T \sim 10^{-5}$.
\par
After release of baryons from the Compton drag of photons at epoch of drag ($z_{\rm d}$, slightly later to $z_{\star}$), the acoustic waves remained frozen in the baryons. A characteristic
length-scale attributed to BAOs is the sound horizon at the epoch of drag ($z_{\rm d}$), defined
as the maximum distance travelled by the acoustic wave before decoupling. BAO, hence,
holds a status of standard ruler for length-scale in Cosmology (\cite{Hu:1995en,  Eisenstein:1997ik}). The primordial density fluctuations in baryons provided the seeds for the formation of cosmic structures,
including galaxies, clusters, and superclusters.
\par
We have used a total of 30 data points of various BAO observables from compilation
in Table B1 of \cite{Ravi:2023nsn}, collected therein from many surveys \cite{Beutler:2011hx,SDSS:2009ocz,Padmanabhan:2012hf,Tojeiro:2014eea,Anderson:2012sa,Bautista:2017wwp,Seo:2012xy,Sridhar:2020czy,deSainteAgathe:2019voe,Carter:2018vce,Ross:2014qpa,Kazin:2014qga,Ata:2017dya,DES:2017rfo,Bautista:2017zgn,duMasdesBourboux:2017mrl,BOSS:2013igd,BOSS:2016wmc,Bautista:2020ahg,Neveux:2020voa}. To compute the epoch of drag ($z_{\rm d}$) and the sound horizon
at the epoch of drag ($r_{\rm d}$), we employ improved fits from \cite{Aizpuru:2021vhd}. The BAO observables
encompass the \emph{Hubble} distance, defined by $D_{\rm H} = c/H(z)$ and other observables, namely,
the transverse comoving distance ($D_{\rm M}(z)$), the angular diameter distance ($D_{\rm A}(z)$), and the
volume-averaged distance ($D_{\rm V}(z)$), are defined as follows:
\begin{equation}
D_{\rm{A}}(z) =\frac{c}{1+z}\int_{0}^{z}\frac{{\rm d}z'}{H(z')}\,,%= d_{\rm C}(z)/(1+z)
\label{D_A}
\end{equation}
%\be
%d_{\rm C}(z) = \frac{c}{H_{0}}\int_{0}^{z}\frac{dz'}{E(z')}
%\ee
\begin{equation}
D_{\rm{M}}(z) = (1+z)D_{\rm{A}}(z)\,,
\label{D_M}
\end{equation}
and,
\begin{equation}
D_{\rm{V}}(z) = \left[(1+z)^{2}D_{\rm{A}}^{2}(z)\frac{z}{H(z)}\right]^{1/3}\,.
\label{D_V}
\end{equation}
Using Eq. \ref{Ez_Pal}, the solution of the ODE given in Eq. \ref{Rode_Pal}, and Eq. \ref{dC_Pal}, all of the above observables can be converted from functions of redshift to functions of $\tcR$. This enables their use in constraining the parameters of any $f(R)$ model within the
Palatini formalism.
\par
To fit any model to the BAO data, the following residuals have been employed for
uncorrelated data points:
\begin{equation}
\chi^{2}_{_{\rm BAO-UC}} = \sum_{i=1}^{20}\left[\frac{A_{\rm th}(z_{i}) - A_{\rm obs}(z_{i})}{\sigma_{i}}\right]^{2},
\label{chiBAO-UC}
\end{equation}
where $A_{\rm obs}(z_{i})$ and $\sigma_{i}$ represent the observed values and their respective measurement
uncertainties at redshift $z_{i}$, and $A_{\rm th}$  signifies the theoretical prediction derived from
the model under consideration. These observed quantities are provided in columns 2 to 4 of aforementioned Table of \cite{Ravi:2023nsn}.
\par
For correlated data points, the relevant residual to be minimized, is given by
\begin{equation}
\chi^{2}_{_{\rm BAO-C}} = \sum_{j=1}^{3}\left[\left({\bf A}_{\rm th} - {\bf A}_{\rm obs}\right)_{j}^{T}{\bf C}_{j}^{-1}
\left({\bf A}_{\rm th} - {\bf A}_{\rm obs}\right)_{j}\right]\,,
\label{chiBAO-C}
\end{equation}
where ${\bf C}_{j}$'s represent covariant matrices for the three distinct data sets, while
$\left({\bf A}_{\rm th} - {\bf A}_{\rm obs}\right)_{j}$ indicates an array of the differences between theoretical and observed
values for each of these three sets. The covariance matrices can be found from the
respective source papers mentioned in the last column of the Table B2 of \cite{Ravi:2023nsn} and/or from 
the Eqs. B1, B2 and B3 of \cite{Ravi:2023nsn}. For whole of the BAO data, we utilize following combined $\chi^{2}$
\begin{equation}
\chi^{2}_{_{\rm BAO}} = \chi^{2}_{_{\rm BAO-UC}} + \chi^{2}_{_{\rm BAO-C}}\,.
\label{chiBAO}
\end{equation}
\par
From CMB observables, we use the so-called `distance-priors' of CMB, namely,
\begin{equation}
\boldsymbol{\rm{y}} = \left(R,\,l_{\rm{A}},\,\omega_{\rm{b}} \right) = \left(\sqrt{\Omega_{m0}}\frac{H_{0}D_{\rm{M}}(z_{\star})}{c},\,\frac{{\pi}D_{\rm{M}}(z_{\star})}{r_{\rm{s}}(z_{\star})},\,\Omega_{b0}h^{2} \right),
\end{equation} 
where $R$ (not to be confused with Ricci scalar) is known as the `shift parameter', and $l_{\rm A}$ is called `acoustic scale'. The
baryonic fraction of total matter density 
($\Omega_{m0}$), at present epoch, is denoted by $\Omega_{b0}$, $h = H_{0}/100$, and $D_{M}(z_{\star})$  is the transverse comoving distance (defined in Eq. \ref{D_M}) at redshift of the epoch of photon decoupling $z_{\star}$. The comoving sound horizon at the epoch of decoupling, $r_{\rm s}(z_{\star})$, defined as
\begin{equation}
 r_{\rm{s}}(z) = \frac{1}{\sqrt{3}}\int_{0}^{1/(1+z)} \frac{{\rm d}a}{a^{2}H(a)\sqrt{1+\left[3\Omega_{b0}/(4\Omega_{r0})\right]a}}\,.
\end{equation}
For computational ease, we calculate $z_{\star}$ from popularly used fitting formula given in
\cite{Hu:1995en}, which is
\begin{equation} \label{eq:zStar}
z_{\star}=1048[1+0.00124(\Omega_{b0} h^2)^{-0.738}][1+g_1(\Omega_{m0} h^2)^{g_2}],
\end{equation}
where
\begin{equation}
g_1=\frac{0.0783(\Omega_{b0} h^2)^{-0.238}}{1+39.5(\Omega_{b0} h^2)^{0.763}}, \quad g_2=\frac{0.560}{1+21.1(\Omega_{b0} h^2)^{1.81}}.
\end{equation}
The most recent estimations of distance priors derived in \cite{Chen:2018dbv} from \emph{Planck Mission} 2018 data \cite{Planck:2018vyg} are
\begin{equation}\label{Pl18:data}
\begin{split}
R^{Pl} &= 1.7428 \pm 0.0053\,,\\
l_{\rm{A}}^{Pl} &= 301.406\pm 0.090\,,\\
\omega_{b}^{Pl} & = 0.02259 \pm 0.00017\,,
\end{split}
\end{equation}
with their covariance matrix given by
\begin{equation}
C_{\rm{CMB}} = ||\tilde{C}_{ij}\sigma_{i}\sigma_{j}||,\quad 
\tilde{C} = \begin{pmatrix}
1 & 0.52 & -0.72\\
0.52 &1 & -0.41\\
-0.72 & -0.41 & 1\\
\end{pmatrix},
\end{equation}
and where $\sigma_{i}$'s denote the standard errors given in the Eq. \ref{Pl18:data}. Finally, for constraining any model with above CMB data, following $\chi^{2}$ function needs be minimized 
\begin{equation}\label{chiCMB}
\chi^{2}_{\rm{CMB}} = \Delta\boldsymbol{\rm{y}}.C_{\rm{CMB}}^{-1}.\left(\Delta\boldsymbol{\rm{y}}\right)^{T},\quad\Delta\boldsymbol{\rm{y}}=\boldsymbol{\rm{y}}-\left(R^{Pl},\,l_{\rm{A}}^{Pl},\,\omega_{\rm{b}}^{Pl} \right).
\end{equation}
%%%%%%%%%%%%%%%%%%%%%%%%%%%%%%%%%%%%%%%%%%% subsection %%%%%%%%%%%%%%%%%%%%%%%%%%%%%%%%%%%%%%%%%%%%%%%%%%%%
\subsection{Statistical analysis}
We estimate model parameters from Markov Chain Monte Carlo (MCMC) samples by maximizing following likelihood function
\begin{equation}
\mathcal{L} = \exp\left(-\sum_{i}\chi^{2}_{i}/2\right).
\label{maxlikeli}
\end{equation}
The subscript $`i'$ in Eq. \ref{maxlikeli} stands for data set combinations from the set: $\{$`SN', `CC', `SH0ES', `H\textsc{ii}G', `BAO', `CMB'$\}$ and corresponding $\chi^{2}_{i}$'s are already defined in Eqs. \ref{chiSN}, \ref{chiCC}, \ref{chiSH0ES}, \ref{chiHIIG}, \ref{chiBAO}, and \ref{chiCMB}, respectively.
We use uniform priors: 
$\Omega_{m0}\in[0,1]$, $H_{0}\in[50,90]$, and $\omega_{b}\in[0.02,0.025]$. Choosing appropriate priors on deviation parameter, $b$, require
some deliberation. So following the logic suggested in \cite{Ravi:2023nsn}, and after some initial pilot
trails, we choose $b\in[0,b_{\rm max}]$, where $b_{\rm max}$ is 1.5, 2, 2, and 2.5 for the models HS1, HS2/ST1, HS3, EXP, and TSUJI, respectively. Here the acronym `HS1' denotes the Hu--Sawicki model with $n_{_{\rm HS}}=1$, and similarly the other acronyms for $f(R)$ models are defined. For the models $f(R) = R - \beta/R^{n} = R - \frac{2\Lambda}{R^{b}}$ (R$\beta$Rn), and $f(R) = R + \alpha\ln(R) - \beta = R - 2\Lambda\left(1 - \frac{b}{2\Lambda}\ln(R)\right)$ (RlnR), we choose $b\in[-1,1]$ and $b\in[-1,3]$, respectively.
\par
For numerical and statistical computations, we have used some of the publicly
available \textsc{python} modules to develop our own \textsc{python} scripts. For solving stiff ODEs
we used `numbalsoda' of \cite{numbalsoda}, for generating MCMC samples we utilized \textsc{emcee} \cite{Foreman-Mackey:2012any},
and the posterior probability distributions of the parameters are plotted with GetDist \cite{Lewis:2019xzd}.%\cite{lewis2019getdist}.
\par
Maximizing above mentioned likelihoods, we generated an initial MCMC sample of
size 875,000 (with 25 walkers, each taking 35,000 steps) for each parameters in each case,
where by `case' we mean any particular model for a given data set combination. These
initial large samples must be tested for convergence and independence. For convergence,
we generously discarded first 5000 steps of each walker (i.e. discarding a sample of size
125,000) to obtained the so-called `burned chains'. Further to ensure the independence
of samples, we thin the burned chains by a factor of 0.75 times the integrated auto-
correlation time. Finally, after burning and thinning we obtained, convergent and independent samples of sizes $\sim 15,000$ to 20,000. From these latter samples only, we made our statistical inferences about model parameters.
%%%%%%%%%%%%%%%%%%%%%%%%%%%%%%%%%%%%%%%%%%%%%%%%%%%%%%
\afterpage{

abcd

\end{comment}

\begin{landscape}
\centering
\fontsize{6}{6}\selectfont 
%\customfontsize % Apply custom font size
%\fontsize{6}{4}\selectfont % Change values as needed
%\captionsetup{width=22cm}
%\captionsetup{width=\linewidth}
\captionof{table}{Results obtained from the MCMC fitting process (for cases without SH0ES prior for $H_{0}$) which include: the median values of model parameters and their 1$\sigma$ (68.26 per cent) confidence intervals, the transition redshift ($z_{\rm t}$), current values of the equation of state parameter for geometric/dark energy component ($w_{\rm DE,0}$) and total equation of state parameter ($w_{\rm tot,0}$), minimum values of $\chi^{2}$, and other quantities essential for statistical inferences, such as 
reduced chi-square values $\chi^{2}_{\rm red} = \chi^{2}_{\rm min}/\nu$  (where the number of degrees of freedom $\nu = N-k$, $N$ represents the total number of data points, and $k$ is the number of model parameters), AIC, BIC, DIC, differences $\Delta$X = X$_{f(R)\rm,\,model}$ $-$ X$_{\Lambda\rm{CDM},\,model}$ (where, $X\equiv$ AIC, BIC and DIC)  -- for all the models and data set combinations considered in this work.}\label{resultstable_Pal}
%\customfontsize % Apply custom font size
%\fontsize{4}{4}\selectfont % Change values as needed
%%%%%%%%%%%%%%%%%%%%%%%%%%%%%%%%%%%%%%%%%%%%%%%%%%%%%%%%
%\begin{longtable}{|c|c|c|c|c|c|c|c|c|c|c|c|c|c|c|c|}
\begin{longtable}{|*{16}{c}|}
\hline
Data & $\Omega_{m0}$ & $b$ & $H_{0}$ & $1000\omega_{b0}$ & $z_{\rm t}$ & $w_{\rm DE0}$ & $w_{\rm tot0}$ &  $\chi^{2}$  & $\chi^{2}_{\rm red}$& AIC & BIC & DIC  & $\Delta$AIC & $\Delta$BIC & $\Delta$DIC\\
%\hline
%\endfirsthead
%\hline
%\multicolumn{2}{c}{{\textit{Table \ref{tab:longtable}: (continued)}}}\\\\
%\multicolumn{16}{c}{{{\small Table \ref{resultstable_Pal}: (continued)}}}\\
%\multicolumn{16}{c}{}\\
%\hline
%Data & $\Omega_{m0}$ & $b$ & $H_{0}$ & $1000\omega_{b0}$ & $z_{\rm t}$ & $w_{\rm DE0}$ & $w_{\rm tot0}$ &  $\chi^{2}$  & $\chi^{2}_{\rm red}$& AIC & BIC & DIC  & $\Delta$AIC & $\Delta$BIC & $\Delta$DIC\\\hline\hline
%\endhead
%\hline
%\multicolumn{2}{|c|}{{\textit{Continued from previous page}}} \\
%\hline
%\endfoot
%\hline
%\endlastfoot
%%%%%%%%%%%%%%%%%%%%%%%%%%%%%%%%%%%%%%%%%%%%%%%%%%%%%%%%
%%%%%%%%%%%%%%%%%%%%%%%%%%%%%%%%%%%%%%%%%%%%%%%%%%%%%%%%
%HS1  &     &        &     &     &     &   &     &     &     &     &     &     &     &   & \\
\hline\multicolumn{16}{|c|}{}\\\multicolumn{16}{|c|}{HS1}\\\multicolumn{16}{|c|}{}\\\hline
SC  &  $0.29_{-0.029}^{+0.040}$  &  $0.75_{-0.355}^{+0.233}$  &  $66.86_{-1.675}^{+1.704}$  &  ---  &  $0.67_{-0.100}^{+0.118}$  &  $-0.86_{-0.070}^{+0.053}$  &  $-0.61_{-0.025}^{+0.028}$  &  1774.82  &  1.03  &  1780.82  &  1797.19  &  1779.67  &  -0.49  &  4.97  &  -1.64 \\
SBC  &  $0.30_{-0.012}^{+0.012}$  &  $0.63_{-0.153}^{+0.144}$  &  $65.18_{-1.226}^{+1.234}$  &  ---  &  $0.64_{-0.032}^{+0.034}$  &  $-0.88_{-0.034}^{+0.035}$  &  $-0.62_{-0.022}^{+0.024}$  &  1801.76  &  1.02  &  1807.76  &  1824.18  &  1807.78  &  -11.63  &  -6.16  &  -11.62 \\
SCH$\textsc{ii}$  &  $0.27_{-0.021}^{+0.034}$  &  $0.80_{-0.307}^{+0.174}$  &  $68.49_{-1.182}^{+1.175}$  &  ---  &  $0.73_{-0.101}^{+0.093}$  &  $-0.85_{-0.059}^{+0.042}$  &  $-0.62_{-0.024}^{+0.026}$  &  2140.86  &  1.12  &  2146.86  &  2163.53  &  2146.63  &  -2.76  &  2.8  &  -2.99 \\
BCH$\textsc{ii}$  &  $0.30_{-0.011}^{+0.012}$  &  $0.21_{-0.139}^{+0.190}$  &  $67.87_{-1.049}^{+0.879}$  &  ---  &  $0.66_{-0.036}^{+0.034}$  &  $-0.97_{-0.022}^{+0.037}$  &  $-0.68_{-0.020}^{+0.030}$  &  408.17  &  1.7  &  414.17  &  424.65  &  414.42  &  1.95  &  5.44  &  2.21 \\
SBCH$\textsc{ii}$  &  $0.31_{-0.010}^{+0.010}$  &  $0.44_{-0.144}^{+0.135}$  &  $66.94_{-0.978}^{+1.003}$  &  ---  &  $0.62_{-0.027}^{+0.028}$  &  $-0.92_{-0.028}^{+0.030}$  &  $-0.64_{-0.020}^{+0.022}$  &  2172.36  &  1.12  &  2178.36  &  2195.08  &  2178.36  &  -5.89  &  -0.32  &  -5.87 \\
SBCH$\textsc{ii}$+CMB  &  $0.31_{-0.006}^{+0.006}$  &  $0.41_{-0.096}^{+0.094}$  &  $67.12_{-0.643}^{+0.654}$  &  $22.73_{-0.136}^{+0.135}$  &  $0.62_{-0.022}^{+0.022}$  &  $-0.93_{-0.020}^{+0.022}$  &  $-0.64_{-0.019}^{+0.020}$  &  2174.15  &  1.12  &  2182.15  &  2204.45  &  2182.18  &  -11.71  &  -6.13  &  -11.69 \\
%ST1/HS2   &     &        &     &     &     &   &     &     &     &     &     &     &     &   & \\
\hline\multicolumn{16}{|c|}{}\\\multicolumn{16}{|c|}{ST1/HS2}\\\multicolumn{16}{|c|}{}\\\hline
SC  &  $0.32_{-0.027}^{+0.028}$  &  $1.11_{-0.400}^{+0.176}$  &  $66.86_{-1.743}^{+1.717}$  &  ---  &  $0.68_{-0.109}^{+0.102}$  &  $-0.87_{-0.086}^{+0.038}$  &  $-0.59_{-0.034}^{+0.027}$  &  1773.32  &  1.03  &  1779.32  &  1795.69  &  1779.49  &  -1.99  &  3.47  &  -1.82 \\
SBC  &  $0.31_{-0.011}^{+0.011}$  &  $1.17_{-0.158}^{+0.116}$  &  $66.21_{-0.798}^{+0.875}$  &  ---  &  $0.74_{-0.044}^{+0.041}$  &  $-0.84_{-0.043}^{+0.018}$  &  $-0.58_{-0.030}^{+0.015}$  &  1799.49  &  1.02  &  1805.49  &  1821.91  &  1805.15  &  -13.9  &  -8.43  &  -14.25 \\
SCH$\textsc{ii}$  &  $0.30_{-0.023}^{+0.026}$  &  $1.17_{-0.245}^{+0.120}$  &  $68.56_{-1.180}^{+1.244}$  &  ---  &  $0.76_{-0.103}^{+0.095}$  &  $-0.85_{-0.067}^{+0.024}$  &  $-0.60_{-0.032}^{+0.023}$  &  2139.25  &  1.12  &  2145.25  &  2161.92  &  2145.61  &  -4.37  &  1.19  &  -4.01 \\
BCH$\textsc{ii}$  &  $0.30_{-0.011}^{+0.012}$  &  $0.51_{-0.341}^{+0.355}$  &  $68.23_{-1.018}^{+0.811}$  &  ---  &  $0.68_{-0.032}^{+0.032}$  &  $-0.98_{-0.021}^{+0.057}$  &  $-0.68_{-0.021}^{+0.045}$  &  408.13  &  1.7  &  414.13  &  424.61  &  413.1  &  1.91  &  5.4  &  0.89 \\
SBCH$\textsc{ii}$  &  $0.31_{-0.010}^{+0.010}$  &  $1.16_{-0.238}^{+0.160}$  &  $67.02_{-0.778}^{+0.859}$  &  ---  &  $0.71_{-0.044}^{+0.039}$  &  $-0.86_{-0.053}^{+0.030}$  &  $-0.59_{-0.037}^{+0.022}$  &  2168.01  &  1.12  &  2174.01  &  2190.73  &  2175.31  &  -10.24  &  -4.67  &  -8.92 \\
SBCH$\textsc{ii}$+CMB  &  $0.31_{-0.004}^{+0.003}$  &  $1.12_{-0.190}^{+0.157}$  &  $66.62_{-0.603}^{+0.680}$  &  $22.70_{-0.132}^{+0.136}$  &  $0.73_{-0.030}^{+0.036}$  &  $-0.85_{-0.051}^{+0.029}$  &  $-0.59_{-0.038}^{+0.020}$  &  2170.8  &  1.12  &  2178.8  &  2201.1  &  2178.75  &  -15.06  &  -9.48  &  -15.12 \\
%HS3   &     &        &     &     &     &   &     &     &     &     &     &     &     &   & \\
\hline\multicolumn{16}{|c|}{}\\\multicolumn{16}{|c|}{HS3}\\\multicolumn{16}{|c|}{}\\\hline
SC  &  $0.35_{-0.022}^{+0.022}$  &  $1.24_{-0.614}^{+0.217}$  &  $66.66_{-1.684}^{+1.740}$  &  ---  &  $0.60_{-0.065}^{+0.071}$  &  $-0.93_{-0.059}^{+0.030}$  &  $-0.61_{-0.027}^{+0.023}$  &  1773.98  &  1.03  &  1779.98  &  1796.35  &  1780.58  &  -1.33  &  4.13  &  -0.73 \\
SBC  &  $0.31_{-0.010}^{+0.010}$  &  $1.36_{-0.138}^{+0.090}$  &  $67.70_{-0.741}^{+0.779}$  &  ---  &  $0.70_{-0.036}^{+0.034}$  &  $-0.90_{-0.026}^{+0.009}$  &  $-0.62_{-0.018}^{+0.011}$  &  1803.13  &  1.02  &  1809.13  &  1825.55  &  1809.53  &  -10.26  &  -4.79  &  -9.87 \\
SCH$\textsc{ii}$  &  $0.33_{-0.021}^{+0.020}$  &  $1.31_{-0.418}^{+0.152}$  &  $68.41_{-1.181}^{+1.205}$  &  ---  &  $0.65_{-0.070}^{+0.069}$  &  $-0.92_{-0.058}^{+0.021}$  &  $-0.62_{-0.026}^{+0.021}$  &  2140.73  &  1.12  &  2146.73  &  2163.4  &  2147.67  &  -2.89  &  2.67  &  -1.95 \\
BCH$\textsc{ii}$  &  $0.30_{-0.011}^{+0.012}$  &  $0.79_{-0.534}^{+0.522}$  &  $68.47_{-0.753}^{+0.717}$  &  ---  &  $0.69_{-0.033}^{+0.036}$  &  $-0.98_{-0.015}^{+0.062}$  &  $-0.69_{-0.019}^{+0.047}$  &  408.17  &  1.7  &  414.17  &  424.65  &  412.52  &  1.95  &  5.44  &  0.31 \\
SBCH$\textsc{ii}$  &  $0.31_{-0.010}^{+0.010}$  &  $1.36_{-0.165}^{+0.103}$  &  $68.07_{-0.660}^{+0.694}$  &  ---  &  $0.70_{-0.035}^{+0.032}$  &  $-0.91_{-0.030}^{+0.011}$  &  $-0.62_{-0.021}^{+0.011}$  &  2169.54  &  1.12  &  2175.54  &  2192.26  &  2176.02  &  -8.71  &  -3.14  &  -8.21 \\
SBCH$\textsc{ii}$+CMB  &  $0.30_{-0.002}^{+0.002}$  &  $1.35_{-0.121}^{+0.081}$  &  $67.41_{-0.560}^{+0.568}$  &  $22.68_{-0.136}^{+0.133}$  &  $0.75_{-0.018}^{+0.014}$  &  $-0.90_{-0.025}^{+0.011}$  &  $-0.63_{-0.018}^{+0.007}$  &  2174.08  &  1.12  &  2182.08  &  2204.37  &  2182.21  &  -11.78  &  -6.21  &  -11.66 \\
\hline
%%%%%%%%%%%%%%%%%%%%%%%%%%%%%%%%%%%%%%%%%%%%%%%%%%%%%%%%
%%%%%%%%%%%%%%%%%%%%%%%%%%%%%%%%%%%%%%%%%%%%%%%%%%%%%%%%
\end{longtable}
\end{landscape}

\begin{landscape}
\centering
\fontsize{6}{6}\selectfont 
%%%%%%%%%%%%%%%%%%%%%%%%%%%%%%%%%%%%%%%%%%%%%%%%%%%%%%%%
%\begin{longtable}{|c|c|c|c|c|c|c|c|c|c|c|c|c|c|c|c|}
\begin{longtable*}{|*{16}{c}|}
\multicolumn{16}{c}{{{\small Table \ref{resultstable_Pal}: (continued)}}}\\
\multicolumn{16}{c}{}\\
\hline
Data & $\Omega_{m0}$ & $b$ & $H_{0}$ & $1000\omega_{b0}$ & $z_{\rm t}$ & $w_{\rm DE0}$ & $w_{\rm tot0}$ &  $\chi^{2}$  & $\chi^{2}_{\rm red}$& AIC & BIC & DIC  & $\Delta$AIC & $\Delta$BIC & $\Delta$DIC\\%\hline\hline
%EXP   &     &        &     &     &     &   &     &     &     &     &     &     &     &   & \\
\hline\multicolumn{16}{|c|}{}\\\multicolumn{16}{|c|}{EXP}\\\multicolumn{16}{|c|}{}\\\hline
SC  &  $0.33_{-0.037}^{+0.033}$  &  $1.50_{-0.403}^{+0.281}$  &  $66.89_{-1.697}^{+1.732}$  &  ---  &  $0.78_{-0.143}^{+0.168}$  &  $-0.81_{-0.092}^{+0.063}$  &  $-0.54_{-0.059}^{+0.032}$  &  1772.38  &  1.02  &  1778.38  &  1794.75  &  1779.71  &  -2.93  &  2.53  &  -1.6 \\
SBC  &  $0.31_{-0.011}^{+0.011}$  &  $1.39_{-0.171}^{+0.267}$  &  $66.34_{-0.940}^{+1.010}$  &  ---  &  $0.84_{-0.057}^{+0.050}$  &  $-0.80_{-0.063}^{+0.050}$  &  $-0.55_{-0.046}^{+0.038}$  &  1800.52  &  1.02  &  1806.52  &  1822.94  &  1806.52  &  -12.87  &  -7.4  &  -12.88 \\
SCH$\textsc{ii}$  &  $0.30_{-0.035}^{+0.035}$  &  $1.59_{-0.317}^{+0.213}$  &  $68.68_{-1.204}^{+1.211}$  &  ---  &  $0.91_{-0.151}^{+0.183}$  &  $-0.77_{-0.080}^{+0.054}$  &  $-0.54_{-0.052}^{+0.029}$  &  2138.16  &  1.12  &  2144.16  &  2160.83  &  2145.25  &  -5.46  &  0.1  &  -4.37 \\
BCH$\textsc{ii}$  &  $0.30_{-0.012}^{+0.015}$  &  $0.67_{-0.464}^{+0.555}$  &  $68.50_{-1.065}^{+0.748}$  &  ---  &  $0.70_{-0.040}^{+0.083}$  &  $-0.99_{-0.011}^{+0.127}$  &  $-0.69_{-0.020}^{+0.098}$  &  408.2  &  1.7  &  414.2  &  424.68  &  413.53  &  1.98  &  5.47  &  1.32 \\
SBCH$\textsc{ii}$  &  $0.32_{-0.010}^{+0.010}$  &  $1.39_{-0.239}^{+0.375}$  &  $67.31_{-0.880}^{+0.947}$  &  ---  &  $0.81_{-0.059}^{+0.046}$  &  $-0.81_{-0.077}^{+0.053}$  &  $-0.55_{-0.057}^{+0.039}$  &  2168.18  &  1.12  &  2174.18  &  2190.9  &  2174.93  &  -10.07  &  -4.5  &  -9.3 \\
SBCH$\textsc{ii}$+CMB  &  $0.30_{-0.004}^{+0.003}$  &  $1.28_{-0.115}^{+0.148}$  &  $66.44_{-0.707}^{+0.744}$  &  $22.68_{-0.133}^{+0.130}$  &  $0.84_{-0.034}^{+0.043}$  &  $-0.84_{-0.042}^{+0.058}$  &  $-0.59_{-0.033}^{+0.043}$  &  2172.23  &  1.12  &  2180.23  &  2202.52  &  2180.64  &  -13.63  &  -8.06  &  -13.23 \\
%TSUJI   &     &        &     &     &     &   &     &     &     &     &     &     &     &   & \\
\hline\multicolumn{16}{|c|}{}\\\multicolumn{16}{|c|}{TSUJI}\\\multicolumn{16}{|c|}{}\\\hline
SC  &  $0.35_{-0.023}^{+0.023}$  &  $1.87_{-0.653}^{+0.222}$  &  $66.66_{-1.727}^{+1.787}$  &  ---  &  $0.61_{-0.078}^{+0.080}$  &  $-0.92_{-0.078}^{+0.029}$  &  $-0.59_{-0.037}^{+0.021}$  &  1773.3  &  1.03  &  1779.3  &  1795.68  &  1779.97  &  -2.01  &  3.46  &  -1.34 \\
SBC  &  $0.31_{-0.010}^{+0.011}$  &  $1.91_{-0.158}^{+0.118}$  &  $68.08_{-0.790}^{+0.824}$  &  ---  &  $0.75_{-0.040}^{+0.038}$  &  $-0.88_{-0.032}^{+0.015}$  &  $-0.61_{-0.023}^{+0.012}$  &  1804.53  &  1.03  &  1810.53  &  1826.95  &  1810.78  &  -8.86  &  -3.39  &  -8.62 \\
SCH$\textsc{ii}$  &  $0.33_{-0.023}^{+0.022}$  &  $1.93_{-0.362}^{+0.163}$  &  $68.48_{-1.227}^{+1.239}$  &  ---  &  $0.68_{-0.082}^{+0.078}$  &  $-0.89_{-0.060}^{+0.022}$  &  $-0.60_{-0.033}^{+0.017}$  &  2139.82  &  1.12  &  2145.82  &  2162.49  &  2146.59  &  -3.8  &  1.76  &  -3.03 \\
BCH$\textsc{ii}$  &  $0.30_{-0.011}^{+0.012}$  &  $1.03_{-0.653}^{+0.698}$  &  $68.68_{-0.709}^{+0.679}$  &  ---  &  $0.70_{-0.036}^{+0.049}$  &  $-1.00_{-0.004}^{+0.074}$  &  $-0.69_{-0.016}^{+0.054}$  &  408.2  &  1.7  &  414.2  &  424.68  &  412.94  &  1.98  &  5.47  &  0.73 \\
SBCH$\textsc{ii}$  &  $0.31_{-0.010}^{+0.010}$  &  $1.92_{-0.180}^{+0.129}$  &  $68.38_{-0.669}^{+0.710}$  &  ---  &  $0.74_{-0.038}^{+0.034}$  &  $-0.88_{-0.034}^{+0.014}$  &  $-0.60_{-0.025}^{+0.011}$  &  2170.0  &  1.12  &  2176.0  &  2192.71  &  2176.32  &  -8.25  &  -2.69  &  -7.91 \\
SBCH$\textsc{ii}$+CMB  &  $0.29_{-0.002}^{+0.002}$  &  $1.89_{-0.140}^{+0.111}$  &  $67.43_{-0.562}^{+0.598}$  &  $22.66_{-0.133}^{+0.132}$  &  $0.81_{-0.027}^{+0.020}$  &  $-0.88_{-0.031}^{+0.022}$  &  $-0.62_{-0.023}^{+0.015}$  &  2177.16  &  1.12  &  2185.16  &  2207.46  &  2185.44  &  -8.7  &  -3.12  &  -8.43 \\
%R$\beta$Rn  &     &        &     &     &     &   &     &     &     &     &     &     &     &   & \\
\hline\multicolumn{16}{|c|}{}\\\multicolumn{16}{|c|}{R$\beta$Rn}\\\multicolumn{16}{|c|}{}\\\hline
SC  &  $0.30_{-0.118}^{+0.095}$  &  $-0.18_{-0.332}^{+0.290}$  &  $66.59_{-1.710}^{+1.708}$  &  ---  &  $0.55_{-0.051}^{+0.059}$  &  $-0.91_{-0.160}^{+0.140}$  &  $-0.64_{-0.021}^{+0.022}$  &  1775.57  &  1.03  &  1781.57  &  1797.94  &  1782.33  &  0.26  &  5.72  &  1.02 \\
SBC  &  $0.30_{-0.014}^{+0.014}$  &  $-0.15_{-0.054}^{+0.055}$  &  $65.51_{-1.513}^{+1.509}$  &  ---  &  $0.58_{-0.032}^{+0.031}$  &  $-0.92_{-0.029}^{+0.029}$  &  $-0.65_{-0.016}^{+0.017}$  &  1807.11  &  1.03  &  1813.11  &  1829.54  &  1813.18  &  -6.28  &  -0.8  &  -6.22 \\
SCH$\textsc{ii}$  &  $0.25_{-0.102}^{+0.095}$  &  $-0.30_{-0.297}^{+0.291}$  &  $68.39_{-1.185}^{+1.193}$  &  ---  &  $0.60_{-0.052}^{+0.056}$  &  $-0.86_{-0.137}^{+0.111}$  &  $-0.65_{-0.020}^{+0.021}$  &  2142.32  &  1.12  &  2148.32  &  2164.99  &  2149.56  &  -1.3  &  4.26  &  -0.06 \\
BCH$\textsc{ii}$  &  $0.30_{-0.011}^{+0.011}$  &  $0.04_{-0.065}^{+0.069}$  &  $69.47_{-1.279}^{+1.290}$  &  ---  &  $0.70_{-0.045}^{+0.042}$  &  $-1.02_{-0.026}^{+0.028}$  &  $-0.71_{-0.019}^{+0.021}$  &  408.05  &  1.7  &  414.05  &  424.53  &  414.13  &  1.83  &  5.32  &  1.92 \\
SBCH$\textsc{ii}$  &  $0.31_{-0.011}^{+0.011}$  &  $-0.08_{-0.044}^{+0.046}$  &  $67.66_{-1.096}^{+1.117}$  &  ---  &  $0.60_{-0.031}^{+0.030}$  &  $-0.96_{-0.023}^{+0.023}$  &  $-0.67_{-0.015}^{+0.016}$  &  2176.76  &  1.12  &  2182.76  &  2199.48  &  2182.84  &  -1.49  &  4.08  &  -1.39 \\
SBCH$\textsc{ii}$+CMB  &  $0.31_{-0.007}^{+0.007}$  &  $-0.07_{-0.020}^{+0.020}$  &  $67.82_{-0.582}^{+0.603}$  &  $22.73_{-0.135}^{+0.134}$  &  $0.60_{-0.030}^{+0.030}$  &  $-0.97_{-0.011}^{+0.012}$  &  $-0.67_{-0.014}^{+0.014}$  &  2178.72  &  1.12  &  2186.72  &  2209.01  &  2186.72  &  -7.14  &  -1.57  &  -7.15 \\
%RlnR   &     &        &     &     &     &   &     &     &     &     &     &     &     &   & \\
\hline\multicolumn{16}{|c|}{}\\\multicolumn{16}{|c|}{RlnR}\\\multicolumn{16}{|c|}{}\\\hline
SC  &  $0.25_{-0.055}^{+0.099}$  &  $1.32_{-1.200}^{+0.614}$  &  $66.97_{-1.732}^{+1.767}$  &  ---  &  $0.65_{-0.118}^{+0.358}$  &  $-0.82_{-0.159}^{+0.106}$  &  $-0.62_{-0.030}^{+0.041}$  &  1774.86  &  1.03  &  1780.86  &  1797.24  &  1781.09  &  -0.45  &  5.02  &  -0.22 \\
SBC  &  $0.29_{-0.013}^{+0.013}$  &  $0.74_{-0.223}^{+0.210}$  &  $64.87_{-1.435}^{+1.493}$  &  ---  &  $0.58_{-0.030}^{+0.031}$  &  $-0.90_{-0.032}^{+0.034}$  &  $-0.64_{-0.019}^{+0.021}$  &  1805.5  &  1.03  &  1811.5  &  1827.93  &  1811.5  &  -7.89  &  -2.41  &  -7.9 \\
SCH$\textsc{ii}$  &  $0.21_{-0.032}^{+0.066}$  &  $1.77_{-0.806}^{+0.355}$  &  $68.72_{-1.192}^{+1.207}$  &  ---  &  $0.88_{-0.263}^{+0.319}$  &  $-0.77_{-0.115}^{+0.071}$  &  $-0.61_{-0.036}^{+0.040}$  &  2140.25  &  1.12  &  2146.25  &  2162.92  &  2147.08  &  -3.37  &  2.19  &  -2.54 \\
BCH$\textsc{ii}$  &  $0.30_{-0.011}^{+0.011}$  &  $-0.13_{-0.304}^{+0.285}$  &  $69.36_{-1.334}^{+1.368}$  &  ---  &  $0.70_{-0.047}^{+0.044}$  &  $-1.01_{-0.027}^{+0.029}$  &  $-0.71_{-0.020}^{+0.022}$  &  408.03  &  1.7  &  414.03  &  424.51  &  414.02  &  1.81  &  5.3  &  1.81 \\
SBCH$\textsc{ii}$  &  $0.31_{-0.011}^{+0.011}$  &  $0.39_{-0.192}^{+0.187}$  &  $67.35_{-1.121}^{+1.146}$  &  ---  &  $0.60_{-0.030}^{+0.030}$  &  $-0.95_{-0.025}^{+0.026}$  &  $-0.66_{-0.016}^{+0.017}$  &  2176.21  &  1.12  &  2182.21  &  2198.93  &  2182.22  &  -2.04  &  3.53  &  -2.01 \\
SBCH$\textsc{ii}$+CMB  &  $0.31_{-0.007}^{+0.007}$  &  $0.31_{-0.086}^{+0.082}$  &  $67.74_{-0.608}^{+0.615}$  &  $22.73_{-0.137}^{+0.135}$  &  $0.60_{-0.030}^{+0.030}$  &  $-0.96_{-0.012}^{+0.014}$  &  $-0.66_{-0.015}^{+0.016}$  &  2178.29  &  1.12  &  2186.29  &  2208.59  &  2186.33  &  -7.57  &  -1.99  &  -7.54 \\
\hline
%%%%%%%%%%%%%%%%%%%%%%%%%%%%%%%%%%%%%%%%%%%%%%%%%%%%%%%%
%%%%%%%%%%%%%%%%%%%%%%%%%%%%%%%%%%%%%%%%%%%%%%%%%%%%%%%%
\end{longtable*}
\end{landscape}

%\end{document}

abcd
\end{comment}

\addtocounter{table}{-1}
\begin{landscape}
\centering
\fontsize{5.5}{6}\selectfont 
%\customfontsize % Apply custom font size
%\fontsize{6}{4}\selectfont % Change values as needed
%\captionsetup{width=22cm}
%\captionsetup{width=\linewidth}
\captionof{table}{Results obtained from the MCMC fitting process (for cases with SH0ES prior for $H_{0}$) which include: the median values of model parameters and their 1$\sigma$ (68.26 per cent) confidence intervals, the transition redshift ($z_{\rm t}$), current values of the equation of state parameter for geometric/dark energy component ($w_{\rm DE,0}$) and total equation of state parameter ($w_{\rm tot,0}$), minimum values of $\chi^{2}$, and other quantities essential for statistical inferences, such as 
reduced chi-square values $\chi^{2}_{\rm red} = \chi^{2}_{\rm min}/\nu$  (where the number of degrees of freedom $\nu = N-k$, $N$ represents the total number of data points, and $k$ is the number of model parameters), AIC, BIC, DIC, differences $\Delta$X = X$_{f(R)\rm,\,model}$ $-$ X$_{\Lambda\rm{CDM},\,model}$ (where, $X\equiv$ AIC, BIC and DIC)  -- for all the models and data set combinations considered in this work.}\label{resultsH0table_Pal}
%\customfontsize % Apply custom font size
%\fontsize{4}{4}\selectfont % Change values as needed
%%%%%%%%%%%%%%%%%%%%%%%%%%%%%%%%%%%%%%%%%%%%%%%%%%%%%%%%
%\begin{longtable}{|c|c|c|c|c|c|c|c|c|c|c|c|c|c|c|c|}
\begin{longtable}{|*{16}{c}|}
\hline
Data & $\Omega_{m0}$ & $b$ & $H_{0}$ & $1000\omega_{b0}$ & $z_{\rm t}$ & $w_{\rm DE0}$ & $w_{\rm tot0}$ &  $\chi^{2}$  & $\chi^{2}_{\rm red}$& AIC & BIC & DIC  & $\Delta$AIC & $\Delta$BIC & $\Delta$DIC\\
%\hline
%\endfirsthead
%\hline
%\multicolumn{2}{c}{{\textit{Table \ref{tab:longtable}: (continued)}}}\\\\
%\multicolumn{16}{c}{{{\small Table \ref{resultsH0table_Pal}: (continued)}}}\\
%\multicolumn{16}{c}{}\\
%\hline
%Data & $\Omega_{m0}$ & $b$ & $H_{0}$ & $1000\omega_{b0}$ & $z_{\rm t}$ & $w_{\rm DE0}$ & $w_{\rm tot0}$ &  $\chi^{2}$  & $\chi^{2}_{\rm red}$& AIC & BIC & DIC  & $\Delta$AIC & $\Delta$BIC & $\Delta$DIC\\\hline\hline
%\endhead
%\hline
%\multicolumn{2}{|c|}{{\textit{Continued from previous page}}} \\
%\hline
%\endfoot
%\hline
%\endlastfoot
%%%%%%%%%%%%%%%%%%%%%%%%%%%%%%%%%%%%%%%%%%%%%%%%%%%%%%%%
%%%%%%%%%%%%%%%%%%%%%%%%%%%%%%%%%%%%%%%%%%%%%%%%%%%%%%%%
%HS1  &     &        &     &     &     &   &     &     &     &     &     &     &     &   & \\
\hline\multicolumn{16}{|c|}{}\\\multicolumn{16}{|c|}{HS1}\\\multicolumn{16}{|c|}{}\\\hline
SC$H_{0}$  &  $0.27_{-0.020}^{+0.032}$  &  $0.78_{-0.297}^{+0.168}$  &  $71.38_{-0.887}^{+0.898}$  &  ---  &  $0.74_{-0.097}^{+0.087}$  &  $-0.86_{-0.055}^{+0.039}$  &  $-0.63_{-0.022}^{+0.024}$  &  1783.26  &  1.03  &  1789.26  &  1805.63  &  1789.21  &  -2.9  &  2.56  &  -2.97 \\
SBC$H_{0}$  &  $0.32_{-0.010}^{+0.010}$  &  $0.14_{-0.091}^{+0.115}$  &  $69.76_{-0.748}^{+0.720}$  &  ---  &  $0.60_{-0.025}^{+0.026}$  &  $-0.98_{-0.015}^{+0.021}$  &  $-0.66_{-0.014}^{+0.016}$  &  1824.39  &  1.04  &  1830.39  &  1846.81  &  1829.9  &  1.48  &  6.95  &  0.99 \\
SCH$\textsc{ii}H_{0}$  &  $0.26_{-0.017}^{+0.027}$  &  $0.79_{-0.255}^{+0.143}$  &  $71.07_{-0.787}^{+0.796}$  &  ---  &  $0.77_{-0.090}^{+0.074}$  &  $-0.86_{-0.048}^{+0.034}$  &  $-0.64_{-0.022}^{+0.023}$  &  2148.35  &  1.12  &  2154.35  &  2171.03  &  2154.82  &  -4.48  &  1.09  &  -4.02 \\
BCH$\textsc{ii}H_{0}$  &  $0.31_{-0.011}^{+0.011}$  &  $0.05_{-0.037}^{+0.072}$  &  $69.77_{-0.595}^{+0.565}$  &  ---  &  $0.65_{-0.030}^{+0.029}$  &  $-0.99_{-0.006}^{+0.011}$  &  $-0.69_{-0.012}^{+0.014}$  &  420.2  &  1.74  &  426.2  &  436.69  &  427.85  &  1.95  &  5.44  &  3.59 \\
SBCH$\textsc{ii}H_{0}$  &  $0.32_{-0.009}^{+0.010}$  &  $0.13_{-0.084}^{+0.109}$  &  $69.73_{-0.658}^{+0.642}$  &  ---  &  $0.61_{-0.024}^{+0.025}$  &  $-0.98_{-0.014}^{+0.019}$  &  $-0.67_{-0.013}^{+0.015}$  &  2189.72  &  1.13  &  2195.72  &  2212.44  &  2195.24  &  1.64  &  7.21  &  1.15 \\
SBCH$\textsc{ii}H_{0}$+CMB  &  $0.30_{-0.005}^{+0.005}$  &  $0.27_{-0.092}^{+0.091}$  &  $68.83_{-0.565}^{+0.563}$  &  $22.70_{-0.137}^{+0.134}$  &  $0.65_{-0.020}^{+0.020}$  &  $-0.96_{-0.017}^{+0.018}$  &  $-0.67_{-0.017}^{+0.018}$  &  2197.05  &  1.13  &  2205.05  &  2227.35  &  2205.06  &  -2.6  &  2.98  &  -2.61 \\
%ST1/HS2    &     &        &     &     &     &   &     &     &     &     &     &     &     &   & \\
\hline\multicolumn{16}{|c|}{}\\\multicolumn{16}{|c|}{ST1/HS2}\\\multicolumn{16}{|c|}{}\\\hline
SC$H_{0}$  &  $0.29_{-0.023}^{+0.026}$  &  $1.17_{-0.274}^{+0.126}$  &  $71.47_{-0.919}^{+0.890}$  &  ---  &  $0.79_{-0.114}^{+0.096}$  &  $-0.85_{-0.071}^{+0.027}$  &  $-0.61_{-0.033}^{+0.022}$  &  1782.09  &  1.03  &  1788.09  &  1804.47  &  1788.37  &  -4.07  &  1.4  &  -3.81 \\
SBC$H_{0}$  &  $0.32_{-0.010}^{+0.010}$  &  $0.52_{-0.308}^{+0.242}$  &  $69.85_{-0.796}^{+0.750}$  &  ---  &  $0.62_{-0.027}^{+0.028}$  &  $-0.97_{-0.022}^{+0.035}$  &  $-0.66_{-0.017}^{+0.025}$  &  1823.5  &  1.04  &  1829.5  &  1845.92  &  1829.01  &  0.59  &  6.06  &  0.1 \\
SCH$\textsc{ii}H_{0}$  &  $0.28_{-0.021}^{+0.023}$  &  $1.19_{-0.199}^{+0.100}$  &  $71.18_{-0.799}^{+0.795}$  &  ---  &  $0.84_{-0.105}^{+0.091}$  &  $-0.84_{-0.057}^{+0.020}$  &  $-0.61_{-0.030}^{+0.020}$  &  2146.72  &  1.12  &  2152.72  &  2169.39  &  2153.04  &  -6.11  &  -0.55  &  -5.8 \\
BCH$\textsc{ii}H_{0}$  &  $0.31_{-0.011}^{+0.011}$  &  $0.20_{-0.140}^{+0.207}$  &  $69.87_{-0.567}^{+0.565}$  &  ---  &  $0.66_{-0.028}^{+0.029}$  &  $-1.00_{-0.003}^{+0.011}$  &  $-0.69_{-0.012}^{+0.014}$  &  420.19  &  1.74  &  426.19  &  436.68  &  426.11  &  1.94  &  5.43  &  1.85 \\
SBCH$\textsc{ii}H_{0}$  &  $0.32_{-0.009}^{+0.010}$  &  $0.49_{-0.296}^{+0.239}$  &  $69.81_{-0.695}^{+0.643}$  &  ---  &  $0.63_{-0.025}^{+0.027}$  &  $-0.98_{-0.019}^{+0.032}$  &  $-0.66_{-0.016}^{+0.023}$  &  2188.92  &  1.13  &  2194.92  &  2211.64  &  2194.27  &  0.84  &  6.41  &  0.18 \\
SBCH$\textsc{ii}H_{0}$+CMB  &  $0.30_{-0.005}^{+0.005}$  &  $0.77_{-0.171}^{+0.142}$  &  $68.73_{-0.581}^{+0.597}$  &  $22.68_{-0.138}^{+0.135}$  &  $0.70_{-0.005}^{+0.007}$  &  $-0.94_{-0.027}^{+0.032}$  &  $-0.66_{-0.023}^{+0.026}$  &  2197.14  &  1.13  &  2205.14  &  2227.43  &  2205.62  &  -2.51  &  3.06  &  -2.05 \\
%HS3     &     &        &     &     &     &   &     &     &     &     &     &     &     &   & \\
\hline\multicolumn{16}{|c|}{}\\\multicolumn{16}{|c|}{HS3}\\\multicolumn{16}{|c|}{}\\\hline
SC$H_{0}$  &  $0.32_{-0.020}^{+0.020}$  &  $1.30_{-0.466}^{+0.162}$  &  $71.35_{-0.888}^{+0.905}$  &  ---  &  $0.67_{-0.073}^{+0.069}$  &  $-0.92_{-0.062}^{+0.023}$  &  $-0.63_{-0.027}^{+0.021}$  &  1783.46  &  1.03  &  1789.46  &  1805.83  &  1790.42  &  -2.7  &  2.76  &  -1.76 \\
SBC$H_{0}$  &  $0.33_{-0.010}^{+0.010}$  &  $1.22_{-0.576}^{+0.276}$  &  $69.84_{-0.691}^{+0.720}$  &  ---  &  $0.65_{-0.038}^{+0.038}$  &  $-0.94_{-0.052}^{+0.038}$  &  $-0.63_{-0.035}^{+0.027}$  &  1820.21  &  1.03  &  1826.21  &  1842.64  &  1828.68  &  -2.7  &  2.78  &  -0.23 \\
SCH$\textsc{ii}H_{0}$  &  $0.31_{-0.019}^{+0.019}$  &  $1.34_{-0.314}^{+0.127}$  &  $71.09_{-0.791}^{+0.808}$  &  ---  &  $0.71_{-0.073}^{+0.067}$  &  $-0.91_{-0.053}^{+0.016}$  &  $-0.64_{-0.025}^{+0.019}$  &  2148.79  &  1.12  &  2154.79  &  2171.46  &  2155.88  &  -4.04  &  1.52  &  -2.96 \\
BCH$\textsc{ii}H_{0}$  &  $0.31_{-0.011}^{+0.011}$  &  $0.39_{-0.268}^{+0.347}$  &  $69.91_{-0.568}^{+0.563}$  &  ---  &  $0.66_{-0.028}^{+0.028}$  &  $-1.00_{-0.002}^{+0.012}$  &  $-0.69_{-0.012}^{+0.015}$  &  420.19  &  1.74  &  426.19  &  436.68  &  425.58  &  1.94  &  5.43  &  1.32 \\
SBCH$\textsc{ii}H_{0}$  &  $0.32_{-0.009}^{+0.010}$  &  $1.21_{-0.607}^{+0.297}$  &  $69.78_{-0.659}^{+0.646}$  &  ---  &  $0.66_{-0.038}^{+0.036}$  &  $-0.94_{-0.050}^{+0.041}$  &  $-0.64_{-0.034}^{+0.029}$  &  2185.25  &  1.13  &  2191.25  &  2207.97  &  2194.17  &  -2.83  &  2.74  &  0.08 \\
SBCH$\textsc{ii}H_{0}$+CMB  &  $0.29_{-0.003}^{+0.002}$  &  $1.23_{-0.231}^{+0.164}$  &  $68.82_{-0.563}^{+0.574}$  &  $22.67_{-0.134}^{+0.132}$  &  $0.74_{-0.021}^{+0.025}$  &  $-0.93_{-0.037}^{+0.032}$  &  $-0.65_{-0.030}^{+0.022}$  &  2197.76  &  1.13  &  2205.76  &  2228.06  &  2205.76  &  -1.89  &  3.69  &  -1.91 \\
\hline
%%%%%%%%%%%%%%%%%%%%%%%%%%%%%%%%%%%%%%%%%%%%%%%%%%%%%%%%
%%%%%%%%%%%%%%%%%%%%%%%%%%%%%%%%%%%%%%%%%%%%%%%%%%%%%%%%
\end{longtable}
\end{landscape}

\begin{landscape}
\centering
\fontsize{5.5}{6}\selectfont 
%%%%%%%%%%%%%%%%%%%%%%%%%%%%%%%%%%%%%%%%%%%%%%%%%%%%%%%%
%\begin{longtable}{|c|c|c|c|c|c|c|c|c|c|c|c|c|c|c|c|}
\begin{longtable}{|*{16}{c}|}
\multicolumn{16}{c}{{{\small Table \ref{resultsH0table_Pal}: (continued)}}}\\
\multicolumn{16}{c}{}\\
\hline
Data & $\Omega_{m0}$ & $b$ & $H_{0}$ & $1000\omega_{b0}$ & $z_{\rm t}$ & $w_{\rm DE0}$ & $w_{\rm tot0}$ &  $\chi^{2}$  & $\chi^{2}_{\rm red}$& AIC & BIC & DIC  & $\Delta$AIC & $\Delta$BIC & $\Delta$DIC\\%\hline\hline
%EXP     &     &        &     &     &     &   &     &     &     &     &     &     &     &   & \\
\hline\multicolumn{16}{|c|}{}\\\multicolumn{16}{|c|}{EXP}\\\multicolumn{16}{|c|}{}\\\hline
SC$H_{0}$  &  $0.29_{-0.039}^{+0.035}$  &  $1.54_{-0.307}^{+0.273}$  &  $71.48_{-0.913}^{+0.901}$  &  ---  &  $0.95_{-0.169}^{+0.215}$  &  $-0.78_{-0.087}^{+0.065}$  &  $-0.56_{-0.052}^{+0.040}$  &  1781.68  &  1.03  &  1787.68  &  1804.05  &  1788.64  &  -4.48  &  0.98  &  -3.54 \\
SBC$H_{0}$  &  $0.33_{-0.010}^{+0.010}$  &  $0.90_{-0.474}^{+0.182}$  &  $69.91_{-0.790}^{+0.741}$  &  ---  &  $0.65_{-0.045}^{+0.053}$  &  $-0.95_{-0.046}^{+0.049}$  &  $-0.64_{-0.029}^{+0.036}$  &  1821.87  &  1.03  &  1827.87  &  1844.3  &  1828.65  &  -1.04  &  4.44  &  -0.26 \\
SCH$\textsc{ii}H_{0}$  &  $0.27_{-0.038}^{+0.035}$  &  $1.59_{-0.269}^{+0.227}$  &  $71.26_{-0.835}^{+0.796}$  &  ---  &  $1.05_{-0.182}^{+0.230}$  &  $-0.76_{-0.080}^{+0.058}$  &  $-0.55_{-0.050}^{+0.037}$  &  2145.85  &  1.12  &  2151.85  &  2168.52  &  2152.69  &  -6.98  &  -1.42  &  -6.15 \\
BCH$\textsc{ii}H_{0}$  &  $0.31_{-0.011}^{+0.011}$  &  $0.39_{-0.266}^{+0.283}$  &  $69.95_{-0.555}^{+0.555}$  &  ---  &  $0.66_{-0.029}^{+0.029}$  &  $-1.00_{-0.000}^{+0.012}$  &  $-0.69_{-0.012}^{+0.014}$  &  420.19  &  1.74  &  426.19  &  436.68  &  424.93  &  1.94  &  5.43  &  0.67 \\
SBCH$\textsc{ii}H_{0}$  &  $0.32_{-0.010}^{+0.010}$  &  $0.88_{-0.466}^{+0.185}$  &  $69.83_{-0.709}^{+0.664}$  &  ---  &  $0.66_{-0.041}^{+0.048}$  &  $-0.96_{-0.043}^{+0.046}$  &  $-0.65_{-0.029}^{+0.035}$  &  2187.14  &  1.13  &  2193.14  &  2209.86  &  2193.88  &  -0.94  &  4.63  &  -0.21 \\
SBCH$\textsc{ii}H_{0}$+CMB  &  $0.29_{-0.004}^{+0.004}$  &  $1.03_{-0.149}^{+0.114}$  &  $68.71_{-0.611}^{+0.632}$  &  $22.66_{-0.136}^{+0.136}$  &  $0.78_{-0.031}^{+0.029}$  &  $-0.92_{-0.035}^{+0.035}$  &  $-0.65_{-0.029}^{+0.028}$  &  2198.76  &  1.13  &  2206.76  &  2229.06  &  2207.33  &  -0.89  &  4.69  &  -0.34 \\
%TSUJI     &     &        &     &     &     &   &     &     &     &     &     &     &     &   & \\
\hline\multicolumn{16}{|c|}{}\\\multicolumn{16}{|c|}{TSUJI}\\\multicolumn{16}{|c|}{}\\\hline
SC$H_{0}$  &  $0.32_{-0.022}^{+0.022}$  &  $1.92_{-0.424}^{+0.174}$  &  $71.41_{-0.886}^{+0.900}$  &  ---  &  $0.71_{-0.092}^{+0.082}$  &  $-0.89_{-0.072}^{+0.024}$  &  $-0.61_{-0.038}^{+0.018}$  &  1782.75  &  1.03  &  1788.75  &  1805.12  &  1789.55  &  -3.41  &  2.05  &  -2.63 \\
SBC$H_{0}$  &  $0.33_{-0.010}^{+0.010}$  &  $1.88_{-0.521}^{+0.226}$  &  $70.08_{-0.696}^{+0.691}$  &  ---  &  $0.69_{-0.059}^{+0.040}$  &  $-0.90_{-0.078}^{+0.024}$  &  $-0.60_{-0.054}^{+0.017}$  &  1819.29  &  1.03  &  1825.29  &  1841.72  &  1826.9  &  -3.62  &  1.86  &  -2.01 \\
SCH$\textsc{ii}H_{0}$  &  $0.31_{-0.021}^{+0.021}$  &  $1.96_{-0.258}^{+0.142}$  &  $71.15_{-0.806}^{+0.806}$  &  ---  &  $0.76_{-0.084}^{+0.077}$  &  $-0.88_{-0.048}^{+0.020}$  &  $-0.61_{-0.029}^{+0.016}$  &  2147.68  &  1.12  &  2153.68  &  2170.35  &  2154.3  &  -5.15  &  0.41  &  -4.54 \\
BCH$\textsc{ii}H_{0}$  &  $0.31_{-0.011}^{+0.011}$  &  $0.76_{-0.464}^{+0.481}$  &  $69.95_{-0.556}^{+0.547}$  &  ---  &  $0.66_{-0.029}^{+0.030}$  &  $-1.00_{-0.000}^{+0.014}$  &  $-0.69_{-0.012}^{+0.015}$  &  420.18  &  1.74  &  426.18  &  436.67  &  424.91  &  1.93  &  5.42  &  0.65 \\
SBCH$\textsc{ii}H_{0}$  &  $0.32_{-0.009}^{+0.009}$  &  $1.94_{-0.478}^{+0.192}$  &  $69.94_{-0.616}^{+0.647}$  &  ---  &  $0.71_{-0.056}^{+0.036}$  &  $-0.89_{-0.073}^{+0.019}$  &  $-0.60_{-0.052}^{+0.014}$  &  2183.91  &  1.12  &  2189.91  &  2206.63  &  2191.69  &  -4.17  &  1.4  &  -2.4 \\
SBCH$\textsc{ii}H_{0}$+CMB  &  $0.29_{-0.003}^{+0.002}$  &  $1.69_{-0.403}^{+0.203}$  &  $68.93_{-0.576}^{+0.624}$  &  $22.65_{-0.138}^{+0.136}$  &  $0.78_{-0.055}^{+0.039}$  &  $-0.92_{-0.057}^{+0.045}$  &  $-0.66_{-0.044}^{+0.033}$  &  2200.05  &  1.13  &  2208.05  &  2230.35  &  2209.08  &  0.4  &  5.98  &  1.41 \\
%R$\beta$Rn     &     &        &     &     &     &   &     &     &     &     &     &     &     &   & \\
\hline\multicolumn{16}{|c|}{}\\\multicolumn{16}{|c|}{R$\beta$Rn}\\\multicolumn{16}{|c|}{}\\\hline
SC$H_{0}$  &  $0.24_{-0.098}^{+0.091}$  &  $-0.32_{-0.287}^{+0.285}$  &  $71.33_{-0.909}^{+0.895}$  &  ---  &  $0.62_{-0.051}^{+0.055}$  &  $-0.85_{-0.127}^{+0.105}$  &  $-0.65_{-0.019}^{+0.019}$  &  1784.44  &  1.03  &  1790.44  &  1806.81  &  1792.01  &  -1.72  &  3.74  &  -0.17 \\
SBC$H_{0}$  &  $0.33_{-0.010}^{+0.010}$  &  $0.02_{-0.038}^{+0.040}$  &  $70.66_{-0.854}^{+0.872}$  &  ---  &  $0.62_{-0.029}^{+0.029}$  &  $-1.01_{-0.018}^{+0.019}$  &  $-0.68_{-0.014}^{+0.014}$  &  1824.75  &  1.04  &  1830.75  &  1847.18  &  1830.77  &  1.84  &  7.32  &  1.86 \\
SCH$\textsc{ii}H_{0}$  &  $0.21_{-0.095}^{+0.091}$  &  $-0.38_{-0.283}^{+0.286}$  &  $71.08_{-0.794}^{+0.803}$  &  ---  &  $0.65_{-0.052}^{+0.053}$  &  $-0.84_{-0.121}^{+0.096}$  &  $-0.66_{-0.018}^{+0.018}$  &  2150.09  &  1.12  &  2156.09  &  2172.76  &  2157.93  &  -2.74  &  2.82  &  -0.91 \\
BCH$\textsc{ii}H_{0}$  &  $0.30_{-0.010}^{+0.010}$  &  $0.14_{-0.053}^{+0.055}$  &  $71.62_{-0.808}^{+0.813}$  &  ---  &  $0.74_{-0.038}^{+0.037}$  &  $-1.05_{-0.018}^{+0.019}$  &  $-0.74_{-0.016}^{+0.017}$  &  413.22  &  1.71  &  419.22  &  429.72  &  419.27  &  -5.03  &  -1.53  &  -4.99 \\
SBCH$\textsc{ii}H_{0}$  &  $0.32_{-0.010}^{+0.010}$  &  $0.02_{-0.036}^{+0.039}$  &  $70.55_{-0.764}^{+0.772}$  &  ---  &  $0.63_{-0.029}^{+0.029}$  &  $-1.01_{-0.017}^{+0.017}$  &  $-0.69_{-0.013}^{+0.014}$  &  2189.84  &  1.13  &  2195.84  &  2212.55  &  2195.86  &  1.76  &  7.32  &  1.77 \\
SBCH$\textsc{ii}H_{0}$+CMB  &  $0.30_{-0.006}^{+0.006}$  &  $-0.05_{-0.020}^{+0.020}$  &  $69.13_{-0.532}^{+0.529}$  &  $22.70_{-0.137}^{+0.136}$  &  $0.63_{-0.029}^{+0.029}$  &  $-0.97_{-0.010}^{+0.011}$  &  $-0.68_{-0.013}^{+0.014}$  &  2197.75  &  1.13  &  2205.75  &  2228.05  &  2205.78  &  -1.9  &  3.68  &  -1.89 \\
%RlnR     &     &        &     &     &     &   &     &     &     &     &     &     &     &   & \\
\hline\multicolumn{16}{|c|}{}\\\multicolumn{16}{|c|}{RlnR}\\\multicolumn{16}{|c|}{}\\\hline
SC$H_{0}$  &  $0.20_{-0.032}^{+0.063}$  &  $1.83_{-0.806}^{+0.360}$  &  $71.48_{-0.904}^{+0.908}$  &  ---  &  $0.91_{-0.268}^{+0.343}$  &  $-0.77_{-0.109}^{+0.070}$  &  $-0.62_{-0.034}^{+0.038}$  &  1782.73  &  1.03  &  1788.73  &  1805.1  &  1789.23  &  -3.43  &  2.03  &  -2.95 \\
SBC$H_{0}$  &  $0.32_{-0.010}^{+0.010}$  &  $-0.06_{-0.173}^{+0.164}$  &  $70.64_{-0.890}^{+0.892}$  &  ---  &  $0.62_{-0.029}^{+0.030}$  &  $-1.01_{-0.019}^{+0.020}$  &  $-0.68_{-0.014}^{+0.015}$  &  1824.74  &  1.04  &  1830.74  &  1847.16  &  1830.7  &  1.83  &  7.3  &  1.79 \\
SCH$\textsc{ii}H_{0}$  &  $0.18_{-0.023}^{+0.044}$  &  $2.02_{-0.565}^{+0.250}$  &  $71.26_{-0.807}^{+0.798}$  &  ---  &  $1.10_{-0.357}^{+0.240}$  &  $-0.74_{-0.088}^{+0.047}$  &  $-0.61_{-0.037}^{+0.034}$  &  2146.92  &  1.12  &  2152.92  &  2169.59  &  2153.52  &  -5.91  &  -0.35  &  -5.32 \\
BCH$\textsc{ii}H_{0}$  &  $0.30_{-0.010}^{+0.010}$  &  $-0.59_{-0.235}^{+0.225}$  &  $71.67_{-0.823}^{+0.834}$  &  ---  &  $0.74_{-0.039}^{+0.039}$  &  $-1.05_{-0.017}^{+0.018}$  &  $-0.73_{-0.015}^{+0.016}$  &  412.78  &  1.71  &  418.78  &  429.27  &  418.79  &  -5.47  &  -1.98  &  -5.47 \\
SBCH$\textsc{ii}H_{0}$  &  $0.32_{-0.010}^{+0.010}$  &  $-0.08_{-0.165}^{+0.159}$  &  $70.54_{-0.779}^{+0.791}$  &  ---  &  $0.63_{-0.029}^{+0.029}$  &  $-1.01_{-0.018}^{+0.019}$  &  $-0.68_{-0.013}^{+0.014}$  &  2189.81  &  1.13  &  2195.81  &  2212.53  &  2195.78  &  1.73  &  7.3  &  1.69 \\
SBCH$\textsc{ii}H_{0}$+CMB  &  $0.30_{-0.006}^{+0.006}$  &  $0.24_{-0.084}^{+0.082}$  &  $69.10_{-0.530}^{+0.533}$  &  $22.70_{-0.137}^{+0.136}$  &  $0.63_{-0.029}^{+0.029}$  &  $-0.97_{-0.011}^{+0.012}$  &  $-0.68_{-0.014}^{+0.014}$  &  2197.7  &  1.13  &  2205.7  &  2227.99  &  2205.71  &  -1.95  &  3.62  &  -1.96 \\
\hline
%%%%%%%%%%%%%%%%%%%%%%%%%%%%%%%%%%%%%%%%%%%%%%%%%%%%%%%%
%%%%%%%%%%%%%%%%%%%%%%%%%%%%%%%%%%%%%%%%%%%%%%%%%%%%%%%%
\end{longtable}
\end{landscape}

%\end{document}

}
\begin{comment}
\afterpage{
\input{Palatini_Table_1A_userdefined_break}
\input{Palatini_Table_2A_userdefined_break}
}
\end{comment}
%%%%%%%%%%%%%%%%%%%%%%%%%%%%%%%%%%%%%%%%%%%%%%%%%%%%%%
\section{Observational constraints}% on $\lowercase{f}(R)$ models}
\label{results_Pal}

Here we give a detailed account of obtained parameters of all seven $f(R)$ models
investigated in this work. For assessments of any 
$f(R)$ models, one also needs, as a
reference, constraints on the $\Lambda$CDM model with same data set. The results of $\Lambda$CDM
model is not presented here as the same can be found in \cite{Ravi:2023nsn}.
\paragraph{Acronyms and colour conventions:}\label{acc}
The six data set combinations explored in this
work are: (i) SNIa+CC (SC), (ii) SNIa+BAO+CC (SBC), (iii) SNIa+CC+H$\textsc{ii}$G (SCH$\textsc{ii}$), (iv) BAO+CC+H$\textsc{ii}$G (BCH$\textsc{ii}$), (v) SNIa+ BAO+CC+H$\textsc{ii}$G (SBCH$\textsc{ii}$), and (vi) SNIa+BAO+CC+H$\textsc{ii}$G+CMB (SBCH$\textsc{ii}$+CMB). To study \emph{Hubble} tension, we also
use SH0ES prior for $H_{0}$ to aforementioned six data sets (so effectively we have explored a
total of 12 data set combinations). In addition to above mentioned acronyms for different
data set combinations, for denoting data sets with SH0ES prior for $H_{0}$ included, we
append these acronyms with $H_{0}$, for example, SC$H_{0}$ for SNIa+CC+$H_{0}$ and SC($H_{0}$) to
mean SNIa+CC or SNIa+CC+$H_{0}$ depending on the context. Unless otherwise specified,
in the figures, the following colour and data set correspondences are there: (i) SC($H_{0}$) -- blue, (ii) SBC($H_{0}$) -- red, (iii) SCH\textsc{ii}($H_{0}$) -- black, (iv) BCH\textsc{ii}($H_{0}$) -- orange, (v)
SBCH\textsc{ii}($H_{0}$) -- green, and (vi) SBCH\textsc{ii}($H_{0}$)+CMB  -- brown.
\par
While with six data sets one can explore a total of 63 data set combinations, but
our goal is to obtain tighter constraints on model parameters and we also need to be
mindful of goodness-of-fit. Thus we decided to explore only six data set combinations
and the inclusion of data combination BCH($H_{0}$) is to emphasize the fact that not all combinations of data give meaningfully acceptable results.
\par
Tables \ref{resultstable_Pal} and \ref{resultsH0table_Pal} present the median values of model parameters along with their
corresponding 1$\sigma$ confidence intervals. The posterior probability distribution of model
parameters are displayed in Figs. \multiref{HS1dist_Pal}{CMBdist_Pal}, where 2D contour plots can be observed to see
possible correlations among parameters, and 1D marginal plots can be observed to
see symmetry or skewness in distribution of each parameter. Including CMB distance
priors in any data combination amounts to constraining a total of four parameters:
$(\Omega_{m0},\,b,\,H_{0},\,\omega_{b0})$. Consequently, for the data sets SBCH\textsc{ii}+CMB and SBCH\textsc{ii}$H_{0}$ +CMB,
for all the $f(R)$ models, the posterior probability distributions are shown in Fig. \ref{CMBdist_Pal}.
When reading the sections regarding constraints on individual models, the readers are
requested to refer Fig. \ref{CMBdist_Pal} also.
%%%%%%%%%%%%%%%%%%%%%%%%%%%%%%%%%%%%%%%%%%%%%%%%%%%%%%%%%%%%%%%%%%%%%%%%%%%%
%%%%%%%%%%%%%%%%%%%%%%%%%%%%%%%%%%%%%%%%%%%%%%%%%%%%%%%%%%%%%%%%%%%%%%%%%%%%
\begin{figure}%[htp]
\centering
\includegraphics[scale=0.4]{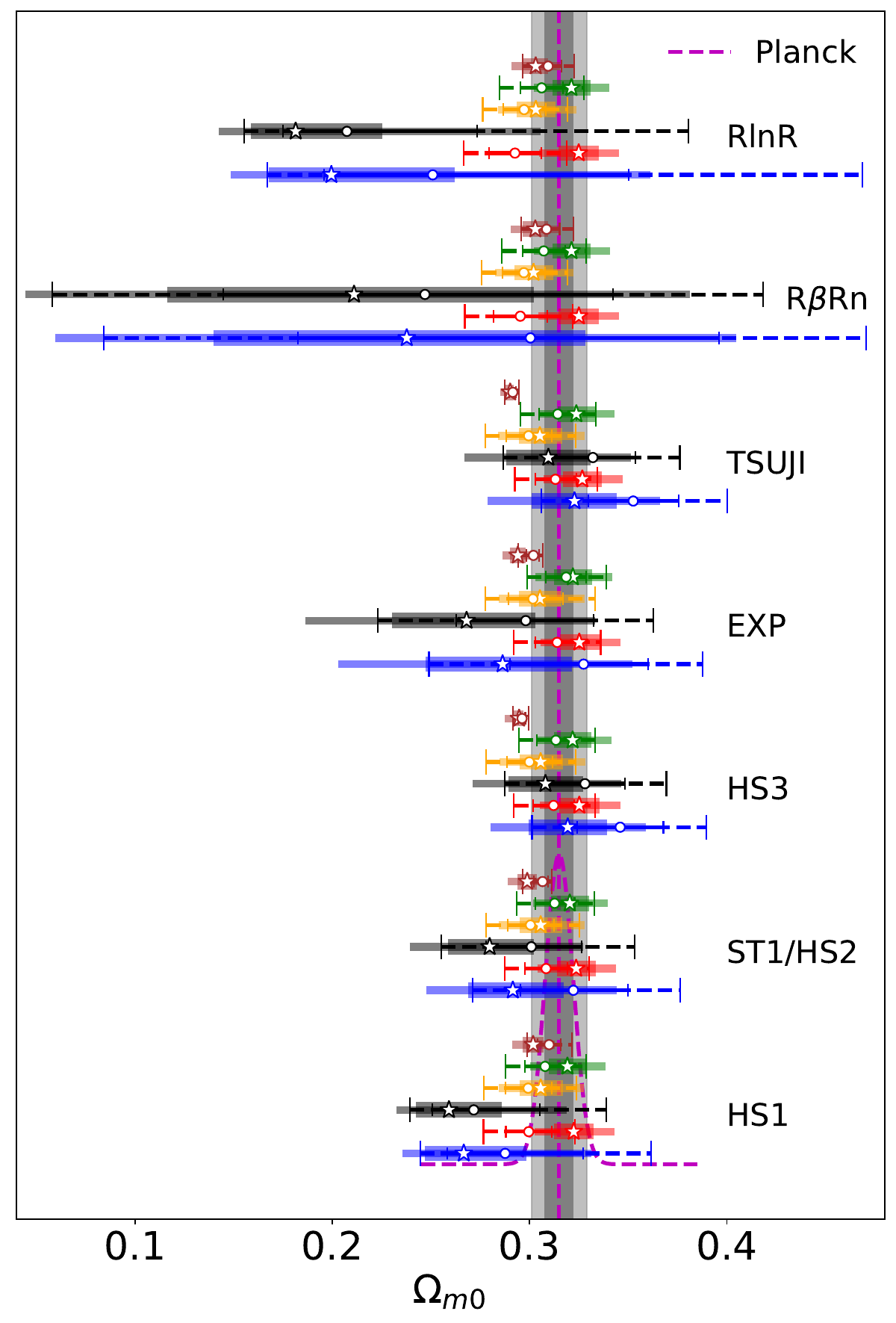}
\caption{The matter density at current epoch ($\Omega_{m0}$) variations across models and data sets are depicted in this figure. The color coding for various data sets corresponds identically to that in any of the parameter distribution plots (e.g., Fig. \ref{HS1dist_Pal}) or can
be referenced from the second paragraph of Section \ref{acc}. The markers `blank star' and `circle' denote median values with and without SH0ES prior for $H_{0}$, respectively. In instances where SH0ES prior wasn't applied, we depicted 1$\sigma$ (68.26 per cent, with shorter caps) and 2$\sigma$ (95.44 per cent with longer caps) confidence intervals using colored continuous/dashed lines. However, for cases with SH0ES prior, we represented 1$\sigma$ and 2$\sigma$ confidence intervals using thick and thin horizontal colored bars, respectively.}
\label{Om0tension_Pal}
\end{figure}

\begin{figure}%[htp]
\centering
\includegraphics[scale=0.4]{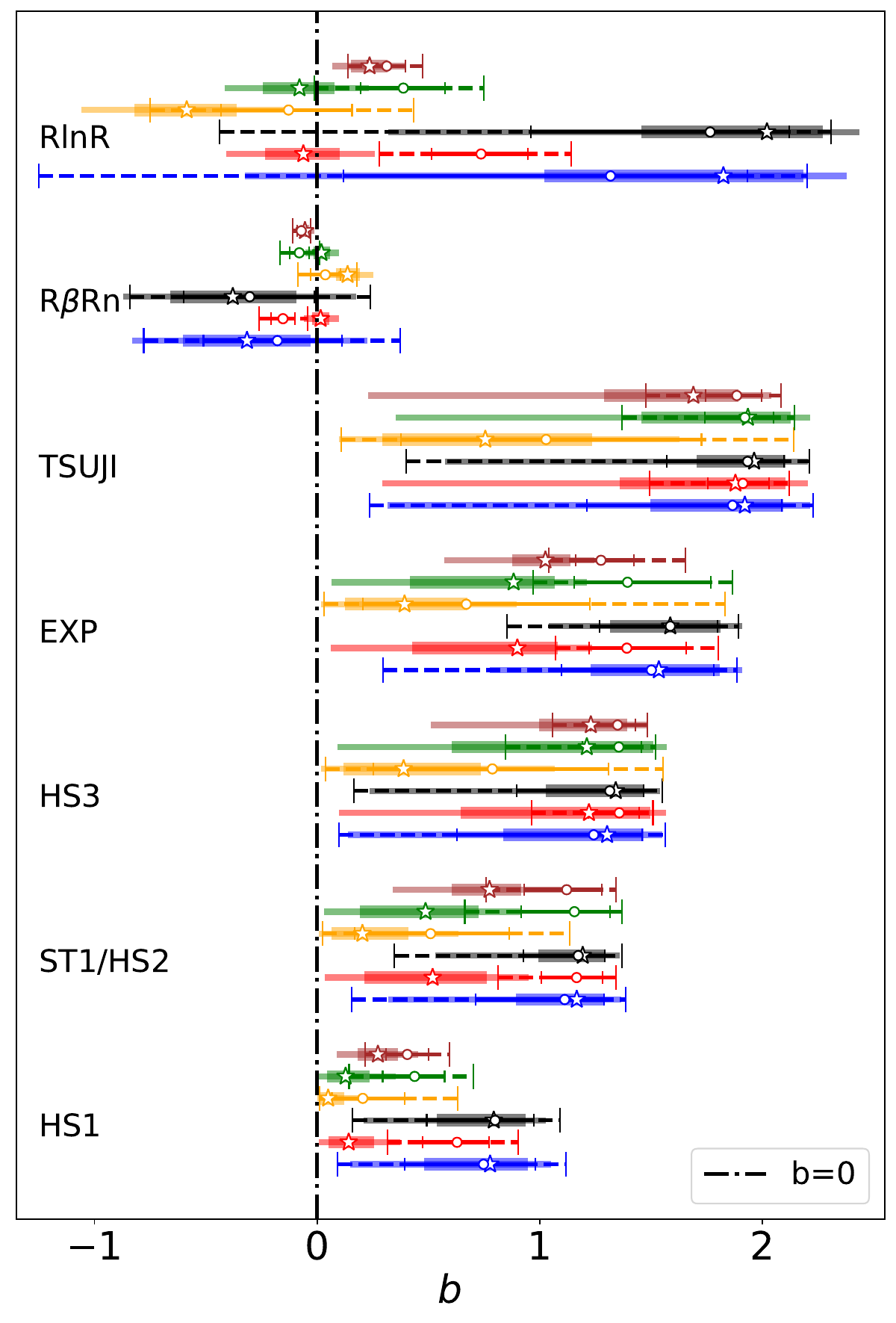}
\caption{The deviation parameter ($b$) variations across models and data sets are depicted in this figure. The color coding for various data sets corresponds identically to that in any of the parameter distribution plots (e.g., Fig. \ref{HS1dist_Pal}) or can
be referenced from the second paragraph of Section \ref{acc}. The markers `blank star' and `circle' denote median values with and without SH0ES prior for $H_{0}$, respectively. In instances where SH0ES prior wasn't applied, we depicted 1$\sigma$ (68.26 per cent, with shorter caps) and 2$\sigma$ (95.44 per cent with longer caps) confidence intervals using colored continuous/dashed lines. However, for cases with SH0ES prior, we represented 1$\sigma$ and 2$\sigma$ confidence intervals using thick and thin horizontal colored bars, respectively.}
\label{btension_Pal}
\end{figure}

\begin{figure}%[htp]
\centering
\includegraphics[scale=0.4]{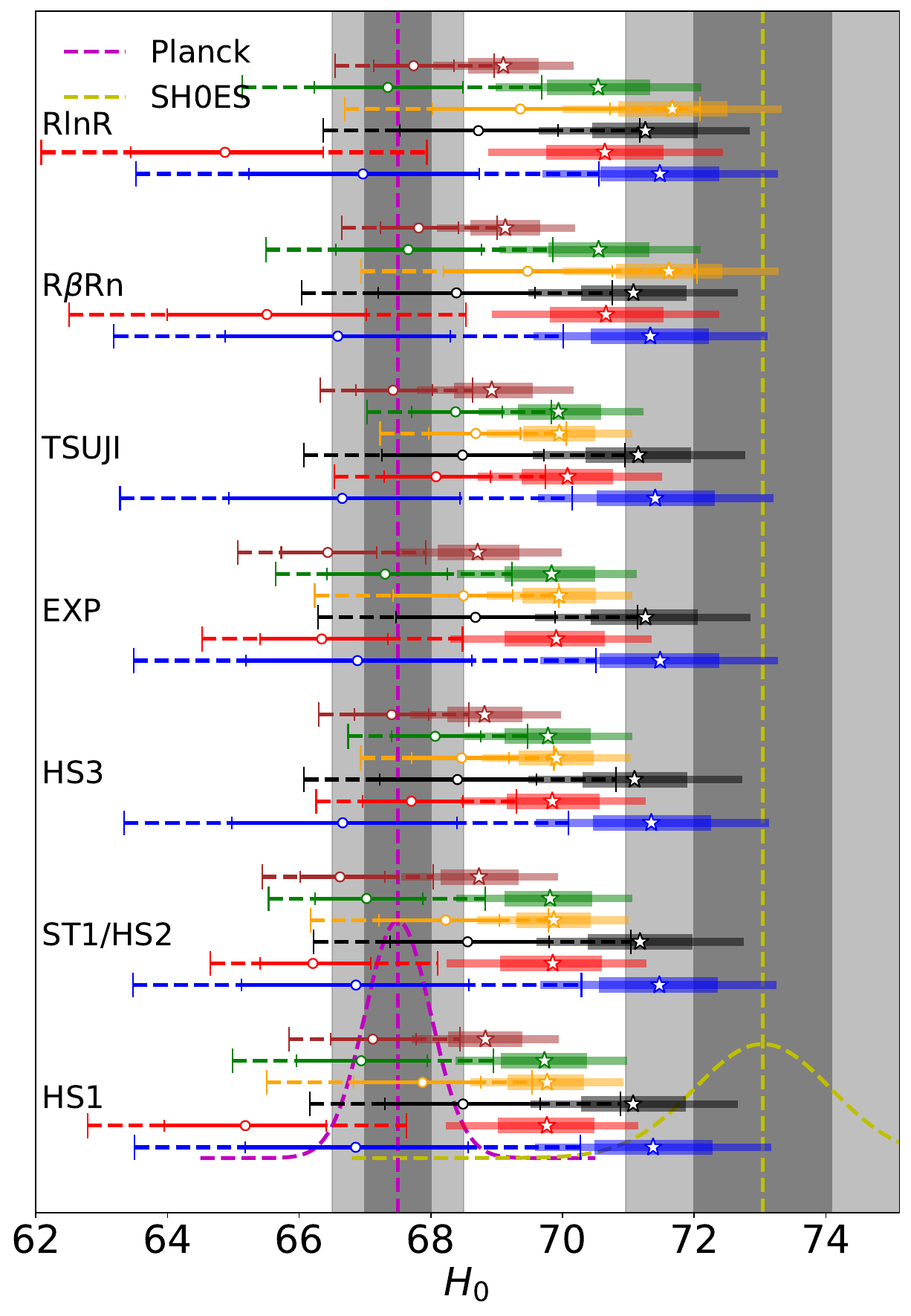}
\caption{The \emph{Hubble} parameter at current epoch ($H_{0}$) variations across models and data sets are depicted in this figure. The color coding for various data sets corresponds identically to that in any of the parameter distribution plots (e.g., Fig. \ref{HS1dist_Pal}) or can
be referenced from the second paragraph of Section \ref{acc}. The markers `blank star' and `circle' denote median values with and without SH0ES prior for $H_{0}$, respectively. In instances where SH0ES prior wasn't applied, we depicted 1$\sigma$ (68.26 per cent, with shorter caps) and 2$\sigma$ (95.44 per cent with longer caps) confidence intervals using colored continuous/dashed lines. However, for cases with SH0ES prior, we represented 1$\sigma$ and 2$\sigma$ confidence intervals using thick and thin horizontal colored bars, respectively.}
\label{H0tension_Pal}
\end{figure}
%%%%%%%%%%%%%%%%%%%%%%%%%%%%%%%%%%%%%%%%%%%%%%%%%%%%%%%%%%%%%%%%%%%%%%%%%%%%
%%%%%%%%%%%%%%%%%%%%%%%%%%%%%%%%%%%%%%%%%%%%%%%%%%%%%%%%%%%%%%%%%%%%%%%%%%%%
\par
First, considering Figs. \ref{Om0tension_Pal}, and \ref{H0tension_Pal}, we present a gist of the model predicted parameters $\Omega_{m0}$ and $H_{0}$, as how they compare to the standard cosmological measurements like $\Omega_{m0,\rm Planck}$, $H_{0,\rm Planck}$ and 
$H_{0,\rm SH0ES}$ -- across the models and across the data sets, then we elaborate more on individual models. For most of the cases, the model predicted median values of $\Omega_{m0}$ are 
within the 1$\sigma$-2$\sigma$ range of the \emph{Planck}  
value ($\Omega_{m0,\rm Planck}=0.315\pm0.007$ \cite{Planck:2018vyg}), while for a few cases $\Omega_{m0,\rm Planck}$ is within the 1$\sigma$-2$\sigma$ range of the former. When compared with the \emph{Planck} value of 
$H_{0}$ ($H_{0,\rm Planck}=67.4\pm0.5$ \cite{Planck:2018vyg}), for almost all the models and for most of the data sets without SH0ES prior, the model predicted median values of $H_{0}$ fall within 1$\sigma$-2$\sigma$ range of the former, while for a few cases its other way round. In contrast, when applying the SH0ES prior on $H_{0}$, there are slight departures towards higher 
values of $H_{0}$ (as expected), but these values do not closely 
align with $H_{0,\rm SH0ES} = 73.04\pm 1.04$ \cite{Riess:2021jrx}. 
It's worth noting that in cases with the SH0ES prior, the bounds 
on the model $H_{0}$ values are generally tighter. Additionally, 
overall there is less tension among model-fitted values of $H_{0}$ compared 
to the tension between $H_{0,\rm Planck}$ and $H_{0,\rm SH0ES}$. Some trends about the model predicted values of $b$ is also worth noting, in light of Fig. \ref{btension_Pal}. When incorporating the SH0ES prior on $H_{0}$, there is a consistent tendency for the $b$ values to shift towards lower values. This trend is observed in the cases of data sets: SBC$(H_{0})$, BCH$\textsc{ii}(H_{0})$, SBCH$\textsc{ii}(H_{0})$, and SBCH$\textsc{ii}(H_{0})$+CMB. Conversely, opposite trends are observed for the SC$(H_{0})$ and SCH$\textsc{ii}(H_{0})$ data sets compared to the corresponding cases without 
the SH0ES prior on $H_{0}$.
%%%%%%%%%%%%%%%%%%%%%%%%%%%%%%%%%%%%%%%%%%%%%%%%%%%%%%%%%%%%%%%%%%%%%%%%%%%%
\subsection{Constraints on Hu--Sawicki and Starobinsky models}
%%%%%%%%%%%%%%%%%%%%%%%%%%%%%%%%%%%%%%%%%%%%%%%%%%%%%%%%%%%%%%%%%%%%%%%%%%%
%%%%%%%%%%%%%%%%%%%%%%%%%%%%%%%%%%%%%%%%%%%%%%%%%%%%%%%%%%%%%%%%%%%%%%%%%%%
\begin{figure*}
\imgtwo{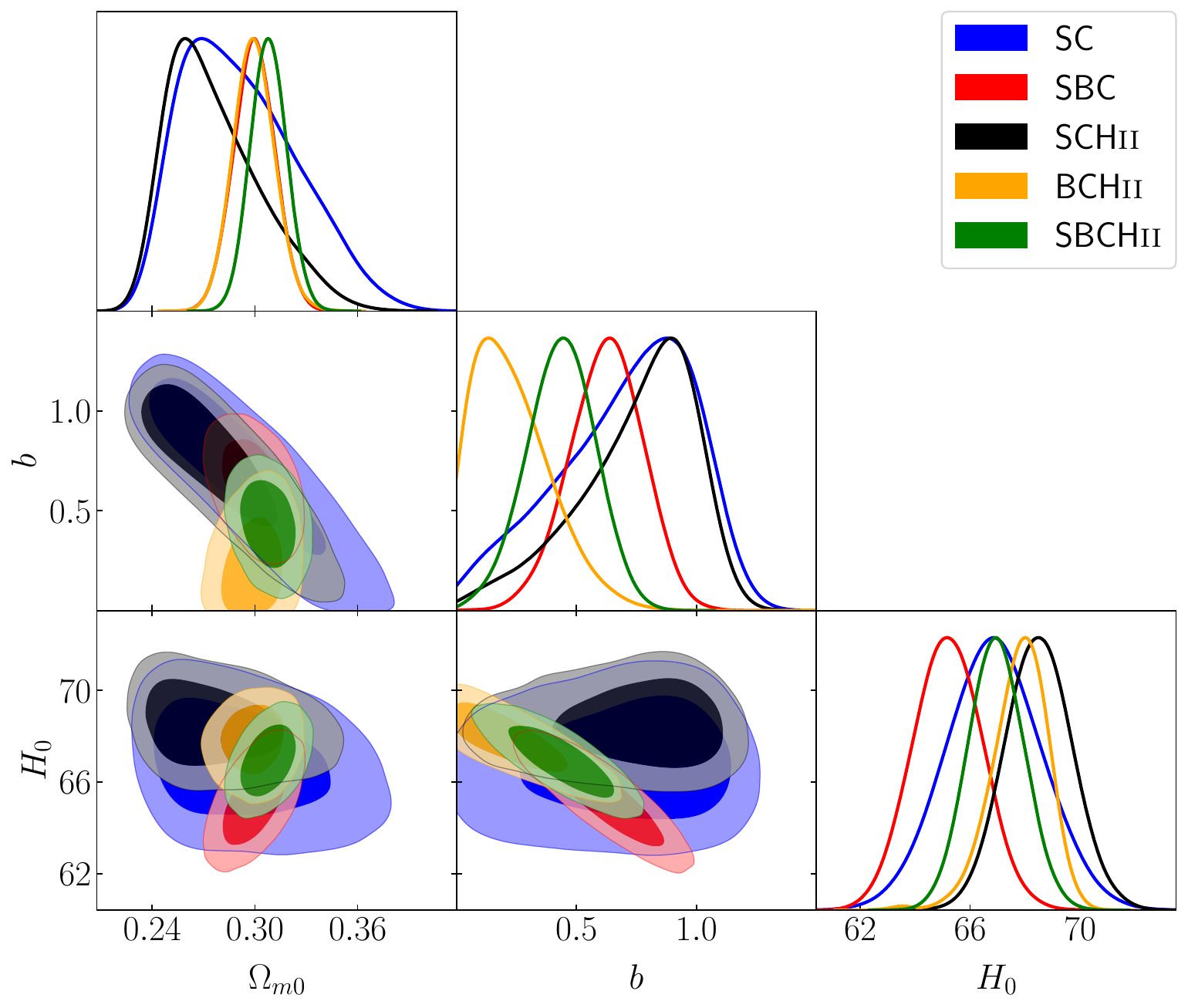}\hfill
\imgtwo{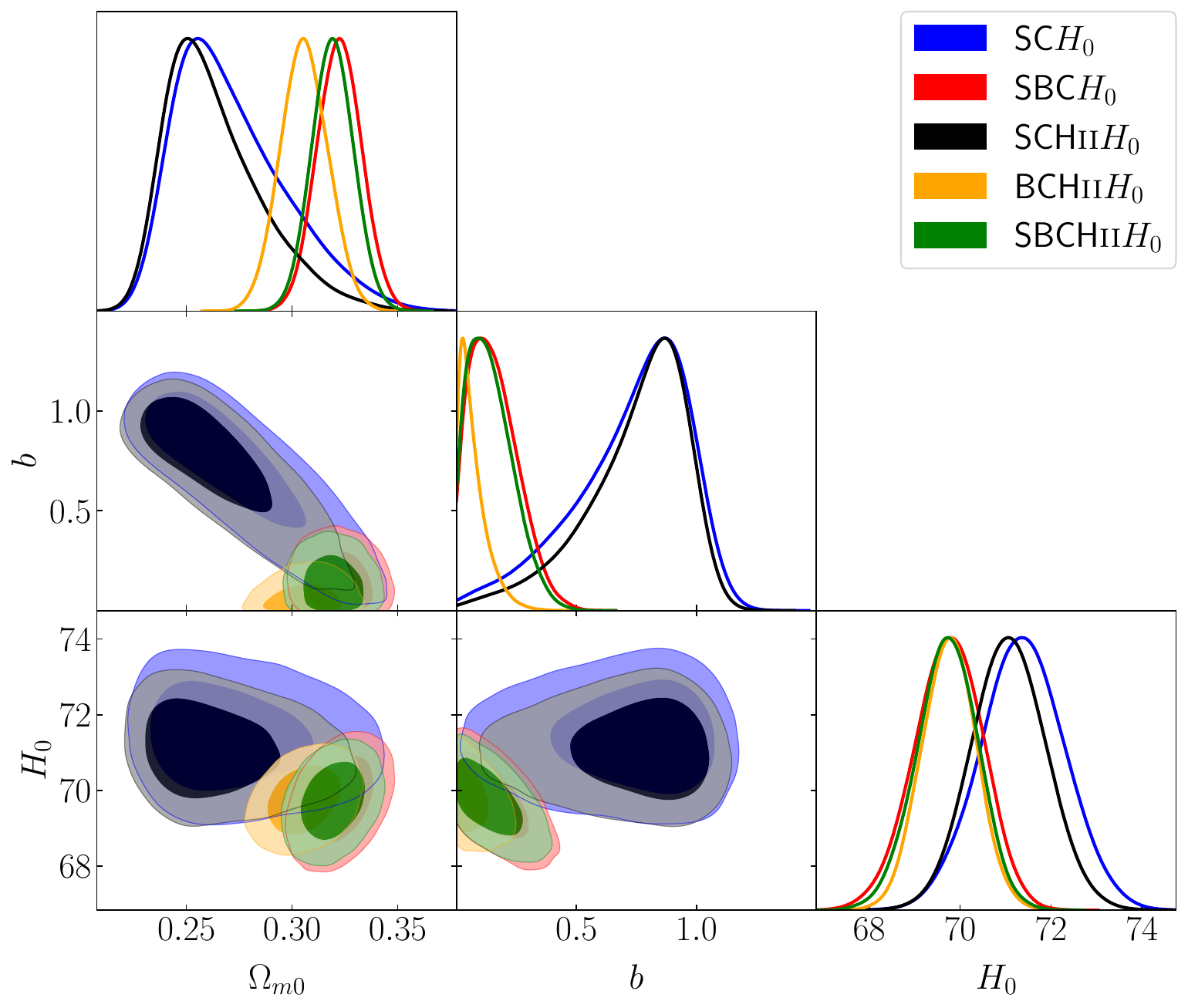}
\caption{The Hu--Sawicki model ($n_{_{\rm HS}} = 1$, \emph{left:} without SH0ES prior on $H_{0}$, \emph{right:} with SH0ES prior for $H_{0}$): the posterior probability distribution plots of fitted parameters. Figure legends show the colour correspondence for different data set combinations. The 1$\sigma$ (68.26 per cent) and 2$\sigma$ (95.44 per cent) confidence intervals are represented by darker and lighter shades of colours, respectively.}
\label{HS1dist_Pal}
\end{figure*}
%%%%%%%%%%%%%%%%%%%%%%%%%%
\begin{figure*}
\imgtwo{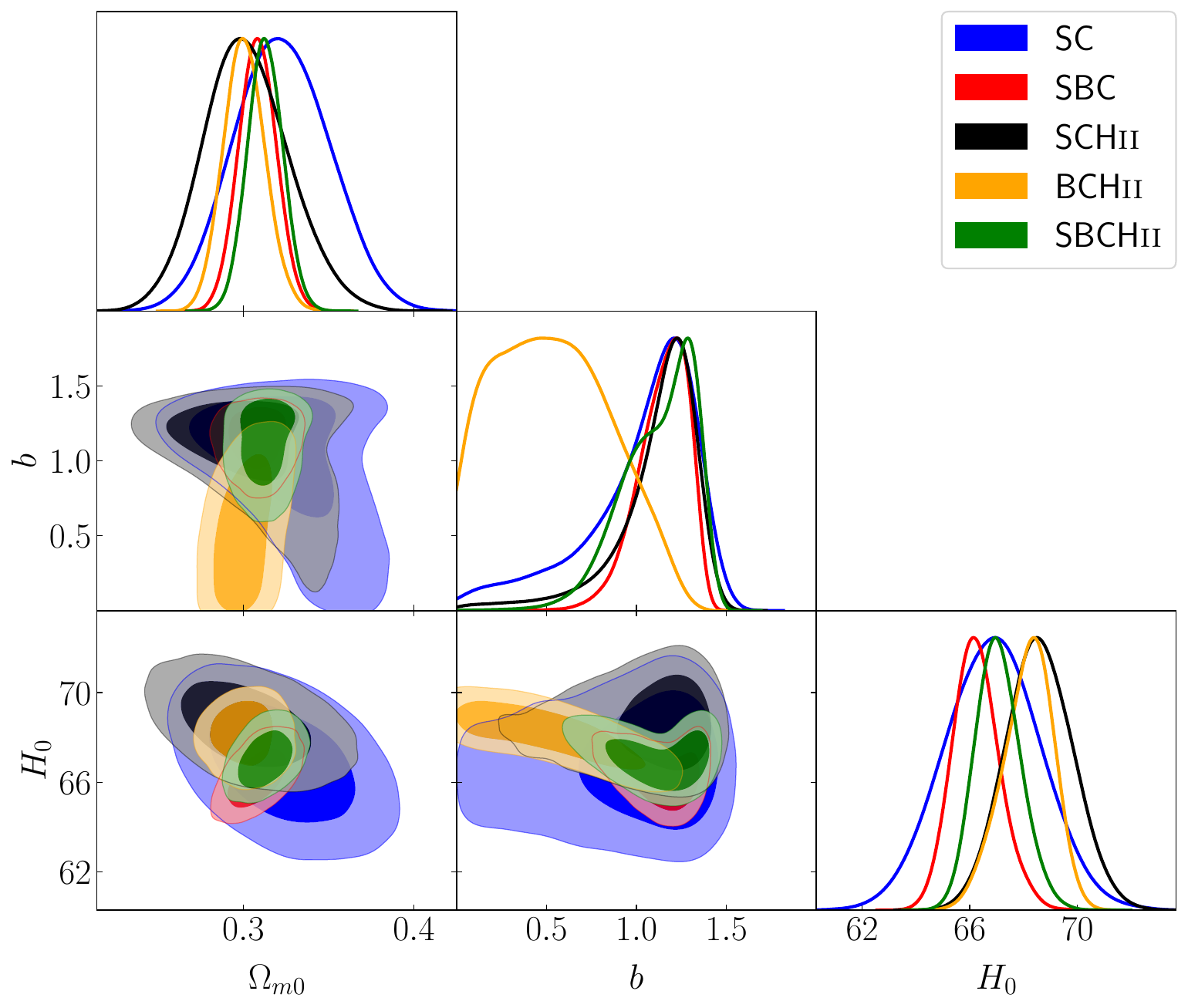}\hfill
\imgtwo{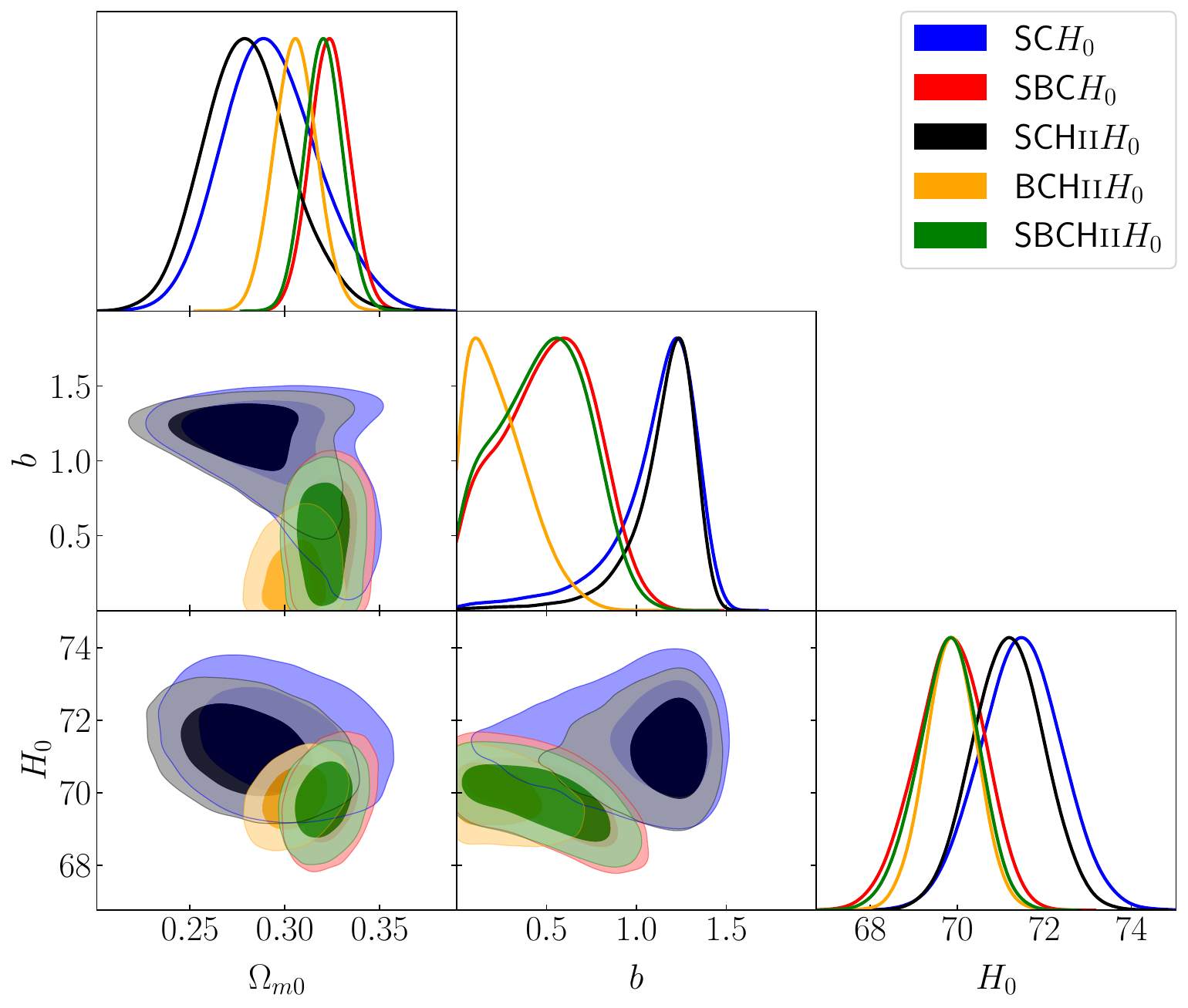}
\caption{The Starobinsky model ($n_{_{\rm S}} = 1$, \emph{left:} without SH0ES prior on $H_{0}$, \emph{right:} with SH0ES prior for $H_{0}$): the posterior probability distribution plots of fitted parameters. Figure legends show the colour correspondence for different data set combinations. The 1$\sigma$ (68.26 per cent) and 2$\sigma$ (95.44 per cent) confidence intervals are represented by darker and lighter shades of colours, respectively.}
\label{HS2dist_Pal}
\end{figure*}
%%%%%%%%%%%%%%%%%%%%%%%%%%
\begin{figure*}
\imgtwo{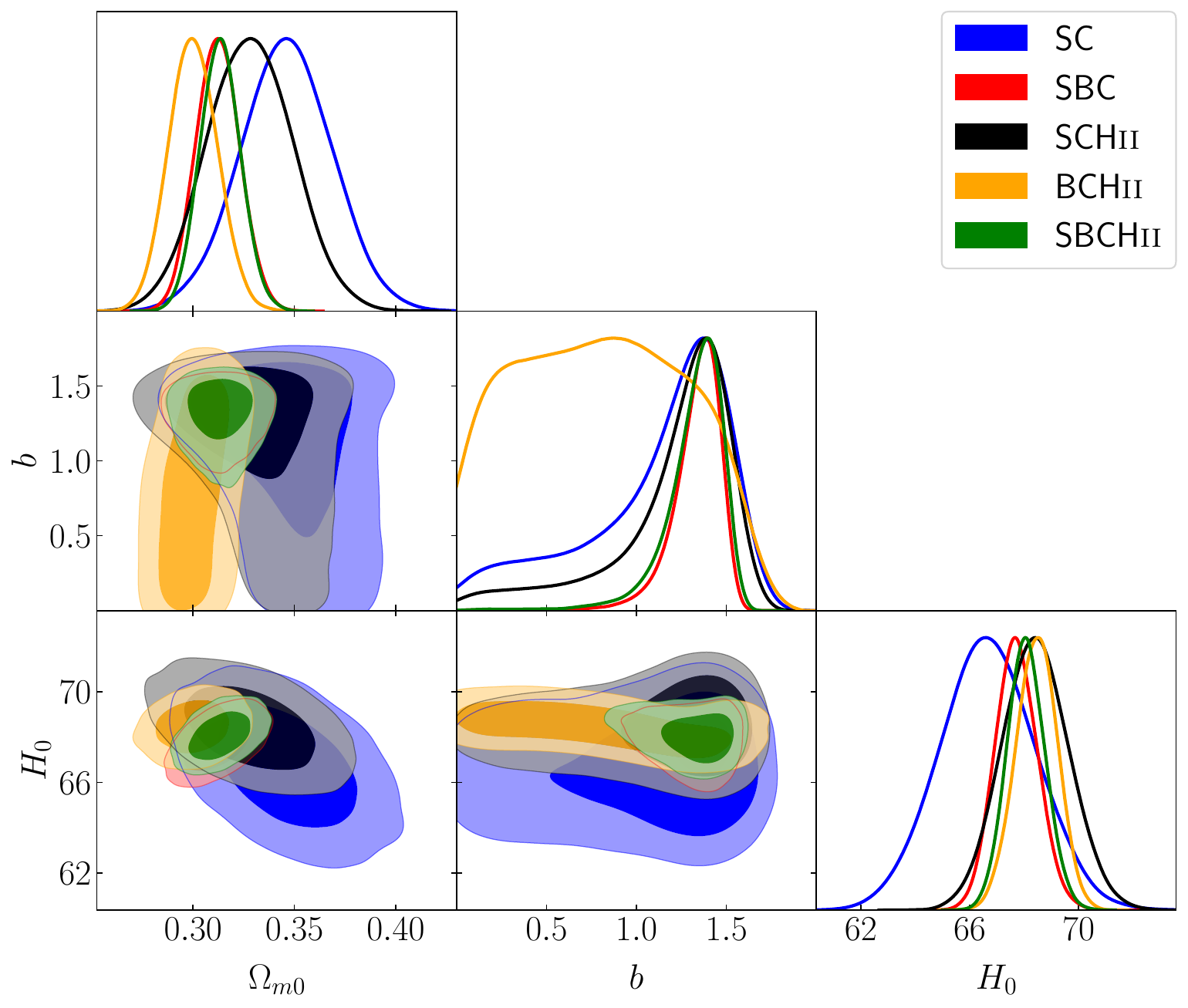}\hfill
\imgtwo{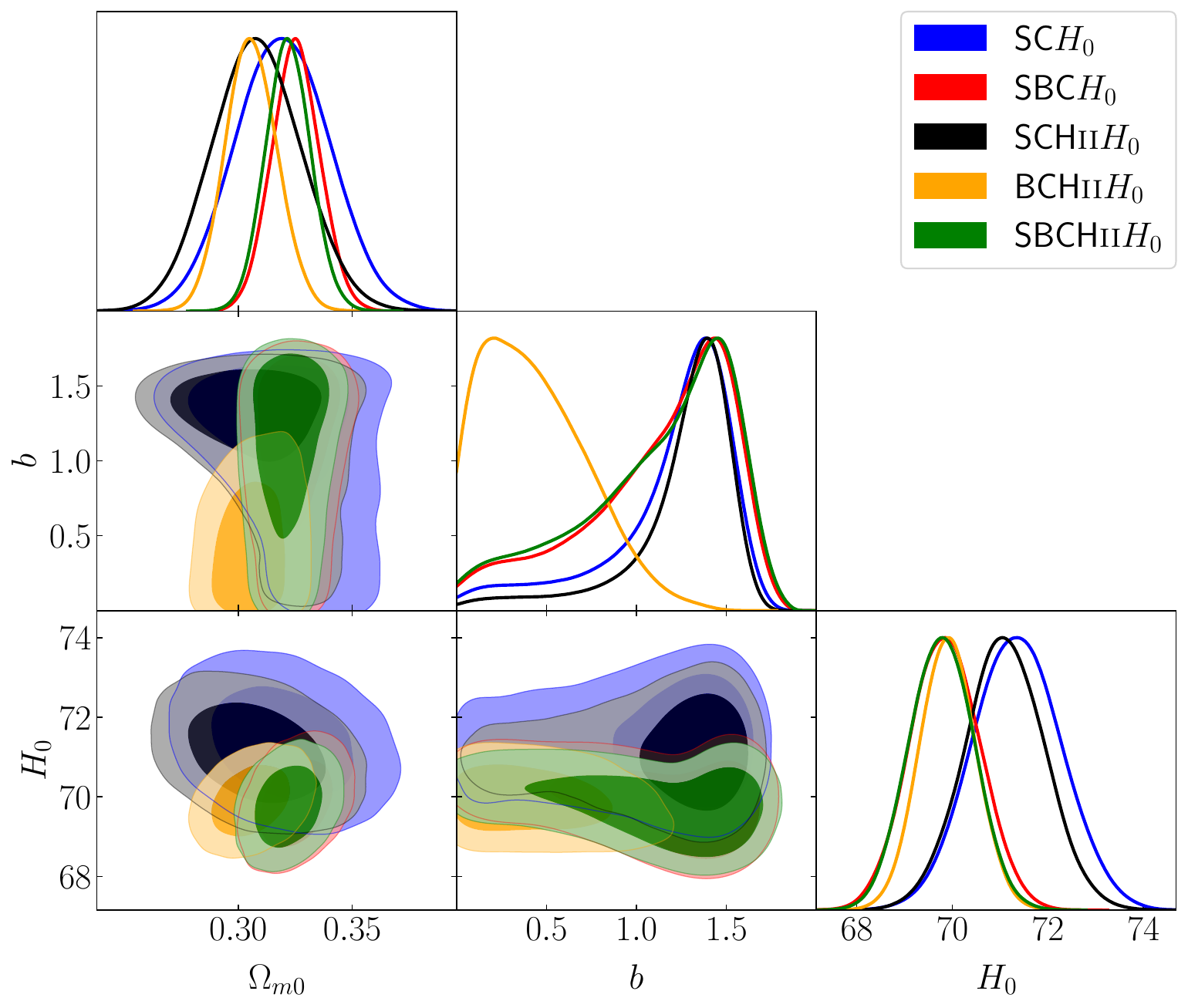}
\caption{The Hu--Sawicki model ($n_{_{\rm HS}} = 3$, \emph{left:} without SH0ES prior on $H_{0}$, \emph{right:} with SH0ES prior for $H_{0}$): the posterior probability distribution plots of fitted parameters. Figure legends show the colour correspondence for different data set combinations. The 1$\sigma$ (68.26 per cent) and 2$\sigma$ (95.44 per cent) confidence intervals are represented by darker and lighter shades of colours, respectively.}
\label{HS3dist_Pal}
\end{figure*}
%%%%%%%%%%%%%%%%%%%%%%%%%%%%%%%%%%%%%%%%%%%%%%%%%%%%%%%%%%%%%%%%%%%%%%%%%%%
%%%%%%%%%%%%%%%%%%%%%%%%%%%%%%%%%%%%%%%%%%%%%%%%%%%%%%%%%%%%%%%%%%%%%%%%%%%
Although these models are widely explored in the metric formalism (see \cite{Ravi:2023nsn} and references therein), there is only a few earlier work which has constrained these models in the Palatini formalism \cite{Santos:2012vs}. For the reasons discussed in \cite{Ravi:2023nsn}, we choose to constraint only cases with $n_{_{\rm{HS}}}=1,\,3$ and $n_{_{\rm{S}}}=1$.
\par
The posterior probability distribution plots for HS1, ST1/HS2 and HS3 models are shown in Figs.  \ref{HS1dist_Pal}, \ref{HS2dist_Pal}, and \ref{HS3dist_Pal}, respectively, while the quantitative results like median values of the parameters and $1\sigma$ confidence bound on them are presented in Tables \ref{resultstable_Pal} and 
\ref{resultsH0table_Pal}. Also see Fig. \ref{CMBdist_Pal} for results from data sets where CMB is included.  Except for model HS3 with data sets SBCH\textsc{ii}+CMB and SBCH\textsc{ii}$H_{0}$+CMB, the matter density parameter, $\Omega_{m0}$, for these three models (i.e. HS1, ST1/HS2, and HS3), from all the data sets, are $1\sigma-2\sigma$ compatible with the \emph{Planck} value of $\Omega_{m0,\,\rm{Planck}} = 0.315\pm0.007$. When SH0ES prior for $H_{0}$ is not taken, the model predicted median values of $H_{0}$ are $1\sigma-2\sigma$ compatible with the \emph{Planck} value of $H_{0,\,\rm{Planck}} = 67.5\pm 0.5$ (covering both values smaller than and bigger than 67.5). From data sets with SH0ES prior, we get relatively tighter constraints on $H_{0}$, these model predicted median values of $H_{0}$ are $1\sigma-3\sigma$ compatible both $H_{0,\,\rm{Planck}}$ and $H_{0,\,\rm{SH0ES}}$. For latter cases (i.e. with SH0ES prior), model predicted values of $H_{0}$ are always on lower side of $H_{0,\,\rm{SH0ES}}$.
\par
Except for the data sets SC and SC$H_{0}$, we can observe a general trend of decrease in model predicted median values of $b$ when SH0ES prior is included, compared to the corresponding data set without SH0ES prior for $H_{0}$. We also observe that for many data sets $b=0$ is only marginally allowed (i.e. $b=0$ is outside the $2\sigma-3\sigma$ limits), which means we are getting instances of distinguishable HS1, ST1/HS2 and HS3 models from the standard $\Lambda$CDM model. Latter in the model comparison chapter we will see that these distinguishable case of HS1, ST1/HS2 and HS3 models, are also (very) strongly supported by AIC, BIC, and/or DIC statistics, so this is an important result of this work. 
As to be expected, the median values of $b$ increases with increase in 
$n_{_{\rm{HS}}}$ from 1 to 3 (i.e. from model HS1, to ST1/HS2, to HS3).
\par
For data sets with CMB distance priors, the model predicted median values of $\omega_{b0}$ decrease -- with the inclusion of SH0ES prior, when compared with cases without SH0ES prior. Also there is tend of decreasing 
$\omega_{b0}$ with increase in $n_{_{\rm{HS}}}$ from 1 to 3, alternatively, we can say that models more dissimilar to the $\Lambda$CDM model are predicting higher values for $\omega_{b0}$.

%%%%%%%%%%%%%%%%%%%%%%%%%%%%%%%%%%%%%%%%%%%%%%%%%%%%%%%%%%%%%%%%%%%%%%%%%%
%\textbf
\subsection{Constraints on exponential model}

\begin{figure*}
\imgtwo{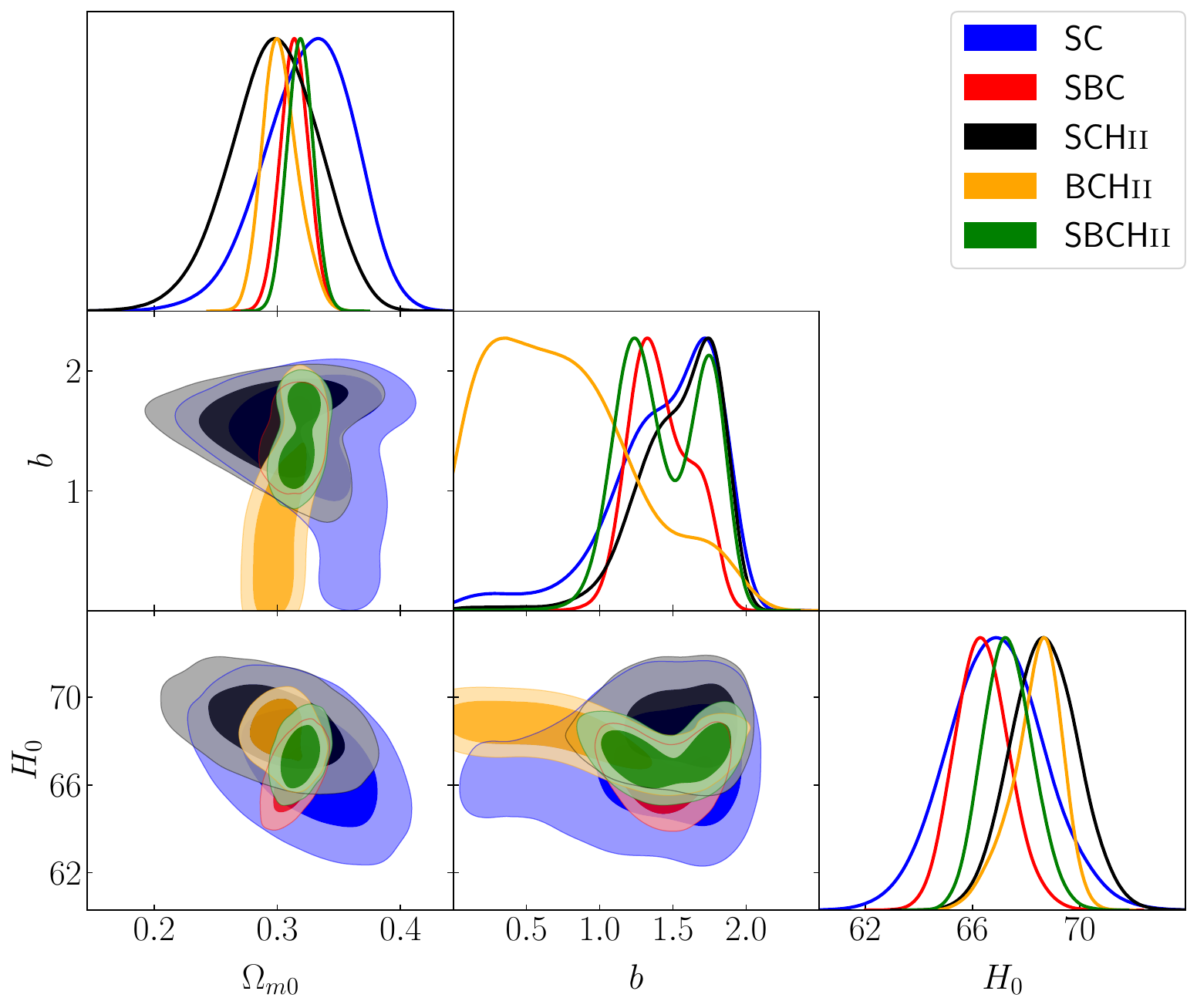}\hfill
\imgtwo{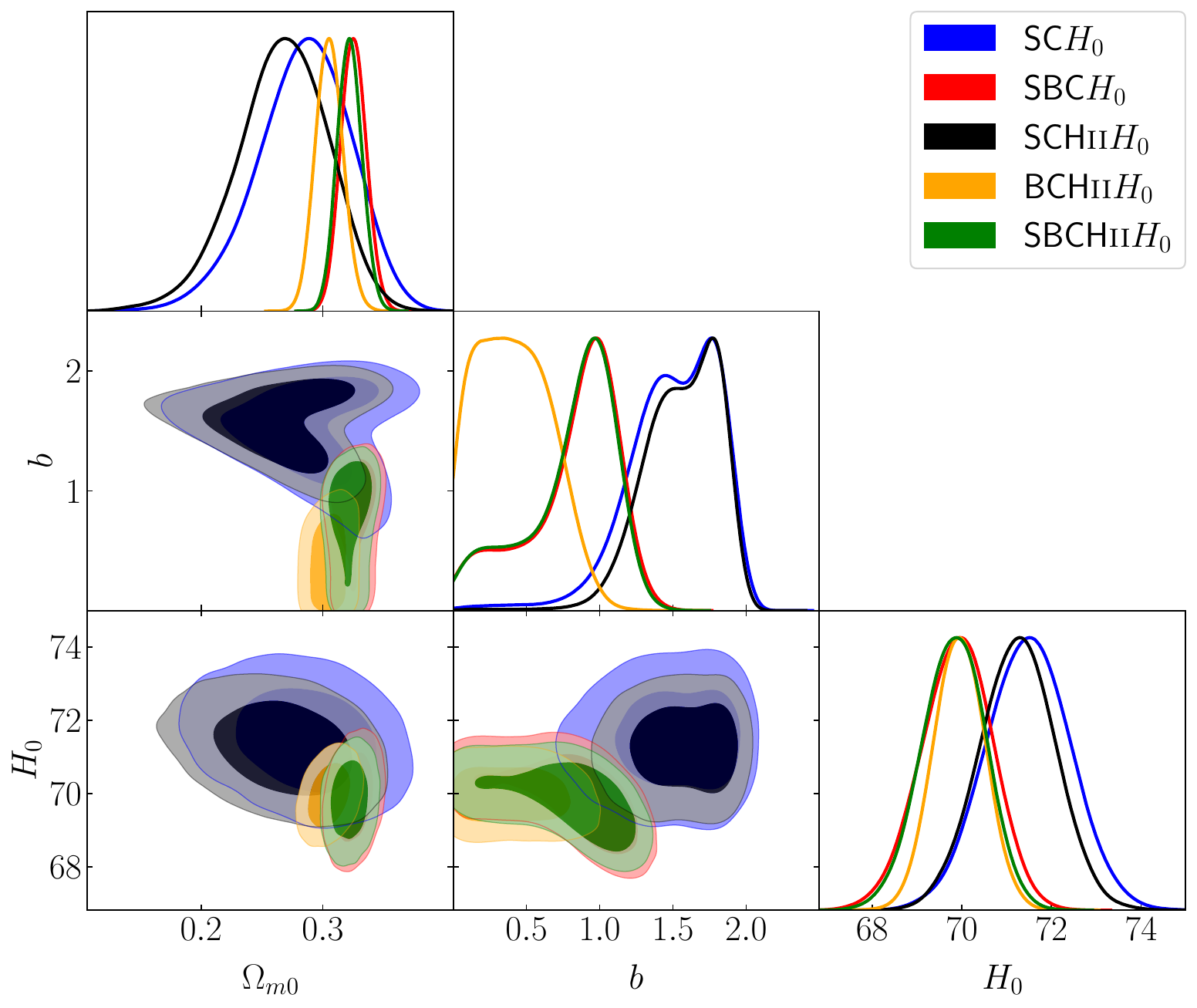}
\caption{The exponential model (\emph{left:} without SH0ES prior on $H_{0}$, \emph{right:} with SH0ES prior for $H_{0}$): the posterior probability distribution plots of fitted parameters. Figure legends show the colour correspondence for different data set combinations. The 1$\sigma$ (68.26 per cent) and 2$\sigma$ (95.44 per cent) confidence intervals are represented by darker and lighter shades of colours, respectively.}
\label{Expdist_Pal}
\end{figure*}
%%%%%%%%%%%%%%%%%%%%%%%%%%%%%%%%%%%%%%%%%%%%%%%%%%%%%%%%%%%%%%%%%
The exponential model is also very extensively constrained in metric formalism (see \cite{Ravi:2023nsn} and references therein), whereas for Palatini formalism we could find a few studies only \cite{Campista:2010jb}. We present the quantitative predictions for model parameters in  Tables \ref{resultstable_Pal} and \ref{resultsH0table_Pal}. The 2D contour plots of the posterior probability distribution of model parameters ($\Omega_{m0},\,b,\,H_{0}$) are displayed in the Fig. \ref{Expdist_Pal}, for the data sets without CMB distance priors, whereas results on parameters ($\Omega_{m0},\,b,\,H_{0},\,\omega_{b0}$), for data sets with CMB, are displayed Fig. \ref{CMBdist_Pal}. 
\par 
It is evident from Figs. \ref{btension_Pal}, \ref{Expdist_Pal}, and \ref{CMBdist_Pal} that aside from the BCH$\textsc{ii}(H_{0})$ dataset, the deviation parameter $b$ is notably non-zero. The possibility of $b=0$ is only marginally or scarcely allowed. The significance of this outcome will become more apparent in the subsequent chapter dedicated to model comparisons. Nonetheless, the values of the parameters $\Omega_{m0}$ and $H_{0}$ (see Figs. \ref{Om0tension_Pal} and \ref{H0tension_Pal}) are reasonably close to the standard values derived from \emph{Planck} constraints \cite{Planck:2018vyg}. When the SH0ES prior on $H_{0}$ is taken into account, the model's predictions for $H_{0}$ hover around approximately 68.5-71.5 km\,s$^{-1}$Mpc$^{-1}$ and always lesser than $H_{0,\,\rm{SH0ES}}$.

%%%%%%%%%%%%%%%%%%%%%%%%%%%%%%%%%%%%%%%%%%%%%%%%%%%%%%%%%%%%%%%%%%%%%%%%%%%
%\textbf
\subsection{Constraints on Tsujikawa model}
%%%%%%%%%%%%%%%%%%%%%%%%%%%%%%%%%%%%%%%%%%%%%%%%%%%%%%%%%%%%%%%%%%%%%%%%%%%
\begin{figure*}
\imgtwo{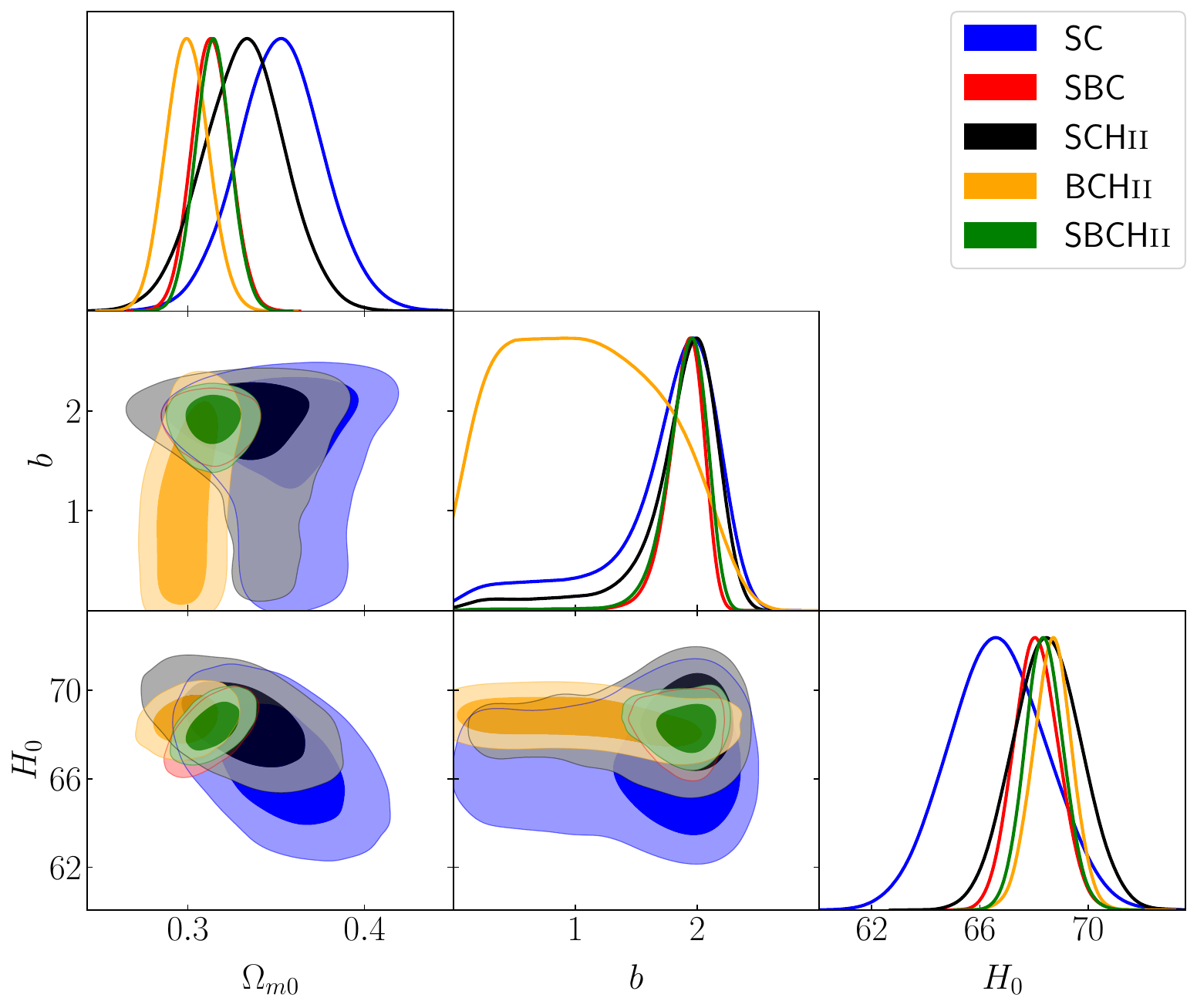}\hfill
\imgtwo{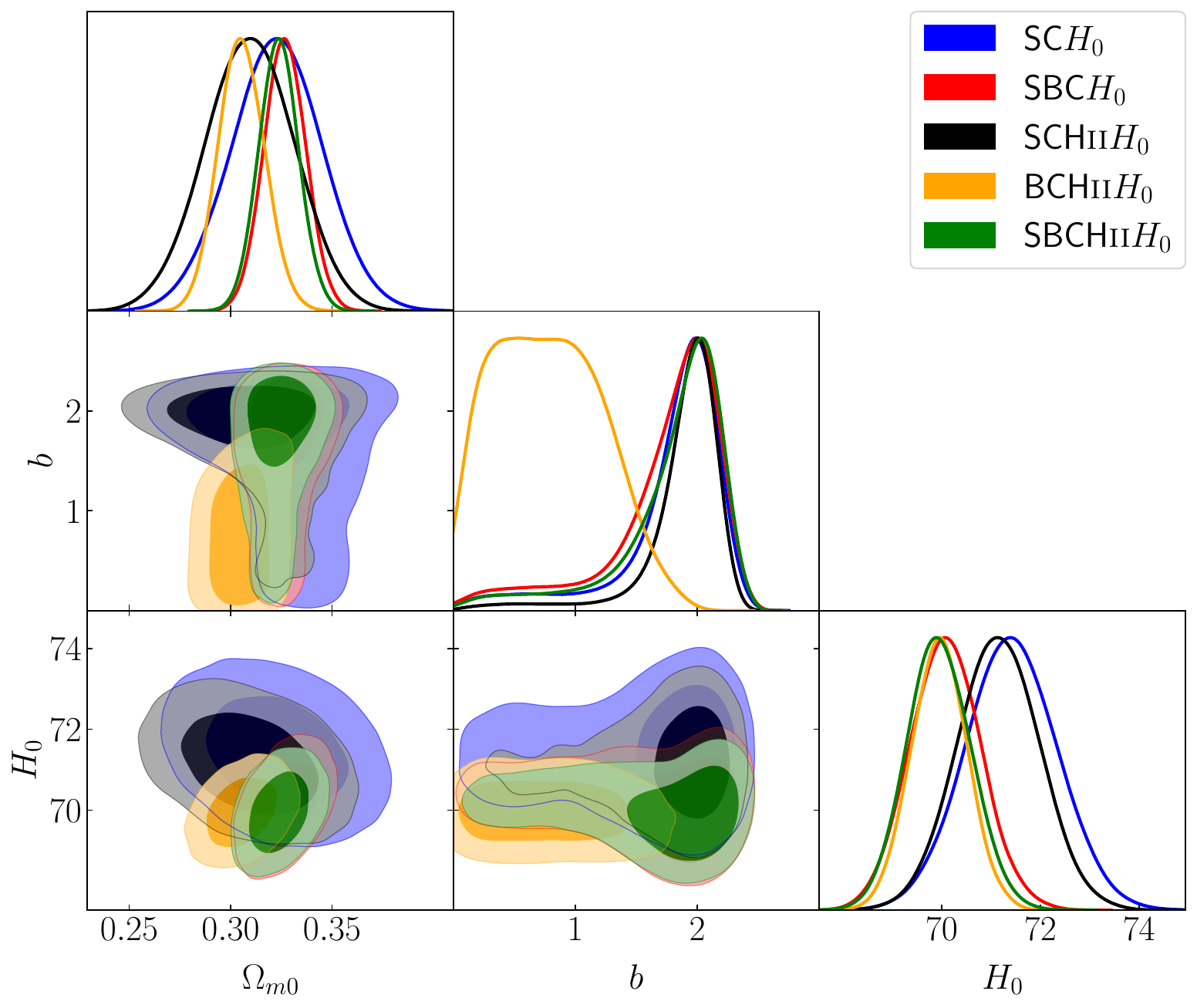}
\caption{The Tsujikawa model (\emph{left:} without SH0ES prior on $H_{0}$, \emph{right:} with SH0ES prior for $H_{0}$): the posterior probability distribution plots of fitted parameters. Figure legends show the colour correspondence for different data set combinations. The 1$\sigma$ (68.26 per cent) and 2$\sigma$ (95.44 per cent) confidence intervals are represented by darker and lighter shades of colours, respectively.}
\label{Tsujidist_Pal}
\end{figure*}
%%%%%%%%%%%%%%%%%%%%%%%%%%%%%%%%%%%%%%%%%%%%%%%%%%%%%%%%%%%%%%%%%%%%%%%%%%%
In Tables \ref{resultstable_Pal} and \ref{resultsH0table_Pal} we present the constraints on parameters of the Tsujikawa model outlined in Eq. \ref{tsujimodel1}. Additionally, Figs. \ref{Tsujidist_Pal}, and \ref{CMBdist_Pal}, respectively for without and with CMB data, depict posterior probability distributions of these parameters. 
Similar to the findings from the exponential model, we observe here that the deviation parameter, $b$, is non-zero, with zero being allowed only to a very limited extent, for all the datasets except BCH$\textsc{ii}(H_{0})$. As for any other models, here also, we disregard the results from BCH$\textsc{ii}(H_{0})$ due to corresponding high reduced $\chi^{2}$ value.
Thus, in this case as well, we obtain an $f(R)$ model that is distinctly different from the $\Lambda$CDM model. The determined values of $\Omega_{m0}$ and $H_{0}$ also fall within reasonable ranges of $\Omega_{m0,{\rm Planck}}$ and $H_{0,{\rm Planck}}$ (or $H_{0,{\rm SH0ES}}$). When considering the SH0ES prior on $H_{0}$, the model's median values for $H_{0}$ range from approximately 69 to 71.5 km\,s$^{-1}$Mpc$^{-1}$. These values exhibit a $2\sigma-3\sigma$ tension with $H_{0,{\rm Planck}}$ or $H_{0,{\rm SH0ES}}$.

\subsection{Constraints on $f(R) = R - \beta/R^{n}$  model}
%%%%%%%%%%%%%%%%%%%%%%%%%%%%%%%%%%%%%%%%%%%%%%%%%%%%%%%%%%%%%%%%%%%%%%%%%%%
\begin{figure*}
\imgtwo{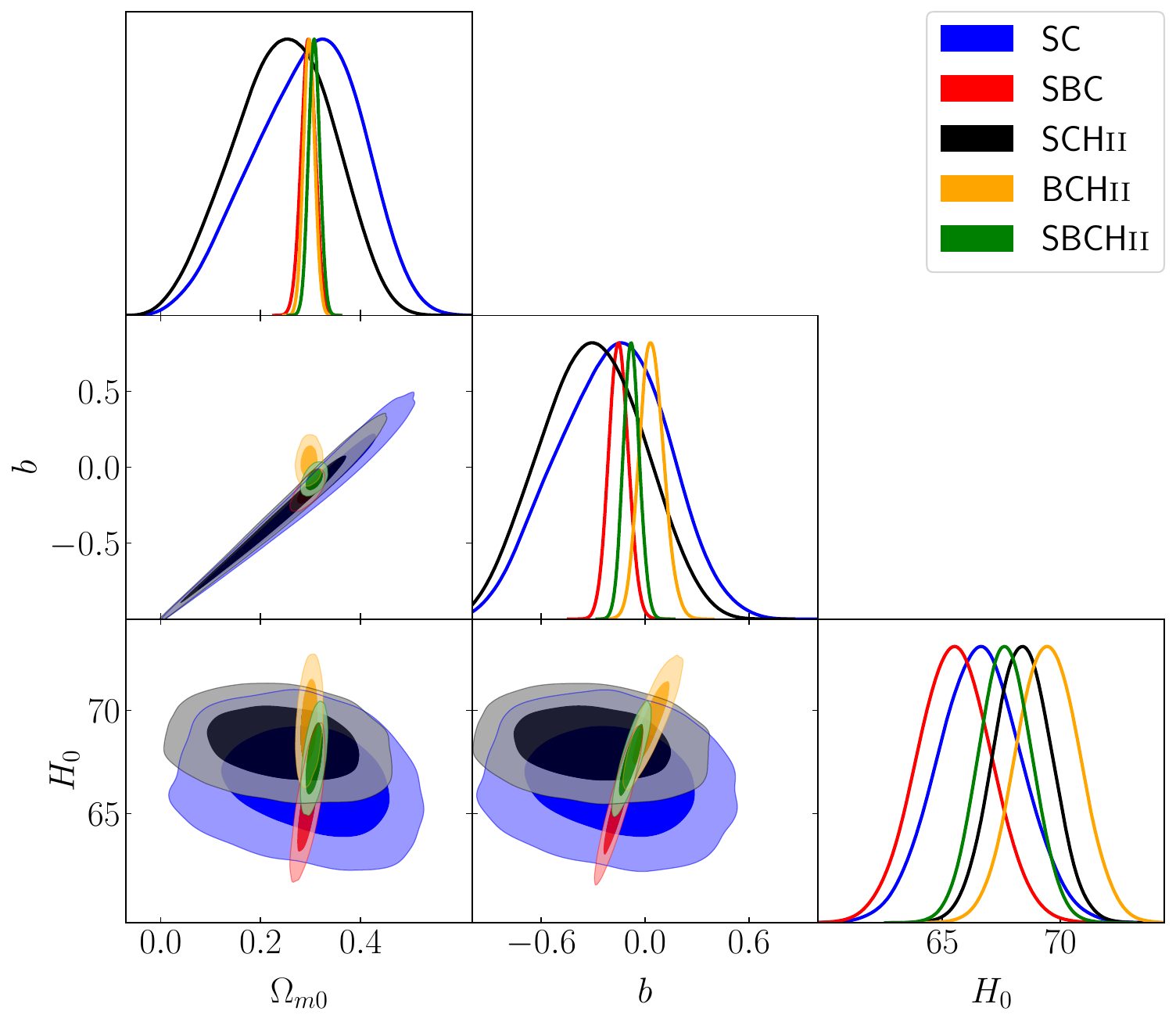}\hfill
\imgtwo{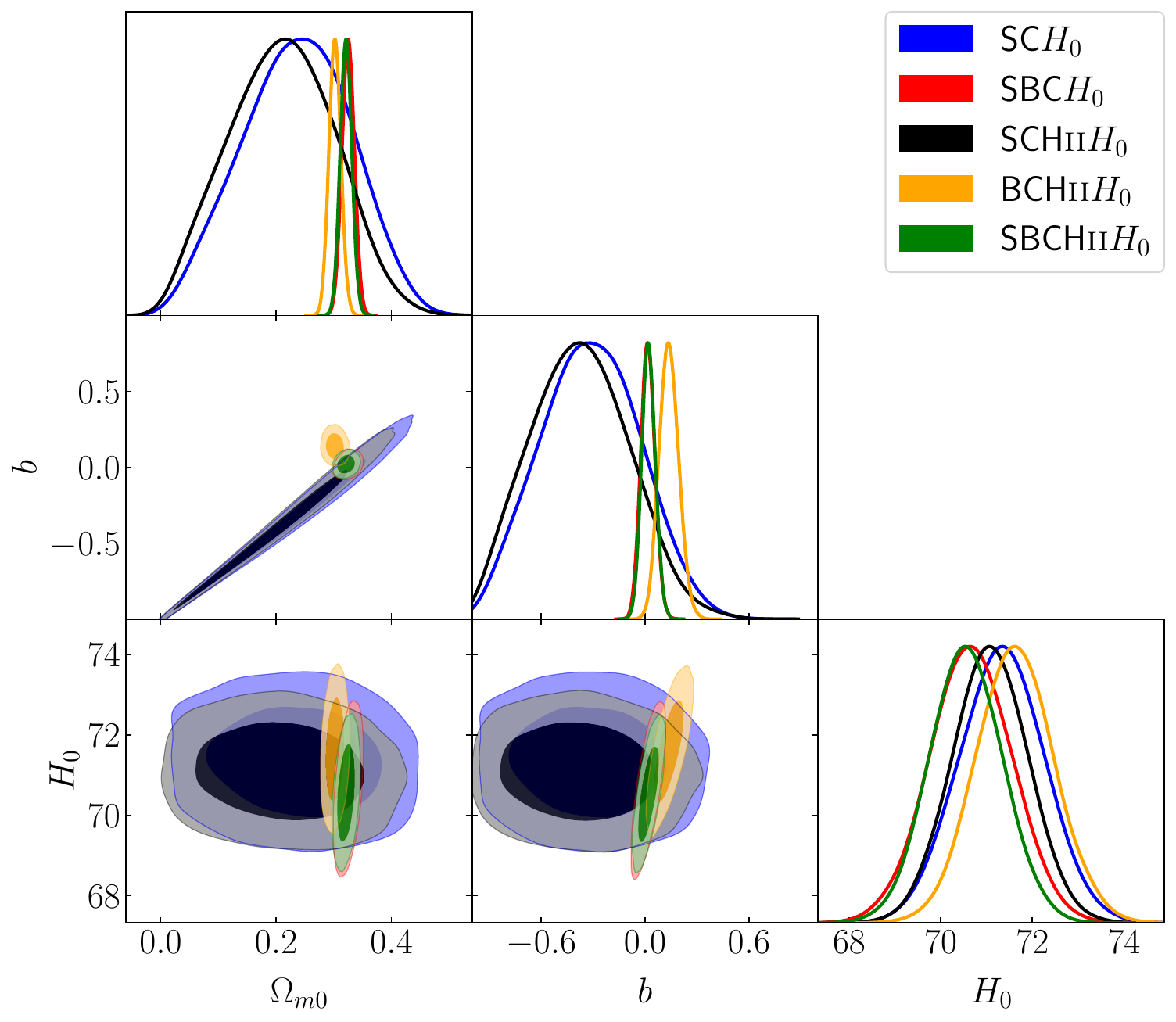}
\caption{The $f(R) = R - \beta/R^{n}$ model (\emph{left:} without SH0ES prior on $H_{0}$, \emph{right:} with SH0ES prior for $H_{0}$): the posterior probability distribution plots of fitted parameters. Figure legends show the colour correspondence for different data set combinations. The 1$\sigma$ (68.26 per cent) and 2$\sigma$ (95.44 per cent) confidence intervals are represented by darker and lighter shades of colours, respectively.}
\label{RbetaRndist_Pal}
\end{figure*}
%%%%%%%%%%%%%%%%%%%%%%%%%%%%%%%%%%%%%%%%%%%%%%%%%%%%%%%%%%%%%%%%%%%%%%%%%%%
Though not recently but this model has been widely explored in the past 
\cite{Amarzguioui:2005zq,Fay:2007gg,Carvalho:2008am,Santos:2008qp}, as it 
successfully produces the latter three phases of the history of the Universe, in Palatini formalism.
The model predicted median values of parameters and corresponding $1\sigma$ confidence bounds, after fitting Eq. \ref{Eq:RbetaRn} to all the data sets, are presented in Tables \ref{resultstable_Pal} and \ref{resultsH0table_Pal}, for data sets without and with SH0ES prior for $H_{0}$, respectively. The posterior probability distribution of parameters are shown in the Figs. \ref{RbetaRndist_Pal}, and \ref{CMBdist_Pal}, respectively for the data sets without and with CMB data. From aforementioned tables and figures (and/or Fig. \ref{Om0tension_Pal}), we can observe that without BAO data, the constraints on $\Omega_{m0}$ is poor, whereas for data sets with BAO included, we are getting reasonably tighter constraints on $\Omega_{m0}$. Anyway the $\Omega_{m0,\,\rm{Planck}}$ is always within $1\sigma-2\sigma$ limits of the model predicted median values. For all the data sets (except SCB, SCBH$\textsc{ii}$($H_{0}$)+CMB), $b=0$ is well within $1\sigma$ limits, that is, we are not getting any instances of  this model being significantly different from the $\Lambda$CDM model. For three data sets, namely, SCB, SCBH$\textsc{ii}$+CMB, and SCBH$\textsc{ii}$$H_{0}$+CMB, $b=0$ is outside $2\sigma/3\sigma$ confidence interval of model predicted $b$. On account of AIC, BIC and/or DIC, for these three data sets, the results regarding $b$, becomes important only for former two data sets. This model estimates $H_{0}$ values $1\sigma-2\sigma$ compatible with $H_{0,\,\rm{Planck}}$, when SH0ES prior is not considered whereas with SH0ES prior, median values of $H_{0}$ are $\sim$ 69.13-71.33, and are $1\sigma-2\sigma$ compatible with  $H_{0,\,\rm{Planck}}$ and/or $H_{0,\,\rm{SH0ES}}$.

%\textbf{Constraints on Rln Model:}
\subsection{Constraints on $f(R) = R +\alpha\ln(R)- \beta$ model}
%%%%%%%%%%%%%%%%%%%%%%%%%%%%%%%%%%%%%%%%%%%%%%%%%%%%%%%%%%%%%%%%%%%%%%%%%%%
\begin{figure*}
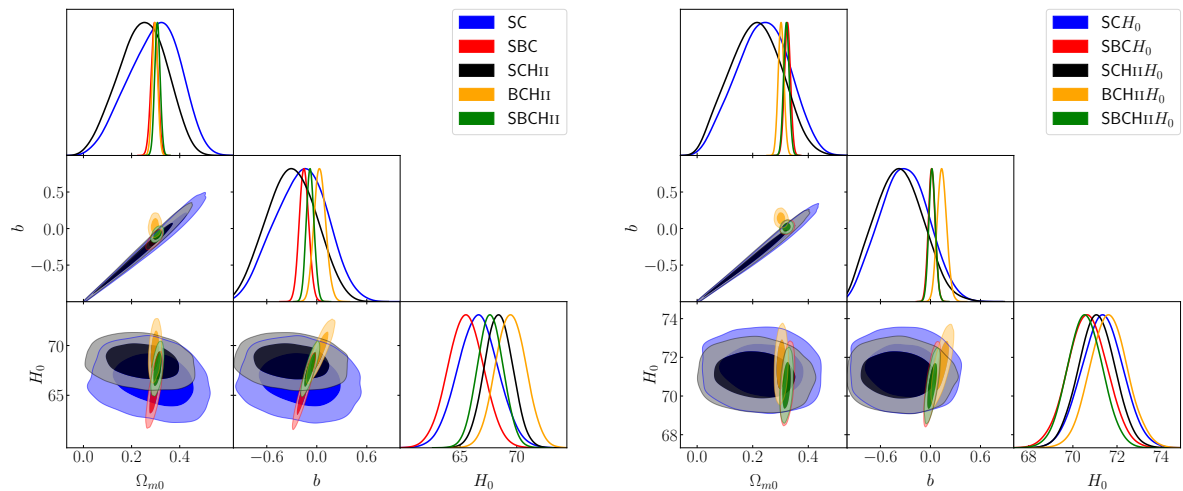

\imgtwo{RbetaRn_Palatini_PUB_Five-in-One_fR_GetDist_samples.pdf}\hfill
\imgtwo{RbetaRn_SH0ES_Palatini_PUB_Five-in-One_fR_GetDist_samples.pdf}
\caption{The $f(R) = R +\alpha\ln(R)- \beta$ model (\emph{left:} without SH0ES prior on $H_{0}$, \emph{right:} with SH0ES prior for $H_{0}$): the posterior probability distribution plots of fitted parameters. Figure legends show the colour correspondence for different data set combinations. The 1$\sigma$ (68.26 per cent) and 2$\sigma$ (95.44 per cent) confidence intervals are represented by darker and lighter shades of colours, respectively.}
\label{RlnRdist_Pal}
\end{figure*}
\begin{figure*}
\imgtwo{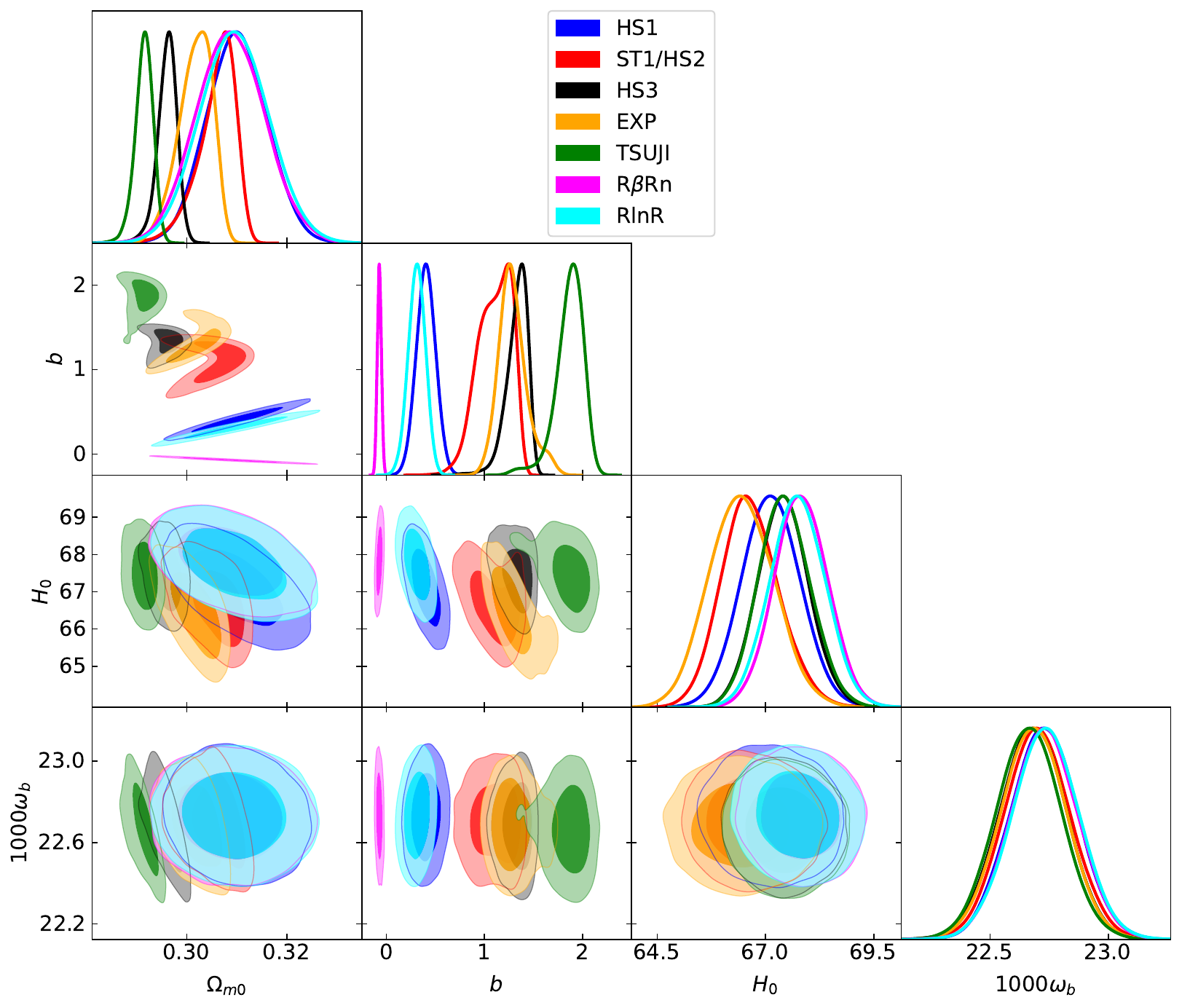}\hfill
\imgtwo{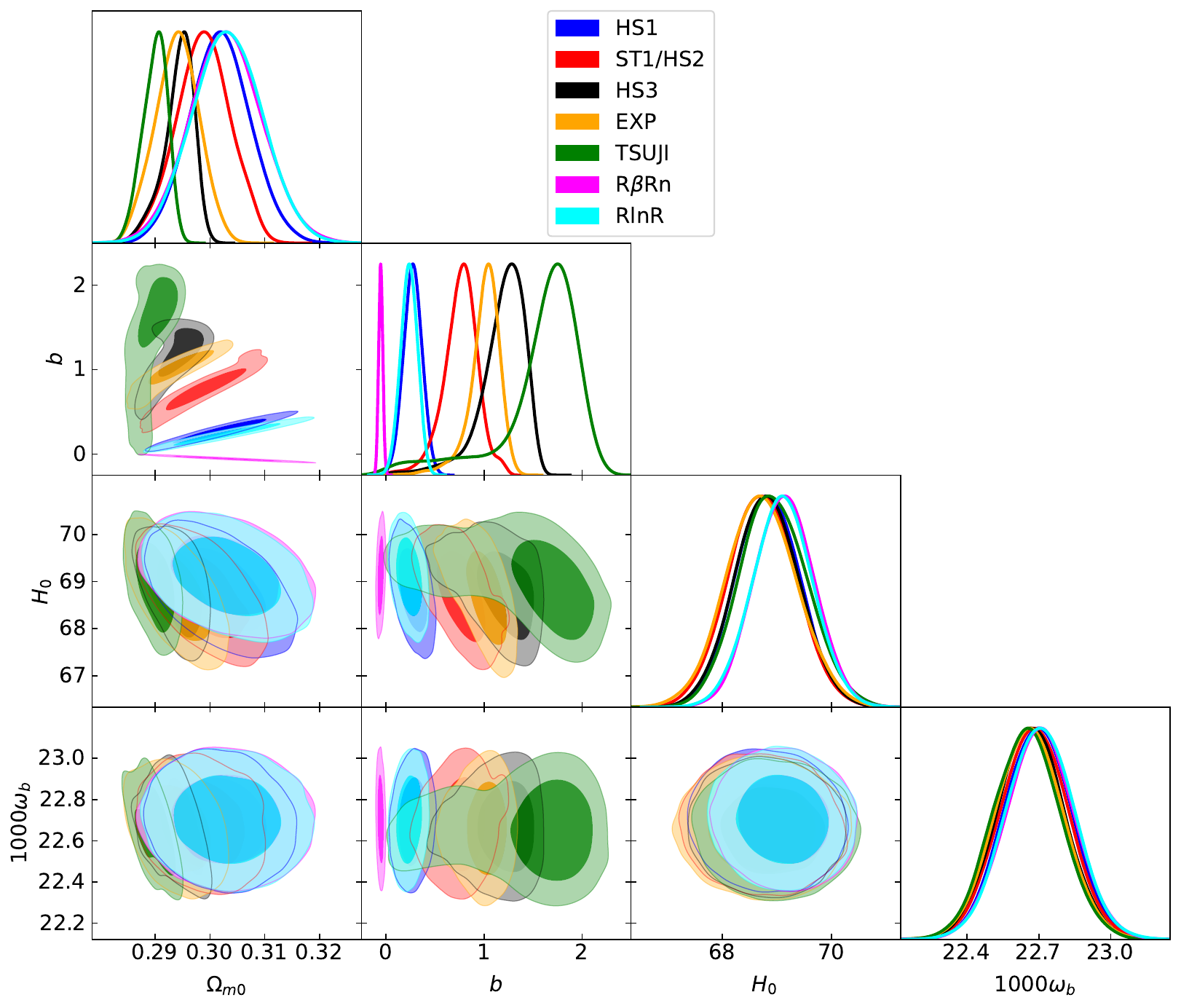}
\caption{For all models (\emph{left:} without SH0ES prior on $H_{0}$, \emph{right:} with SH0ES prior for $H_{0}$): the posterior probability distribution plots of fitted parameters for the data sets SBCH\textsc{ii}+CMB and SBCH\textsc{ii}$H_{0}$+CMB.. Figure legends show the colour correspondence for different data set combinations. The 1$\sigma$ (68.26 per cent) and 2$\sigma$ (95.44 per cent) confidence intervals are represented by darker and lighter shades of colours, respectively.}
\label{CMBdist_Pal}
\end{figure*}

%%%%%%%%%%%%%%%%%%%%%%%%%%%%%%%%%%%%%%%%%%%%%%%%%%%%%%%%%%%%%%%%%%%%%%%%%%%
This model has been studied earlier in \cite{Fay:2007gg} and was shown to be able to produce latter three phase of the Universe and so we have investigated this model with latest data sets.
The quantitative results from fitting Eq. \ref{Eq:RlnR} to all 12 data sets mentioned earlier, are presented in Tables \ref{resultstable_Pal} and \ref{resultsH0table_Pal}, for data sets without and with SH0ES prior for $H_{0}$, respectively. For the data sets excluding and including CMB data,  the posterior probability distribution of parameters are shown in the Figs. \ref{RlnRdist_Pal}, and \ref{CMBdist_Pal}, respectively. Like the last model, here also the constraints on $\Omega_{m0}$ are poor for the data sets without BAO data, though $1\sigma-2\sigma$ compatible with $\Omega_{m0,\,\rm{Planck}}$, which can be seen from Fig. \ref{Om0tension_Pal} and above mentioned tables and figures. The only data sets which provides $b=0$ as outside of $2\sigma-3\sigma$ limits of model predicted median value of $b$ (i.e. a case of being significantly distinguishable from the $\Lambda$CDM model), are SCBH$\textsc{ii}$+CMB and SCBH$\textsc{ii}$$H_{0}$+CMB. Without SH0ES prior, the model predicted median values of $H_{0}$ are $1\sigma-2\sigma$ compatible with $H_{0,\,\rm{Planck}}$, whereas with SH0ES priors they are $2\sigma-3\sigma$ compatible with $H_{0,\,\rm{SH0ES}}$ and/or $H_{0,\,\rm{Planck}}$.

%%%%%%%%%%%%%%%%%%%%%%%%%%%%%%%%%%%%%%%%%%%%%%%%%%%%%%%%%%%%%%%%%%%%%%%%%%%
%%%%%%%%%%%%%%%%%%%%%%%%%%%%%%%%%%%%%%%%%%%%%%%%%%%%%%%%%%%%%%%%%%%%%%%%%%%%
%%%%%%%%%%%%%%%%%%%%%%%%%%%%%%%%%%%%%%%%%%%%%%%%%%%%%%%%%%%%%%%%%%%%%%%%%%%%
%%%%%%%%%%%%%%%%%%%%%%%%%%%%%%%%%%%%%%%%%%%%%%%%%%%%%%%%%%%%%%%%%%%%%%%%%%%%
%%%%%%%%%%%%%%%%%%%%%%%%%%%%%%%%%%%%%%%%%%%%%%%%%%%%%%%%%%%%%%%%%%%%%%%%%%%%
\section{Model Comparison}
\label{comparison}
The standard statistical tools, more frequently used in cosmology for the purpose of
assessing quality of model fitting and model comparisons, are the reduced chi-square
statistics ($\chi^{2}_{\nu}$), the Akaike Information Criterion (AIC, \cite{AIC}), the Bayesian Information Criterion (BIC, \cite{BIC}), and the Deviance Information Criterion (DIC, \cite{Spiegelhalter2002}) (also see \cite{Liddle:2004nh, Liddle:2007fy, Burnham2004}). We have presented these quantities and other such related quantities in the last eight columns of Tables \ref{resultstable_Pal} and \ref{resultsH0table_Pal}. While the reduced chi-square statistics is simply
$\chi^{2}_{\nu}=\chi^{2}_{\rm min}/\nu$, the AIC, the BIC and the DIC are defined by the following equations
\begin{equation}
{\rm AIC} = -2\ln\mathcal{L}_{\rm max} + 2k\,,
\label{AICeq}
\end{equation}
\begin{equation}
{\rm BIC} = -2\ln\mathcal{L}_{\rm max} + k\ln{N}\,,
\label{BICeq}
\end{equation}
and,
\begin{equation}
{\rm DIC} = \chi^{2}(\hat{\theta}) + 2p_{\mathrm{D}}\,.
\label{DICeq}
\end{equation}
In the above definitions, the number of degrees of freedom, $\nu$, is obtained by subtracting
the number of model parameters ($k$) from the total number of data points ($N$). The minimum value of $\chi^{2}$, $\chi^{2}_{\rm min}$, is related to the maximum likelihood ($\mathcal{L}_{\rm max}$) as $\chi^{2}_{\rm min} = -2\mathcal{L}_{\rm max}$. In the definition of DIC, $p_{\mathrm{D}}=\overline{\chi^{2}(\theta)} - \chi^{2}(\hat{\theta})$ is called `an effective
number of parameters of a model' or `Bayesian complexity', where an overbar denotes
the mean value of the corresponding quantity. While in general $\hat{\theta}$ is to be taken as an array of mean of parameter values, but if the likelihood has definite peak then in that case an
array of the maximum likelihood estimates of parameters would be a more appropriate
choice for $\hat{\theta}$ \cite{Liddle:2007fy}. It has been observed that the DIC $\rightarrow$ the AIC, when the parameters
of a model are well-constrained and/or in case of large data set (both of these situations
seem applicable in present work).
\par
Although a model with the lowest value of $\chi^{2}_{\rm min}$ is considered to be more favoured by the data but still considerations of AIC, BIC, and DIC are imperative. In general,
for a nested model, as the number of parameters increases, so does the corresponding
value of $\chi^{2}_{\rm min}$ decreases. But the guiding principle in this context is `the principle of Occam's razor', which demands a trade-off between the quality of fit (i.e. lower value of $\chi^{2}_{\rm min}$) and the model complexity (i.e. number of parameters of a model). Actually taking this principle in consideration the AIC, the BIC and the DIC are defined.
\par
To make qualitative statements regarding preference for a model (say 2) over
another reference model (say 1), usually we do employ $\Delta{X} = X_{2} - X_{1}$ , where $X$
is AIC, DIC or BIC. A rule of thumb which is common in practice to indicate the
degree of strength of the evidence with which the model 2 (model 1) is favoured over the
other, is as follows: (i) $0<\Delta{X}\leqslant 2$ ($-2\leqslant\Delta{X}<0$): weak evidence, (ii) $2<\Delta{X}\leqslant 6$ ($-6\leqslant\Delta{X}<-2$): positive evidence, (iii) $6<\Delta{X}\leqslant 10$ ($-10\leqslant\Delta{X}<-6$): strong
evidence, (iv) $\Delta{X}> 10$ ($\Delta{X}<-10$): very strong evidence. 
\par
For all the models studied in this work, we can observe that corresponding to the data sets BCH$\textsc{ii}(H_{0})$, we are getting a high value (i.e. not so close to 1) of the $\chi^{2}_{\rm{red}}$. Therefore, in our subsequent discussions of model comparison we will discard the case of these two data sets for all the models. As mentioned earlier, the inclusion of data sets BCH$\textsc{ii}(H_{0})$, were only to demonstrate that while one can have a total of $2^{6}-1$ data set combinations from six data sets, but not all these possible combinations are worth exploring.
\par
For the data sets without SH0ES prior, based on $\Delta$AICs and $\Delta$DICs, we can observe the cases of weak to very strong evidence in support of the models HS1, ST1/HS2, HS3, and EXP, whereas for the models TSUJI, R$\beta$Rn and RlnR, there are cases of weak to strong evidence 
(where the $\Lambda$CDM model is taken as a reference model). When we consider $\Delta$BICs for the assessment, for the above mentioned cases, the overall support for $f(R)$ models deteriorates, now only weak to strong supports and even weak to positive evidence against.
\par
With SH0ES prior for $H_{0}$ included data sets, the overall support for $f(R)$ models against the $\Lambda$CDM model, reduces drastically, with the maximum support being strong only and that too for fewer cases, when we use $\Delta$AICs and/or $\Delta$DICs as criteria. From $\Delta$BICs, we can infer that there are cases of weak, positive and a few cases of strong, supports in favour of $f(R)$ models as well as weak to even a few cases of strong support against the $f(R)$ models.
\par 
One important result of this work is that whenever there is strong or very strong support for any $f(R)$ model (based on $\Delta$AICs, $\Delta$DICs, and/or $\Delta$BICs) is there, the same cases also correspond to $b=0$ being only marginally or very marginally contained in the model predicted values of $b$. That is to say, some of the data sets explored in this study supports $f(R)$ model significantly distinguishable from the standard $\Lambda$CDM model.
%%%%%%%%%%%%%%%%%%%%%%%%%%%%%%%%%%%%%%%%%%%%%%%%%%%%%%%%%%%%%%%%%%%%%%%%%%%%
%%%%%%%%%%%%%%%%%%%%%%%%%%%%%%%%%%%%%%%%%%%%%%%%%%%%%%%%%%%%%%%%%%%%%%%%%%%%
%%%%%%%%%%%%%%%%%%%%%%%%%%%%%%%%%%%%%%%%%%%%%%%%%%%%%%%%%%%%%%%%%%%%%%%%%%%%
%%%%%%%%%%%%%%%%%%%%%%%%%%%%%%%%%%%%%%%%%%%%%%%%%%%%%%%%%%%%%%%%%%%%%%%%%%%%
%\section{Radiation, Matter and Late-time Accelerated Expansion Epochs of the Universe}
\section{Radiation dominated, matter dominated, and late-time accelerated expansion: epochs of the Universe}
\label{accel}
It is expected from any viable model of the Universe to reproduce the four phases
of evolution, namely, the early accelerated expansion phase (inflation), followed by the
radiation dominated epoch, followed by matter dominated epoch, finally to the current
accelerated expansion. Although early inflation is beyond the scope of this work, we will
explore latter three periods with the obtained constrains on parameters of $f(R)$ models
studied in this work.
\par
\begin{figure*}
\imgtwo{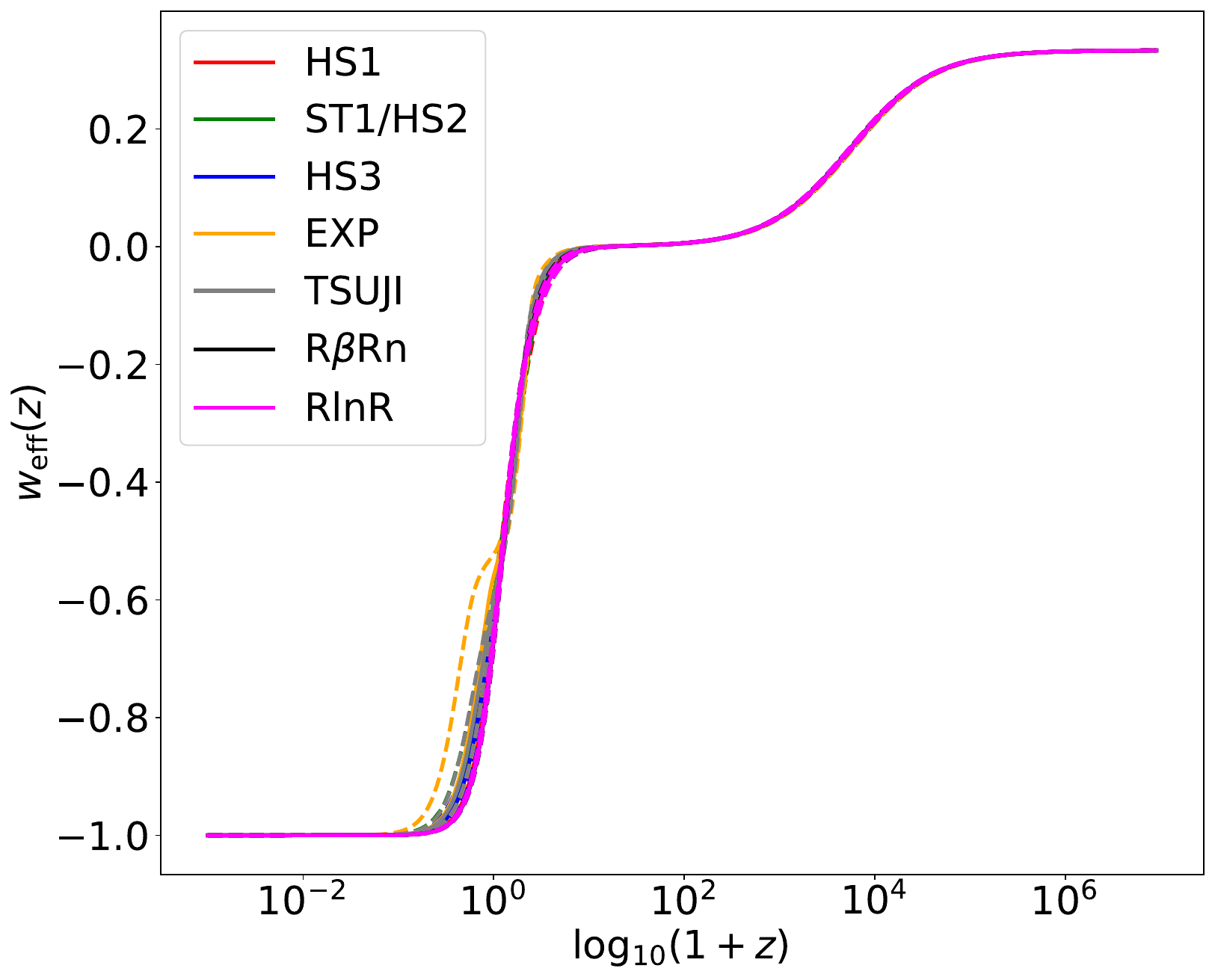}\hfill
\imgtwo{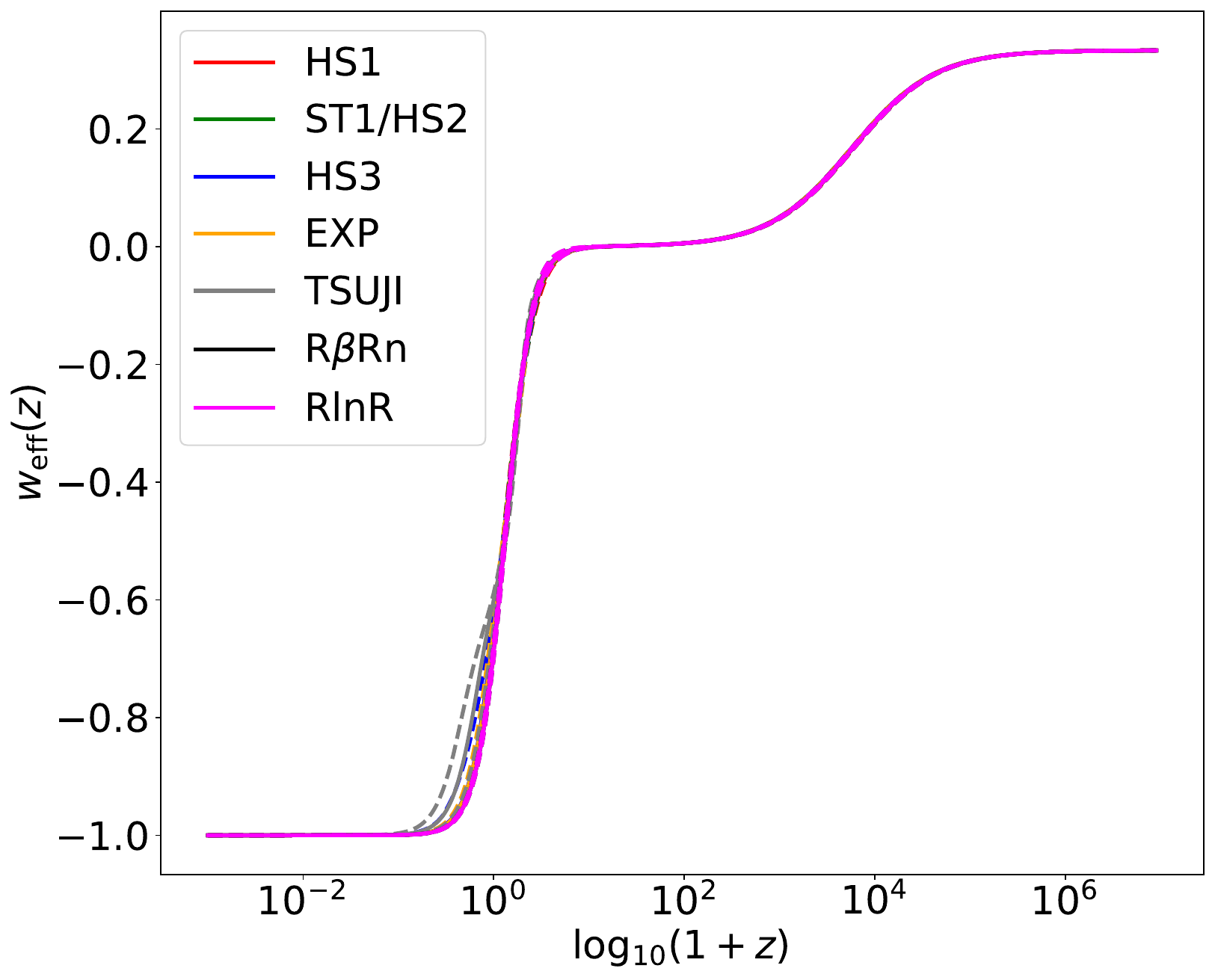}
\caption{Evolution of the total EoS parameter with redshift: for all $f(R)$ models
obtained from the data sets SBCH\textsc{ii} (\emph{left}) and SBCHii$H_{0}$ (\emph{right}). Solid lines of various
colors denote median values for different models, as indicated in the legends, while
dashed lines of corresponding colors represent the 1$\sigma$ confidence interval.}
\label{weff_evol_Pal}
\end{figure*}

The evolution of the Universe, transitioning through radiation-dominated, matter-
dominated to final de-Sitter phase (in distant future), is characterized by the total
equation-of-state (EoS) parameter ($w_{\rm eff}$) as a function of redshift, exhibiting values
$w_{\rm eff}\sim 1/3, 0$, and $-1$ in these respective phases. Whereas for the current accelerated
expansion phase we expect $w_{\rm eff}\sim 0.7$ (at $z \simeq 0$) or more appropriately we should
explore $w_{\rm DE}$ for this purpose. From Fig. \ref{weff_evol_Pal} we can see that all the models considered in present work are able to reproduce aforementioned three phases of the evolution of the Universe, both for the data sets without and with SH0ES prior on $H_{0}$ . Also the model
predicted values of $w_{\rm tot0}$ (i.e. $w_{\rm eff}$ at $z = 0$) are reasonably close to 0.7, as can be seen from the eighth column of Tables \ref{resultstable_Pal} and \ref{resultsH0table_Pal}, for data sets without and with SH0ES prior,
respectively.
\par
\begin{figure}%[htp]
\centering
\includegraphics[scale=0.4]{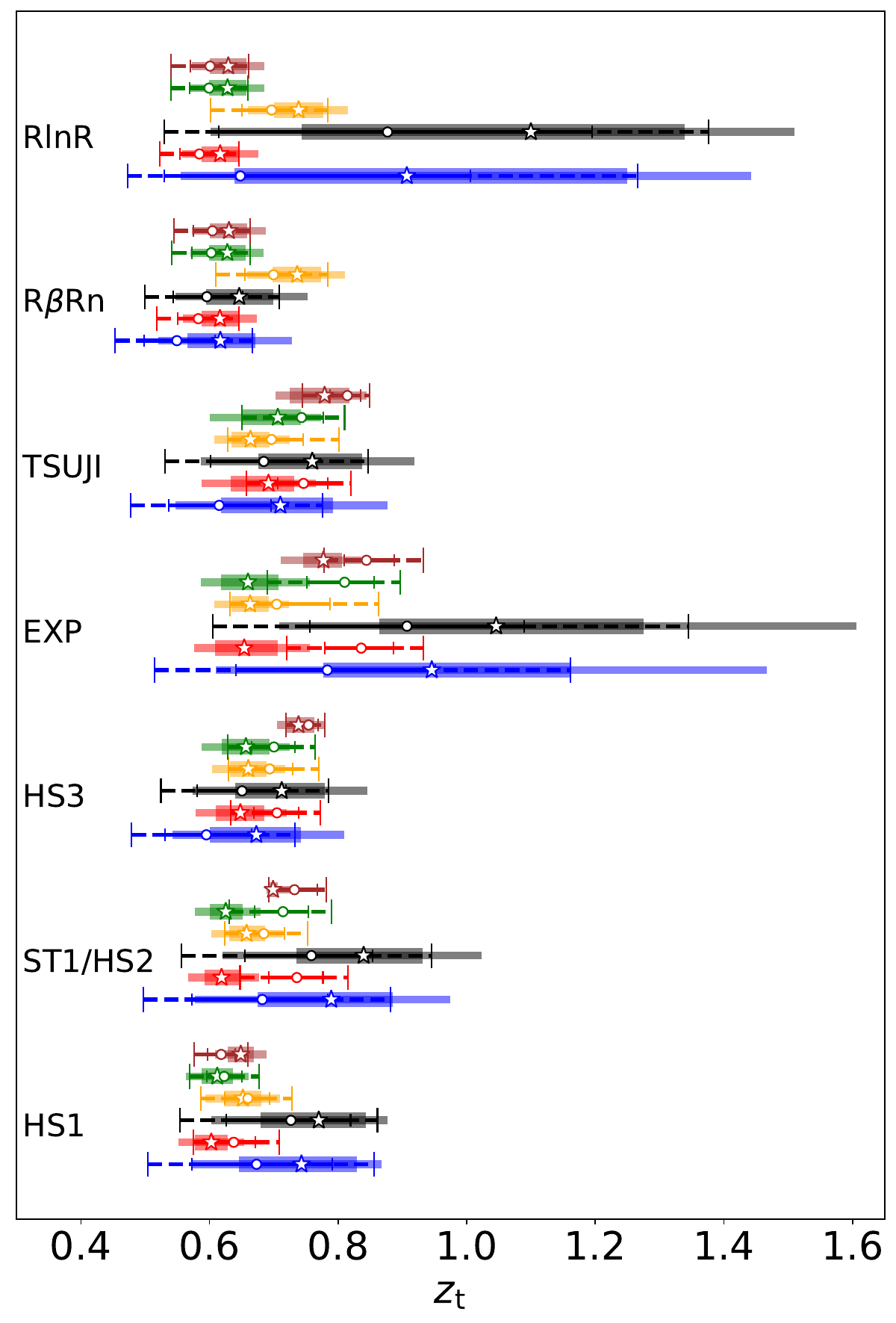}
\caption{The transition redshift ($z_{\rm t}$) variations across models and data sets are depicted in this figure. The color coding for various data sets corresponds identically to that in any of the parameter distribution plots (e.g., Fig. \ref{HS1dist_Pal}) or can
be referenced from the second paragraph of Section \ref{acc}. The markers `blank star' and `circle' denote median values with and without SH0ES prior for $H_{0}$, respectively. In instances where SH0ES prior wasn't applied, we depicted 1$\sigma$ (68.26 per cent, with shorter caps) and 2$\sigma$ (95.44 per cent with longer caps) confidence intervals using colored continuous/dashed lines. However, for cases with SH0ES prior, we represented 1$\sigma$ and 2$\sigma$ confidence intervals using thick and thin horizontal colored bars, respectively.}
\label{z_trans_Pal}
\end{figure}

Now let us explore the nature of late-time accelerated expansion, and  prediction
for the distant future. The relevant derived quantities that describe the current accelerated
expansion of the Universe is the deceleration parameter ($q(z)$), defined as 
$q(z) \equiv -a\ddot{a}/\dot{a}^{2}=-\ddot{a}/(H^{2}a)$, indicates whether and when the expansion of the Universe
experiences acceleration ($q < 0$) or deceleration ($q > 0$). Many measurements of
$q(z)$, which are independent of any specific cosmological model (in addition to model-
dependent ones), indicate that $q(z = 0) < 0$ and $q(z > z_{\rm t}) > 0$ \cite{Rani:2015lia,Jesus:2019nnk}. Here, $z_{\rm t}$ is
referred to as the transition redshift, signifying the redshift at which the Universe transitioned 
from a phase of decelerated expansion to the current accelerated expansion. As can be
seen from Fig. \ref{z_trans_Pal} and sixth column of Tables  \ref{resultstable_Pal} and \ref{resultsH0table_Pal}, that the model predicted values
of transition redshift are mostly within the range $z_{\rm t}\sim 0.5\,-\,1.0$, and these values are in agreement with model independent estimates from \cite{Rani:2015lia} and \cite{Jesus:2019nnk}. 
\par
\begin{figure*}
\imgtwo{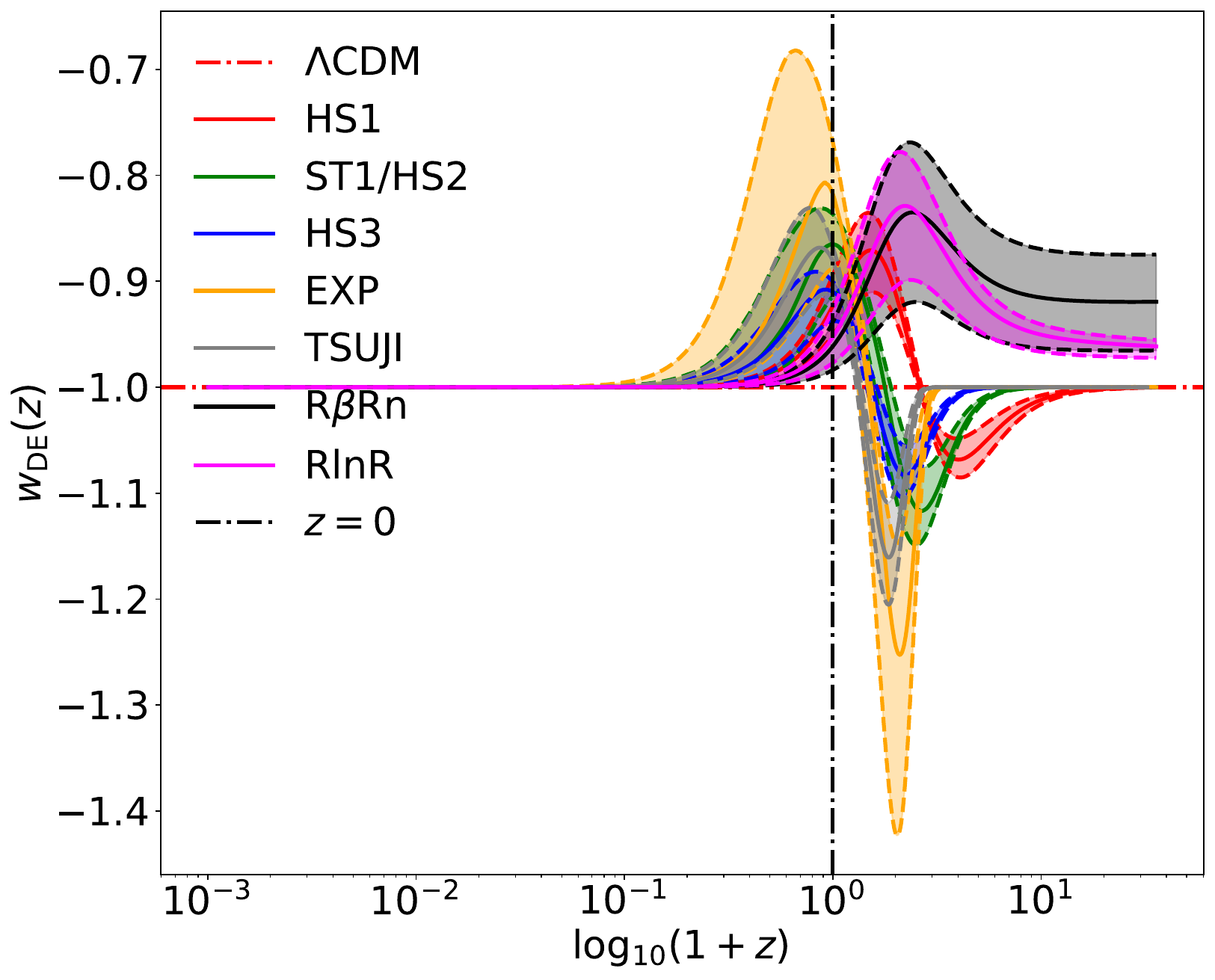}\hfill
\imgtwo{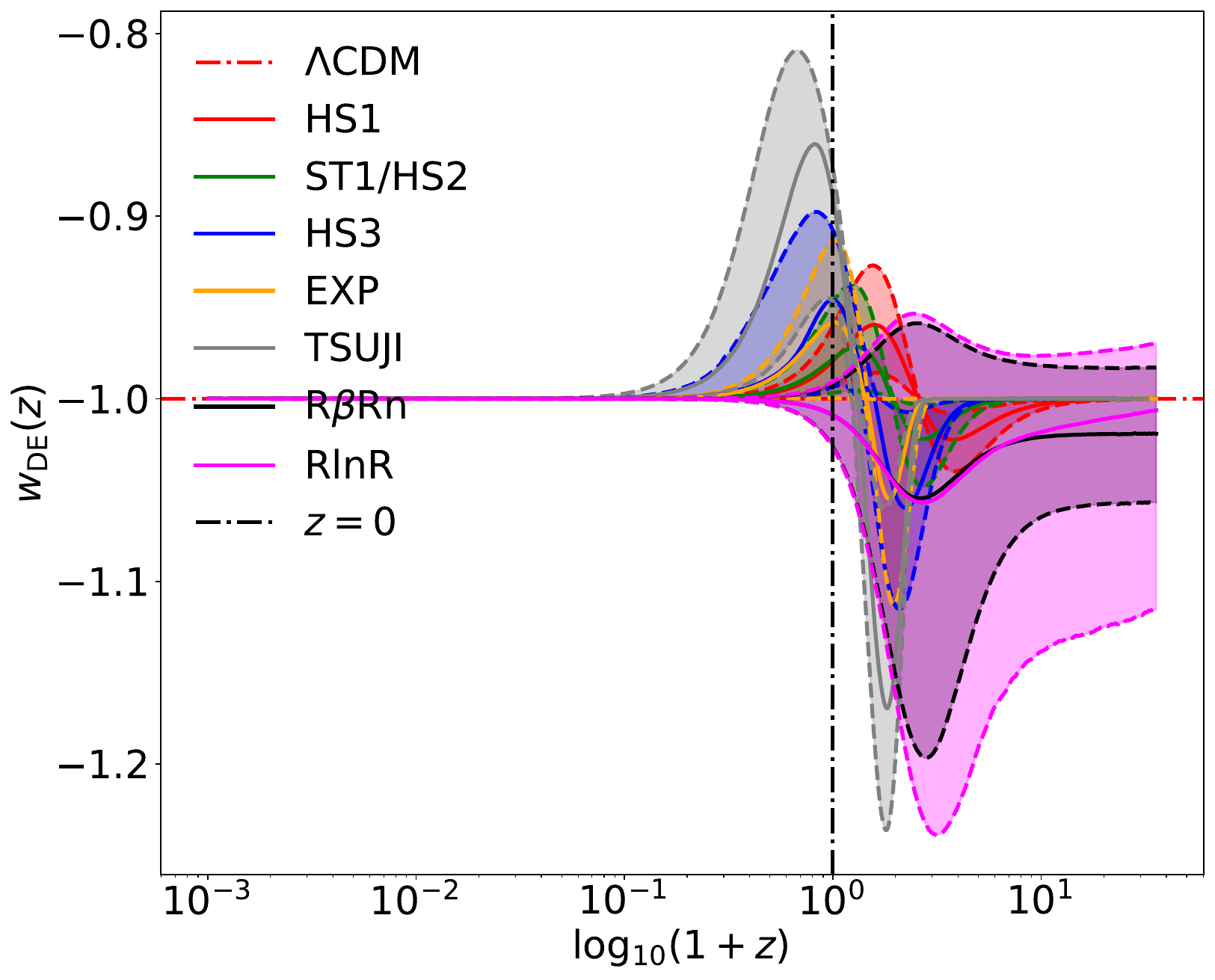}
\caption{Evolution of the `effective geometric dark energy' parameter with redshift:
for all $f(R)$ models obtained from the data sets SBCH\textsc{ii} (\emph{left}) and SBCH\textsc{ii}$H_{0}$ (\emph{right}).
Solid lines of various colors denote median values for different models, as indicated in
the legends, while dashed lines containing shaded areas of corresponding colors represent
the 1$\sigma$ confidence interval.}
\label{wDE_evol_Pal}
\end{figure*}

In Fig. \ref{wDE_evol_Pal} we have plotted the evolution of $w_{\rm DE}(z)$ (define in Eq. \ref{Eq:wDE_Pal} by the name $w_{\rm geo}$) for $z \sim 30$ (past) to distant future $z \sim -1$, from the constraints on $f(R)$
models from data sets SCBH\textsc{ii} and 
SCBH\textsc{ii}$H_{0}$. These $w_{\rm DE}(z)$ curves, for all the models,
are supposed to converge to $-1$ in the distant past for successfully producing matter
dominated epoch, the trends for which are clearly visible in Fig. \ref{wDE_evol_Pal}. Except for the
models R$\beta$Rn and RlnR, corresponding to data set SCBH\textsc{ii}, for all models we can observe the crossing of phantom-divide 
line ($w_{\rm DE}(z)=-1$) by model predicted $w_{\rm DE}(z)$, in recent
past from the so-called phantom 
regime ($w_{\rm DE}(z)<-1$) to the so-called quintessence
regime ($w_{\rm DE}(z)>-1$). Also can we see that $w_{\rm DE}(z)$ all the $f(R)$ models are converging
to a value of $-1$ in distant future, that is, successfully reproducing the de-Sitter phase
of the Universe.

%%%%%%%%%%%%%%%%%%%%%%%%%%%%%%%%%%%%%%%%%%%%%%%%%
%%%%%%%%%%%%%%%%%%%%%%%%%%%%%%%%%%%%%%%%%
%%%%%%%%%%%%%%%%%%%%%%%%%%%%%%%%%%%%%%%%%
%%%%%%%%%%%%%%%%%%%%%%%%%%%%%%%%%%%%%%%%%%%%%%%%%%%%%%%%%%%%%%%%%%%%%%%%%%%%
%%%%%%%%%%%%%%%%%%%%%%%%%%%%%%%%%%%%%%%%%%%%%%%%%%%%%%%%%%%%%%%%%%%%%%%%%%%%
%%%%%%%%%%%%%%%%%%%%%%%%%%%%%%%%%%%%%%%%%%%%%%%%%%%%%%%%%%%%%%%%%%%%%%%%%%%%
%%%%%%%%%%%%%%%%%%%%%%%%%%%%%%%%%%%%%%%%%%%%%%%%%%%%%%%%%%%%%%%%%%%%%%%%%%%%
\section{Concluding Remarks}
\label{concl}

With SNIa data from Pantheonplus compilation, CC data, BAO data, H \textsc{ii} starburst
galaxies data, distance priors of the CMBR and local measurement of $H_{0}$, we have
investigated seven $f(R)$ models in the Palatini formalism, namely, the 
Hu--Sawicki model ($n_{_{\rm HS}} = 1, 3$), the Starobisnky model ($n_{_{\rm S}} = 1$), the exponential model, the Tsujikawa
model, $f (R) = R - \beta/R^{n}$, and $f(R) = R - \beta - \alpha\ln(R)$. To address the so-called \emph{Hubble} tension we included data set combinations without vis-a-vis with SH0ES prior on $H_{0}$. Although in their original forms, these $f(R)$ models appear unrelated to the
standard $\Lambda$CDM model, but when re-parameterized in terms of deviation parameter
($b$), it becomes apparent that all these models reduce to the $\Lambda$CDM model with $b \rightarrow 0$.
Therefore, all these $f(R)$ models are characterized with three parameters: ($\Omega_{m0},\,b,\,H_{0}$), except when studied with CMB data, we need one more parameter, namely, $\omega_{b0}$.
\par
In most cases, the \emph{Planck} constrained value of matter density at present epoch,
$\Omega_{m0,{\rm Planck}} = 0.315 \pm 0.007$, lie within $1\sigma-2\sigma$ limits of the model predicted median
values of $\Omega_{m0}$ or vice-versa, both for the data sets with or without SH0ES prior on $H_{0}$.
Same is true for $H_{0,{\rm Planck}} = 67.4 \pm 0.5$ vis-a-vis the model predicted median values of
$H_{0}$, but for the data sets without SH0ES prior. For the data sets with SH0ES prior, the
SH0ES measured value $H_{0,{\rm SH0ES}} = 73.04 \pm 1.04$ fall within $2\sigma -3\sigma$ limits of the model
predicted median values of $H_{0}$ but on the higher side, when CMB data is not included.
Whereas with inclusion of CMB data, the tensions between model predicted values of $H_{0}$ and $H_{0,{\rm SH0ES}}$ enhance, as for these cases, even with SH0ES prior, the $H_{0}$ values are
more nearer to $H_{0,{\rm Planck}}$.
\par
The derived parameters essential for characterizing the accelerated expansion of
the Universe, namely, $z_{\rm t}$, $w_{\rm eff,0}$, and $w_{\rm DE,0}$, estimated in this study, are closer to 
and compatible with their estimations from cosmological model-independent methods by earlier
studies. This work predicts that the transition from decelerated phase of expansion into
accelerated phase of expansion happened around $z_{\rm t}\sim0.5-1$, according to almost all
the models investigated here. Additionally, the model predicted values of $w_{\rm DE,0}$ fall in
the quintessential region, having recently transitioned from the phantom region, with
exception of a few cases.
\par
Our analysis unveils instances across all examined $f(R)$ models where the deviation
parameter $b$ significantly differs from zero. It is noteworthy that these instances also
coincide with situations where analyses based on $\Delta$AIC, $\Delta$DIC, and/or $\Delta$BIC strongly
favor the $f(R)$ models over the $\Lambda$CDM model. Previous studies have generally shown
weak or positive support for $f(R)$ models based on $\Delta$AIC, $\Delta$DIC, and/or $\Delta$BIC.
However, this work demonstrates cases of (very) strong support for $f(R)$ models.
\par
With cases of (very) strong support for $f(R)$ models, non-zero $b$ supported by AIC, DIC and/or BIC, and the quantities $z_{\rm t}$, $w_{\rm eff,0}$, and $w_{\rm DE,0}$ being compatible with
their model-independent predicted values, we find that the cosmological data analyzed in this study do not yet warrant dismissing the $f(R)$ models in Palatini formalism. Instead, our study advocates for further exploration of these models within the Palatini formalism as plausible contenders for explaining the evolutionary trajectory of the Universe, with observed cosmological data sets to come in the future.
%%%%%%%%%%%%%%%%%%%%%%%%%%%%%%%%%%%%%%%%%%%%%%%%%%%%%%%%%%%%%%%%%%%%%%%%%%%%
%%%%%%%%%%%%%%%%%%%%%%%%%%%%%%%%%%%%%%%%%%%%%%%%%%%%%%%%%%%%%%%%%%%%%%%%%%%%
%%%%%%%%%%%%%%%%%%%%%%%%%%%%%%%%%%%%%%%%%%%%%%%%%%%%%%%%%%%%%%%%%%%%%%%%%%%%
%%%%%%%%%%%%%%%%%%%%%%%%%%%%%%%%%%%%%%%%%%%%%%%%%%%%%%%%%%%%%%%%%%%%%%%%%%%%
\section*{Data Availability}
The observed cosmological data such as SNIa, CC, BAO, H \textsc{ii} starburst galaxies, CMB
distance priors, and local measurement of $H_{0}$ are publicly available --- the references to
which are cited in the text. The simulated data sets generated in this work are available
from the corresponding author upon reasonable request.

\section*{Acknowledgements}
I would like to thank HoD, Dept. of Comp. Sc., RKMVERI, for providing computational
facilities. I would also like to thank Dr. Abhijit Bandyopadhyay for helpful discussions.
\\\\
\section*{References}

\bibliographystyle{JHEP}
\bibliography{references_pal_metric}

\end{document}